\documentclass[prd,twocolumn,notitlepage,superscriptaddress,nofootinbib,preprintnumbers]{revtex4-1}

\usepackage{amsmath,amssymb}
\usepackage{amsfonts}
\usepackage{graphicx}
\usepackage[utf8]{inputenc}
\usepackage{slashed}
\usepackage{multirow}
\usepackage[normalem]{ulem}
\graphicspath{ {figures/} }
\usepackage{hyperref}
\usepackage{pgffor}
\hypersetup{colorlinks, linkcolor = [rgb]{0,0.0,0.75}, citecolor = [rgb]{0,0.0,0.75}, urlcolor = [rgb]{0,0.0,0.75}}

\makeatletter
\g@addto@macro\bfseries{\boldmath}
\makeatother

\newcommand{\MSb}{\overline{\text{MS}}}

\newcommand{\numax}{\nu_{\textrm{max}}}
\newcommand{\zmax}{z_{\textrm{max}}}
\newcommand{\Nmax}{N_{\textrm{max}}}
\newcommand{\Fbar}{\overline{\mathcal{F}}}
\newcommand{\Hbar}{\overline{\mathcal{H}}}
\newcommand{\Ebar}{\overline{\mathcal{E}}}
\newcommand{\Fval}{\overline{\mathcal{F}}_v}
\newcommand{\Hval}{\overline{\mathcal{H}}_v}
\newcommand{\Eval}{\overline{\mathcal{E}}_v}
\newcommand{\Fvs}{\overline{\mathcal{F}}_{v2s}}
\newcommand{\Hvs}{\overline{\mathcal{H}}_{v2s}}
\newcommand{\Evs}{\overline{\mathcal{E}}_{v2s}}
\newcommand{\chidof}{\chi^2/{\rm dof}}

\begin{document}
\preprint{LA-UR-24-23903}

\title{Generalized parton distributions from the pseudo-distribution approach on the lattice}

\author{Shohini Bhattacharya}
\affiliation{Theoretical Division, Los Alamos National Laboratory, Los Alamos, New Mexico 87545, USA}
\author{Krzysztof~Cichy}
\email{kcichy@amu.edu.pl}
\affiliation{Faculty of Physics, Adam Mickiewicz University, ul.\ Uniwersytetu Pozna\'nskiego 2, 61-614 Poznań, Poland}
\author{Martha Constantinou}
\affiliation{Department of Physics,  Temple University,  Philadelphia,  PA 19122 - 1801,  USA}
\author{Andreas Metz}
\affiliation{Department of Physics,  Temple University,  Philadelphia,  PA 19122 - 1801,  USA}
\author{Niilo Nurminen}
\affiliation{Faculty of Physics, Adam Mickiewicz University, ul.\ Uniwersytetu Pozna\'nskiego 2, 61-614 Poznań, Poland}
\author{Fernanda Steffens}
\affiliation{Institut f\"ur Strahlen- und Kernphysik, Rheinische Friedrich-Wilhelms-Universit\"at Bonn,\\ Nussallee 14-16, 53115 Bonn}

\date{\today}

\begin{abstract}
Generalized parton distributions (GPDs) are key quantities for the description of a hadron's three-dimensional structure.
They are the current focus of all areas of hadronic physics -- phenomenological, experimental, and theoretical, including lattice QCD.
Synergies between these areas are desirable and essential to achieve
precise quantification and understanding of the structure of, particularly, nucleons, as the basic ingredients of matter.
In this paper, we investigate, for the first time, the numerical implementation of the pseudo-distribution approach for the extraction of zero-skewness GPDs for unpolarized quarks.
Pseudo-distributions are Euclidean parton correlators computable in lattice QCD that can be perturbatively matched to the light-cone parton distributions of interest.
Although they are closely related to the quasi-distributions and come from the same lattice-extracted matrix elements, they are, however, subject to different systematic effects.
We use the data previously utilized for quasi-GPDs and extend it with other momentum transfers and nucleon boosts, in particular a higher one ($P_3=1.67$ GeV) with eight-fold larger statistics than the largest one used for quasi-distributions ($P_3=1.25$ GeV).
We renormalize the matrix elements with a ratio scheme and match the resulting Ioffe time distributions to the light cone in coordinate space.
The matched distributions are then used to reconstruct the $x$-dependence with a fitting ansatz.
We investigate some systematic effects related to this procedure, and we also compare the results with the ones obtained in the framework of quasi-GPDs.
Our final results involve the invariant four-momentum transfer squared ($-t$) dependence of the flavor non-singlet ($u-d$) $H$ and $E$ GPDs.
\end{abstract}

\maketitle

\section{Introduction}
\label{sec:intro}
Since the realization that nucleons possess an internal structure, physicists across theoretical, experimental, phenomenological, and lattice QCD communities have collectively endeavored to unravel and quantify this structure. Synergies among these disciplines have been pivotal in the notable progress achieved thus far. However, the quest remains immensely complex. Precise quantitative descriptions often rely on the ``simplest'' functions, such as form factors (FFs) and specific types of parton distribution functions (PDFs). FFs and PDFs encapsulate our understanding in terms of functions of a single variable, namely the invariant four-momentum transfer squared ($-t$) and the longitudinal momentum fraction ($x$) of the partons. It is natural to extend this framework to encompass more general functions dependent on multiple variables. Generalized parton distributions (GPDs) represent a prominent example of such functions, where FFs and PDFs emerge as their moments or forward limits, respectively. Introduced nearly three decades ago~\cite{Mueller:1998fv, Ji:1996ek, Radyushkin:1996nd}, GPDs probe light-cone correlations between hadron states under momentum transfer. Beyond their dependence on momentum fraction, GPDs are sensitive to both the total momentum transfer and its longitudinal momentum component, characterized by the skewness variable ($\xi$).

The inclusion of momentum transfer enables a more comprehensive exploration of hadron structure. GPDs offer three-dimensional portrayals of hadrons~\cite{Burkardt:2000za, Ralston:2001xs, Diehl:2002he, Burkardt:2002hr}, granting access to parton angular momenta~\cite{Ji:1996ek} and insights into internal pressure and shear forces~\cite{Polyakov:2002wz, Polyakov:2002yz, Polyakov:2018zvc}. Recent discoveries have brought to light chiral and trace anomaly poles within GPDs, which may offer insights into phenomena such as mass generation, chiral symmetry breaking, and confinement~\cite{Tarasov:2020cwl, Tarasov:2021yll,Bhattacharya:2022xxw,Bhattacharya:2023wvy}. A plethora of review articles extensively explore the physics of GPDs, providing additional clarity on their significance and implications~\cite{Goeke:2001tz,Diehl:2003ny, Ji:2004gf, Belitsky:2005qn, Boffi:2007yc, Guidal:2013rya, Mueller:2014hsa, Kumericki:2016ehc}.

Experimental knowledge about GPDs is gleaned from hard exclusive scattering processes like deep virtual Compton scattering~\cite{Mueller:1998fv, Ji:1996ek, Radyushkin:1996nd, Ji:1996nm, Collins:1998be} and hard exclusive meson production~\cite{Radyushkin:1996ru,Collins:1996fb,Mankiewicz:1997uy}. However, extracting GPDs from these reactions in a model-independent manner is intricate, primarily due to the integration over the momentum fraction $x$ in observable quantities such as Compton form factors. Recent detailed analyses of this issue are available in Refs.~\cite{Bertone:2021yyz, Moffat:2023svr}. Parameterizing GPDs and fitting them from global experiments has been attempted in various studies; see, for instance, Refs.~\cite{Polyakov:2002wz,Guidal:2004nd,Goloskokov:2005sd,Mueller:2005ed,Kumericki:2009uq,Goldstein:2010gu,Kriesten:2021sqc,Hashamipour:2021kes,Guo:2022upw,Guo:2023ahv,Hashamipour:2022noy,Cuic:2023mki,Irani:2023lol}. However, such studies are challenging mainly due to the multi-dimensional nature of the GPDs and the sparsity of experimental data for some of the key processes. Therefore, acquiring information on GPDs directly from first principles in lattice QCD is highly desirable, especially concerning their dependence on $x$.

Lattice QCD has long explored hadron structure, initially focusing on moments of parton distributions expressed through matrix elements of local operators. While it is theoretically possible to reconstruct $x$-dependent distributions from moments, practical challenges arise due to the power-divergent mixing of higher moments and diminishing signal-to-noise ratio. Early attempts to overcome these obstacles were proposed in the 1990s and 2000s, but limited computing power hindered their application. The resurgence of interest came with Ji's seminal papers \cite{Ji:2013dva,Ji:2014gla}, which introduced quasi-distributions. These exploit the infrared structure of matrix elements while addressing mixing issues by comparing light-front and spatial correlations. This approach sparked extensive research and alternative proposals \cite{Chambers:2017dov,Radyushkin:2017cyf,Ma:2014jla,Ma:2017pxb}, reinvigorating earlier ideas. 
We refer to the reviews \cite{Cichy:2018mum,Radyushkin:2019mye,Ji:2020ect,Constantinou:2020pek,Cichy:2021lih,Cichy:2021ewm} for more details and to original theoretical and lattice papers, see, e.g., Refs.~\cite{Xiong:2013bka,Lin:2014zya,Gamberg:2014zwa, Alexandrou:2015rja,Ji:2015jwa,Ji:2015qla,Xiong:2015nua,Chen:2016utp,Alexandrou:2016jqi,Alexandrou:2017huk,Orginos:2017kos,Ishikawa:2017faj,Ji:2017oey,Radyushkin:2018cvn,Alexandrou:2018pbm,Chen:2018fwa,Alexandrou:2018eet,Liu:2018uuj,Karpie:2018zaz,Bhattacharya:2018zxi,Zhang:2018diq,Li:2018tpe,Braun:2018brg,Sufian:2019bol,Karpie:2019eiq,Liu:2019urm,Alexandrou:2019lfo,Wang:2019tgg,Chen:2019lcm,Izubuchi:2019lyk,Cichy:2019ebf,Wang:2019msf,Joo:2019jct,Radyushkin:2019owq,Bhattacharya:2019cme,Joo:2019bzr,Balitsky:2019krf,Son:2019ghf,Ma:2019agv,Sufian:2020vzb,Green:2020xco,Chai:2020nxw,Lin:2020ssv,Braun:2020ymy,Joo:2020spy,Bhattacharya:2020cen,Bhat:2020ktg,Bhattacharya:2020xlt,Zhang:2020gaj,Bhattacharya:2020jfj,Zhang:2020dbb,Fan:2020cpa,Chen:2020ody,Can:2020sxc,Gao:2020ito,Ji:2020brr,Alexandrou:2020zbe,Alexandrou:2020uyt,Bringewatt:2020ixn,Liu:2020rqi,DelDebbio:2020rgv,Alexandrou:2020qtt,Liu:2020krc,Hua:2020gnw,LatticePartonCollaborationLPC:2021xdx,Detmold:2021uru,Fan:2021bcr,Bhattacharya:2021boh,Karpie:2021pap,Karthik:2021sbj,Ji:2021znw,Alexandrou:2021oih,Li:2021wvl,Bhattacharya:2021moj,Egerer:2021ymv,HadStruc:2021wmh,Shanahan:2021tst,Alexandrou:2021bbo,Detmold:2021qln,CSSMQCDSFUKQCD:2021lkf,HadStruc:2021qdf,Balitsky:2021qsr,Chirilli:2021euj,Balitsky:2021cwr,Gao:2021dbh,Bhattacharya:2021oyr,Xu:2022krn,Son:2022qro,LPC:2022ibr,Chou:2022drv,Bhat:2022zrw,Deng:2022gzi,HadStruc:2022yaw,Su:2022fiu,Shastry:2022obb,Bhattacharya:2022aob,Ji:2022ezo,Xu:2022guw,Khan:2022vot,Gao:2022uhg,Yao:2022vtp,LPC:2022zci,Chu:2023jia,Deng:2023csv,Cichy:2023dgk,Chirilli:2023wii,Ji:2023pba,Bhattacharya:2023ays,Alexandrou:2023ucc,Bhattacharya:2023nmv,LatticePartonLPC:2023pdv,Hu:2023bba,Han:2023xbl,Delmar:2023agv,Bhattacharya:2023jsc, Dutrieux:2023zpy,Radyushkin:2023ref,Han:2024ucv}. 

The cited literature primarily focuses on PDFs as a starting point for direct lattice investigations into $x$-dependence. However, several studies also explore the extraction of GPDs. Matching papers for quasi-GPDs surfaced in 2015 \cite{Ji:2015qla,Xiong:2015nua}, with subsequent contributions \cite{Liu:2019urm,Yao:2022vtp}. Radyushkin extended the pseudo-PDF approach to GPDs \cite{Radyushkin:2019owq,Radyushkin:2019mye,Radyushkin:2023ref}. Model investigations \cite{Bhattacharya:2018zxi,Bhattacharya:2019cme,Luo:2020yqj,Shastry:2022obb,Jia:2024atq} and lattice extractions for pions \cite{Chen:2019lcm} and nucleons \cite{Alexandrou:2020zbe} emerged, later extending to transversity \cite{Alexandrou:2021bbo} and axial GPDs \cite{Bhattacharya:2023nmv}. Initially, symmetric frames like the Breit frame were employed, necessitating separate computations for each momentum transfer. The latest breakthrough involves asymmetric frames \cite{Bhattacharya:2022aob,Cichy:2023dgk,Bhattacharya:2023ays,Bhattacharya:2023jsc}, allowing multiple momentum transfers in a single calculation. Leveraging this advancement, our study utilizes asymmetric-frame data within the pseudo-GPD framework, building upon preliminary findings presented in Ref.~\cite{Nurminen:2023qok}. We expand this analysis to various momentum transfers, capitalizing on increased statistics, especially at high nucleon boosts.

The paper follows this structure:
In Section \ref{sec:setup}, we elaborate on our theoretical and lattice setup.
Relevant details of the pseudo-GPD approach are outlined in Section \ref{sec:pseudo}.
Our numerical results are presented in Section \ref{sec:results}.
Lastly, we offer concluding remarks and discuss future prospects in Section \ref{sec:summary}.

\section{Theoretical and lattice QCD setup}
\label{sec:setup}
In this work, we follow the formulation of Ref.~\cite{Bhattacharya:2022aob}, and we refer to this paper for an extensive discussion of GPD definitions and their parametrizations in different frames of reference (see also Ref.~\cite{Braun:2023alc} for related recent work).
Here, we recall the main definitions to establish our notation.

Since light-front correlations are inaccessible on a Euclidean lattice, we calculate matrix elements (MEs) of the following form:
\begin{align}
\label{eq:MEs}
F^\mu (z, P_f, P_i) = \langle N(P_f) | \bar\psi(z) \gamma^\mu W(0, z) \psi (0) | N (P_i) \rangle,\,\,\,
\end{align}
where $|N(P_{i/f})\rangle$ are the nucleon's initial/final states with four-momenta $P_i$/$P_f$.
The four-vector $z$ is taken as $(0,0,0,z^3)$ and for brevity, we will denote $z^3=z$ to indicate the length of the 
straight Wilson line $W(0,z)$, i.e.\ it will be taken purely along the $z$-direction.
The Dirac matrix $\gamma^\mu$ corresponds to the case of unpolarized quarks, and we use $\mu=0,\,1,2\,$ to construct the twist-2 GPDs $H$ and $E$. 
We also introduce the momentum transfer $\Delta=P_f-P_i$, and the average momentum $P=(P_i+P_f)/2$.

The above MEs can be parametrized in terms of eight Lorentz-invariant amplitudes~\cite{Bhattacharya:2022aob}, $A_i \equiv A_i (z\cdot P, z \cdot \Delta, \Delta^2, z^2)$, which in Minkowski metric read
\begin{align}
    F^\mu (z, P_f, P_i) = \,&\bar{u}(P_f,\lambda')\Bigg[\frac{P^\mu}{m} A_1 + mz^\mu A_2 + \frac{\Delta^\mu}{m} A_3 \nonumber\\&+ im\sigma^{\mu z} A_4  + \frac{i\sigma^{\mu\Delta}}{m} A_5 + \frac{iP^{\mu} \sigma^{z\Delta} }{m} A_6 \\&+ imz^\mu\sigma^{z\Delta} A_7 + \frac{i \Delta^\mu \sigma^{z\Delta} }{m} A_8\Bigg]u(P_i,\lambda),\nonumber
\end{align}
where $\sigma^{\mu \nu} \equiv \tfrac{i}{2} (\gamma^\mu \gamma^\nu - \gamma^\nu \gamma^\mu)$,  
$\sigma^{\mu z} \equiv \sigma^{\mu \rho} z_\rho$, 
$\sigma^{\mu \Delta} \equiv \sigma^{\mu \rho} \Delta_\rho$, $\sigma^{z \Delta} \equiv \sigma^{\rho \tau} z_\rho \Delta_\tau$, and $m$ is the nucleon mass.

Our bare MEs are computed in asymmetric frames of reference, with a fixed final state momentum $P_f=(P_f^0,0,0,P_f^3)$.
Thus, all the momentum transfer is attributed to the initial state, $P_i=(P_f^0-\Delta^0, -\Delta^1,-\Delta^2, P_f^3)$.
Moreover, we concentrate on the zero skewness case, i.e.\ no longitudinal momentum transfer, $\Delta^3=0$, leading to $P_i^3=P_f^3=P^3$. In what follows, we give expressions in Euclidean metric, and thus, we use lower indices.

We use four parity projectors, the unpolarized one, $\Gamma_0 = \left(1 + \gamma_0\right)/4$, and the three polarized in the $k$-direction, $\Gamma_k = \left(1 + \gamma_0\right) i \gamma_5 \gamma_k/4$ for each of the Dirac matrix $\gamma^\mu$.
For $\mu=0,\,1,\,2$, this gives rise to 12 MEs, $\Pi_\mu(\Gamma_\kappa$), six of which are independent upon averaging equivalent contributions with reversed roles of $\Delta^1$ and $\Delta^2$.
In this way, one obtains a system of six equations containing six of the Lorentz-invariant amplitudes ($A_2$ and $A_7$ enter only for $\mu=3$, not used in this work to avoid operator mixing under renormalization~\cite{Constantinou:2017sej}).

Upon the extraction of the amplitudes in such a setup, one can construct the unpolarized GPDs $H$ and $E$. 
In Ref.~\cite{Bhattacharya:2022aob}, two definitions were proposed, referred to as the ``standard'' definition and the ``Lorentz-invariant'' (LI) one.
In the infinite momentum frame, both definitions are equivalent, but at any finite boost, they differ by power-suppressed higher-twist effects (HTEs).
As such, they can have different convergence properties towards physical GPDs.
It was observed in Ref.~\cite{Cichy:2023dgk} that the LI variant leads to a mild improvement in convergence for the $H$ GPD, whereas the improvement in the $E$ function is significant.
This can be interpreted as meaning that the different combinations of amplitudes that contribute to both GPDs profit from some cancellation of HTEs, particularly in the $E$ GPD.
We note that this cancellation is largely accidental and is not related to the feature of Lorentz invariance.
Nevertheless, the better convergence properties of the LI GPDs led us to choose this definition for our work.
In terms of amplitudes, the $H$ and $E$ GPDs in possition space read:
\begin{align}
\label{eq:H}
H=& \,A_1,\\
\label{eq:E}
E=& \,-A_1 + 2A_5 + 2 P^3 zA_6.
\end{align}
It is also instructive to observe which MEs contribute in the asymmetric frame used here (cf.\ Eqs.~(117)-(118) of Ref.~\cite{Bhattacharya:2022aob}): $\Pi_0(\Gamma_0)$, $\Pi_0(\Gamma_{1/2})$ (depending on the direction of momentum transfer), $\Pi_{1/2}(\Gamma_0)$ and $\Pi_{1/2}(\Gamma_{1/2})$, $\Pi_{1/2}(\Gamma_3)$.
This can be contrasted with the contributions to the ``standard'' definition from only $\Pi_0(\Gamma_0)$ and $\Pi_0(\Gamma_{1/2})$.
Thus, the effect of including MEs with the $\gamma_{1/2}$ insertions, which define twist-3 vector GPDs, is to alter the twist-3 and higher-twist contamination in $H$ and $E$ twist-2 GPDs.
Interestingly, in the language of amplitudes, this contamination follows from $A_3$, $A_4$ and $A_8$, which are removed by $\Pi_1$ and $\Pi_2$ MEs, and the effect from $A_6$ is suppressed in $E$ and entirely removed in $H$.

The lattice methodology is thoroughly discussed in Ref.~\cite{Bhattacharya:2022aob}.
Here, we recapitulate the main points.
The MEs given by Eq.~(\ref{eq:MEs}) are computed in a setup consisting of two degenerate light quarks and non-degenerate strange and charm quarks ($N_f=2+1+1$) using the twisted mass (TM) fermion discretization with clover improvement and Iwasaki-improved gluons \cite{Alexandrou:2018egz}. 
The quark masses are chosen such that they correspond to a pion mass of around 260 MeV, with strange and charm quarks tuned to their physical masses.
The lattice size is $L^3\times T=32^3\times64$, and the lattice spacing amounts to $a\approx0.093$ fm, giving a physical lattice extent of around 3 fm in the spatial directions.
MEs are extracted at a source-sink separation of $t_s=10a$, which guarantees sufficient suppression of excited states at this level of precision.

We calculate the MEs at several values of the nucleon boost in the $z$-direction, $P_3=\{0,\,1,\,2,\,3,\,4\}(2\pi/L)$, which corresponds to $P_3=\{0,\,0.42,\,0.83,\,1.25,\,1.67\}$ GeV in physical units.
All results presented here pertain to the zero skewness case, which implies momentum transfer only in the transverse directions ($x$ and $y$).
We consider all permutations and changes of direction of the momentum transfer components, $(\Delta_1,\Delta_2)$, as well as both directions of the longitudinal momentum boost, $P_3$.
Details of the setup are shown in Table \ref{tab:stat}, where we give the employed values of the invariant momentum transfer\footnote{The quoted values of $-t$ pertain to $P_3=1.25$ GeV and are slightly different for other nucleon boosts due to different energies of the initial and final states in the asymmetric frame and the discreteness of $(\Delta_1,\Delta_2)$ components in units of $2\pi/L$. However, the effect of combining data at slightly different $-t$ values is subleading with respect to the current precision of the data.}, the nucleon boost, permutations of the transverse momentum transfer vector, and the numbers of these permutations multiplied by the two directions of $P_3$, the used gauge field configurations, and the source positions per configuration.
The last column displays the total number of measurements for each case.

\begin{table}[t!]
\begin{center}
\renewcommand{\arraystretch}{1.0}
\begin{tabular}{|ccc|cccc|}
\hline
$-t$ & $P_3$ & $(\Delta_1,\Delta_2)$ &  \multirow{2}{*}{$N_\Delta$} & \multirow{2}{*}{$N_{\rm confs}$} & \multirow{2}{*}{$N_{\rm src}$} & \multirow{2}{*}{$N_{\rm meas}$} \\\relax
[GeV$^2$] & [GeV] & $[2\pi/L]$&&&&\\ 
\hline
0  & $\pm$0.42 &(0,0)  &2   &100 &8  &1600\\
0  & $\pm$0.83 &(0,0)  &2   &100 &8  &1600\\
0  & $\pm$1.25 &(0,0)  &2   &269 &16  &8608\\
0  & $\pm$1.67 &(0,0)  &2   &506 &32  &32384\\
\hline
0.17  & $\pm$0.42 &($\pm$1,0), (0,$\pm$1)  &8   &100 &8  &6400\\
0.17  & $\pm$0.83 &($\pm$1,0), (0,$\pm$1)  &8   &100 &8  &6400\\
0.17  & $\pm$1.25 &($\pm$1,0), (0,$\pm$1)  &8   &269 &8  &17216\\
0.17  & $\pm$1.67 &($\pm$1,0), (0,$\pm$1)  &8   &506 &32  &129536\\
\hline
0.34 & $\pm$0.42 &$(\pm 1,\pm 1)$  &8   &100 &8  &6400 \\
0.34 & $\pm$0.83 &$(\pm 1,\pm 1)$  &8   &100 &8  &6400 \\
0.34 & $\pm$1.25 &$(\pm 1,\pm 1)$  &8   &195 &8  &12480 \\
0.34 & $\pm$1.67 &$(\pm 1,\pm 1)$  &8   &506 &32  &129536\\
\hline
0.65  & $\pm$0.42 &($\pm$2,0), (0,$\pm$2)  &8   &100 &8  &6400\\
0.65  & $\pm$0.83 &($\pm$2,0), (0,$\pm$2)  &8   &100 &8  &6400\\
0.65  & $\pm$1.25 &($\pm$2,0), (0,$\pm$2)  &8   &269 &8  &17216\\
0.65  & $\pm$1.67 &($\pm$2,0), (0,$\pm$2)  &8   &506 &32  &129536\\
\hline
0.81 & $\pm$0.42 &($\pm$1,$\pm$2), ($\pm$2,$\pm$1)  &16   &100 &8  &12800 \\
0.81 & $\pm$0.83 &($\pm$1,$\pm$2), ($\pm$2,$\pm$1)  &16   &100 &8  &12800 \\
0.81 & $\pm$1.25 &($\pm$1,$\pm$2), ($\pm$2,$\pm$1)  &16   &195 &8  &24960 \\
0.81 & $\pm$1.67 &($\pm$1,$\pm$2), ($\pm$2,$\pm$1)  &16   &506 &32  &259072 \\
\hline
1.24 & $\pm$0.42 &$(\pm 2,\pm 2)$  &8   &100 &8  &6400 \\
1.24 & $\pm$0.83 &$(\pm 2,\pm 2)$  &8   &100 &8  &6400 \\
1.24 & $\pm$1.25 &$(\pm 2,\pm 2)$  &8   &195 &8  &12480 \\
1.24 & $\pm$1.67 &$(\pm 2,\pm 2)$  &8   &506 &32  &129536\\
\hline
1.38  & $\pm$0.42 &($\pm$3,0), (0,$\pm$3)  &8   &100 &8  &6400\\
1.38  & $\pm$0.83 &($\pm$3,0), (0,$\pm$3)  &8   &100 &8  &6400\\
1.38  & $\pm$1.25 &($\pm$3,0), (0,$\pm$3)  &8   &269 &8  &17216\\
1.38  & $\pm$1.67 &($\pm$3,0), (0,$\pm$3)  &8   &506 &32  &129536\\
\hline
1.52 & $\pm$0.42 &($\pm$1,$\pm$3), ($\pm$3,$\pm$1)  &16   &100 &8  &12800 \\
1.52 & $\pm$0.83 &($\pm$1,$\pm$3), ($\pm$3,$\pm$1)  &16   &100 &8  &12800 \\
1.52 & $\pm$1.25 &($\pm$1,$\pm$3), ($\pm$3,$\pm$1)  &16   &195 &8  &24960 \\
1.52 & $\pm$1.67 &($\pm$1,$\pm$3), ($\pm$3,$\pm$1)  &16   &506 &32  &259072 \\
\hline
2.29  & $\pm$0.42 &($\pm$4,0), (0,$\pm$4)  &8   &100 &8  &6400\\
2.29  & $\pm$0.83 &($\pm$4,0), (0,$\pm$4)  &8   &100 &8  &6400\\
2.29  & $\pm$1.25 &($\pm$4,0), (0,$\pm$4)  &8   &269 &8  &17216\\
2.29  & $\pm$1.67 &($\pm$4,0), (0,$\pm$4)  &8   &506 &32  &129536\\
\hline
\end{tabular}
\caption{\small Details of our lattice setup. The numbers in the right part of the table correspond to: $N_\Delta$ -- number of permutations of $(\Delta_1,\Delta_2)$ times two (signs of $P_3$),  $N_{\rm confs}$ -- number of employed gauge field configurations, $N_{\rm src}$ -- number of source positions for each configuration, $N_{\rm meas}$ -- total number of measurements ($N_{\rm meas}=N_\Delta N_{\rm confs} N_{\rm src}$).}
\label{tab:stat}
\end{center}
\end{table}

We note that MEs for the $P_3=1.25$ GeV case were calculated in Ref.~\cite{Bhattacharya:2022aob} and analyzed within the quasi-distribution approach.
In the latter, the nucleon boost needs to be as large as possible to suppress higher-twist contamination.
In the pseudo-distribution method, data at smaller boosts are also valuable as long as they pertain to sufficiently small $z$.
For this reason, we supplemented the data of Ref.~\cite{Bhattacharya:2022aob} with ones at $P_3=0.42,\,0.83$ GeV.
These cases are straightforward from the point of view of the data's precision, as a considerably smaller number of measurements already leads to statistical errors present at $P_3=1.25$ GeV.
However, we also decided to add data at larger $P_3=1.67$ GeV.
In this instance, a significantly larger number of measurements is needed to match the precision of $P_3=1.25$ GeV, amounting roughly to an order of magnitude increase.
Thus, the calculations for $P_3=1.67$ GeV constituted the bulk of new computations for this work.

\vspace*{3mm}
\section{Pseudo-GPDs}
\label{sec:pseudo}
Bare coordinate-space MEs of GPDs, constructed as implied by Eqs.~(\ref{eq:H})-(\ref{eq:E}), are the starting point for both the quasi- and pseudo-distribution approaches. In both cases, one needs to first renormalize the standard logarithmic and Wilson-line-induced power divergences.
The resulting renormalized MEs are Euclidean objects that can be translated into physical distributions by an appropriate matching procedure. The key difference between quasi- and pseudo-distribution approaches lies in the space in which the matching is performed -- momentum space (quasi) or coordinate space (pseudo). This implies different practical procedures with both methods and possibly different associated systematic effects.
The latter should vanish upon elimination of lattice effects (e.g.\ finite lattice spacing effects or finite volume effects) and other attendant systematics, such as HTEs induced by a finite hadron boost or truncation effects in the perturbative matching.
Thus, it is highly desirable to analyze the bare lattice data through both approaches, with the logic that potential differences in the final distributions give estimates of the unquantified systematics that are still unavoidable at the present stage of these computations.

Below, we summarize the main parts of the pseudo-GPD analysis procedure, concentrating on the three main steps: renormalization of the divergences in a ratio scheme, matching of the coordinate-space distributions to their light-cone counterparts and the reconstruction of the $x$-dependence to obtain the results ultimately in momentum space.
For theoretical details of the pseudo-distribution approach, we refer to the original papers by Radyushkin \cite{Radyushkin:2016hsy,Radyushkin:2017cyf,Radyushkin:2017lvu,Radyushkin:2017sfi,Radyushkin:2018cvn,Radyushkin:2018nbf,Radyushkin:2019owq} and the review \cite{Radyushkin:2019mye}. 

Bare MEs of GPDs, generically denoted here as $F(P_3,z)$ ($F=\{H,\,E\})$, are renormalized in a ratio scheme by forming a double ratio \cite{Orginos:2017kos} with zero-momentum unpolarized PDFs, $f(0,z)$, with the second ratio involving unpolarized PDFs at $z=0$, $f(P_3,0)$ and $f(0,0)$, canceling additional systematics and ensuring the desired normalization:
\begin{equation}
\label{eq:DR}
\mathcal{F}(P_3,z)=\frac{F(P_3,z)}{f(0,z)}\,\frac{f(0,0)}{f(P_3,0)},    
\end{equation}
where $\mathcal{F}=\{\mathcal{H},\mathcal{E}\}$ is called a pseudo-ITD (Ioffe time distribution) or a reduced ITD.
It has been conjectured that, in addition to removing the divergences, the double ratio can remove some part of HTEs and other systematic effects \cite{Orginos:2017kos}.
The double ratio is renormalization group invariant and defines a nonperturbative renormalization scheme, with $1/z$ playing the role of a kinematic scale suppressing HTEs. 

Pseudo-ITDs differ from light-cone ITDs in the ultraviolet regime and thus, their difference can be calculated in perturbation theory and subtracted \cite{Radyushkin:2018cvn,Zhang:2018ggy,Izubuchi:2018srq,Radyushkin:2018nbf,Li:2020xml}, with the procedure commonly called the matching.
In Ref.~\cite{Bhat:2022zrw}, we tested the effects of 2-loop formulae for the matching of ITDs and found the effect of the second order in $\alpha_s$ to be negligible even for unpolarized PDF data analyzed in this paper, comparatively much more precise than our current GPD data.
Hence, for simplicity, we restrict the current work to 1-loop matching.
The latter consists of an action of two perturbative kernels, $B(u)$ that evolves the reduced ITDs from the scales $1/z$ to a common scale $\mu$, and $L(u)$ that performs the actual translation from Euclidean ratio-scheme-renormalized to Minkowski $\MSb$-renormalized observables.
We will show the effects of both kernels and therefore, we introduce the intermediate evolved ITDs, $\mathcal{F}'(P_3 z)$.
Defining additionally
\begin{equation}
\mathcal{F}_u(P_3,z)=\mathcal{F}(uP_3,z)-\mathcal{F}(P_3,z),    
\end{equation}
the evolved ITDs are given by:
\begin{align}
\label{eq:evolved}
\mathcal{F}'(P_3,z) =& \mathcal{F}(P_3,z)\\
-&\frac{\alpha_s C_F}{2\pi} \int_0^1  du\,\, B(u) \ln\frac{z^2\mu^2e^{2\gamma_E+1}}{4} \mathcal{F}_u(P_3,z).\nonumber
\end{align}
The matched (light-cone) ITDs, $\Fbar(P_3,z)$ are then obtained via:
\begin{equation}
\label{eq:matched}
\Fbar(P_3,z) = \mathcal{F}'(P_3,z)-\frac{\alpha_s C_F}{2\pi} \int_0^1  du\,\, L(u) \mathcal{F}_u(P_3,z).
\end{equation}
The perturbative kernels read \cite{Radyushkin:2018cvn,Zhang:2018ggy,Izubuchi:2018srq,Radyushkin:2018nbf,Li:2020xml}:
\begin{equation}
B(u)=\frac{1+u^2}{u-1},
\end{equation}
\begin{equation}
L^{(1)}(u) = 4 \frac{\ln(1-u)}{u-1} - 2(u-1).
\end{equation}
The matching procedure is performed separately for each nucleon boost $P_3$ and Wilson line length $z$. However, the $u$-integral over perturbative kernels necessitates access to data at all continuous values of the boost up to $P_3$ through the object $\mathcal{F}_u(P_3,z)$.
This is commonly done by employing interpolation, which involves fitting a low-order polynomial to all available data at a given value of $z$ (we show examples of such fits in the next section).
Thus, even though the matching proceeds separately for each $(P_3,z)$ pair (we choose to emphasize this fact by explicit arguments of the functions $\mathcal{F}$, $\mathcal{F}'$ and $\Fbar$), the results depend also on all reduced ITDs $\mathcal{F}(P_3,z)$ at a fixed $z$.

Light-cone ITDs are functions of two Lorentz invariants, $\nu\equiv P_3z$ (the Ioffe time) and $z^2$ (length of the 4-vector $z$; we use $z^\mu=(0,0,0,z_3)$ and abbreviate $z_3$ as $z$ in this work).
However, generically, pseudo-ITDs at a given Ioffe time and Wilson line length depend on $P_3$, as indicated by our notation.
The final matched ITDs, $\Fbar(P_3,z)$, should, in turn, be equal regardless of the initial value of $P_3$, as long as they correspond to the same Ioffe time.
In practice, it provides the key check for the robustness of the matching procedure and the maximum value of $z$ that can be used in reconstructing physical distributions.
In principle, $z$ should be in the perturbative regime, i.e.\ $z\lesssim0.2-0.3$ fm.
Such a condition is clearly too prohibitive and would limit the accessible values of the Ioffe time to $\nu\lesssim2$.
However, the validity of the matching holds up to $\mathcal{O}(z^2\Lambda_{\rm QCD}^2)$ HTEs, and these effects may, in practice, be relatively small at intermediate values of $z$ due to possible cancellation in the double ratio and their magnitude being possibly small with respect to statistical errors (see next section for numerical evidence).
In this way, the practical criterion that we adopt for the choice of the maximum value of $z$ ($\zmax$) entering the reconstruction of physical distributions is such that matched ITDs derived from all pairs of $(P_3,z)$ with the same $P_3z$ are compatible with each other. When this is the case, we average results coming from different nucleon boosts that correspond to the same Ioffe time, and we use the notation $\Fbar(\nu,\mu)$ to indicate that this averaging was performed, with the second argument revealing the renormalization scale of the $\MSb$ scheme. We will take the latter to be the standard 2 GeV.

The outcome of the matching procedure is light-cone ITDs in coordinate space, related to momentum-space GPDs, $\Fbar(x,\mu)$, by a Fourier transform:
\begin{equation}
\label{eq:FT}
\Fbar(\nu,\mu)=\int_{-1}^1 dx\, e^{i\nu x} \Fbar(x,\mu).
\end{equation}
Splitting the above equation into real and imaginary parts, one can also express it as:
\begin{equation}
\label{eq:ReFT}
{\rm Re}\,\Fbar(\nu,\mu)=\int_0^1 dx\, \cos(\nu x) \Fbar_v(x,\mu),
\end{equation}
\begin{equation}
\label{eq:ImFT}
{\rm Im}\,\Fbar(\nu,\mu)=\int_0^1 dx\, \sin(\nu x) \Fbar_{v2s}(x,\mu).
\end{equation}
The different distributions appearing in these equations are:
\begin{itemize}
\item $\Fbar(x,\mu)$ -- the full distribution encompassing valence and sea quarks ($\Fbar=(\Fbar_v+\Fbar_{v2s})/2$; i.e.\ involves both real and imaginary parts of ITDs),
\item $\Fbar_s(x,\mu)$ -- only sea quarks ($\Fbar_s=(\Fbar_v-\Fbar_{v2s})/2$; i.e.\ involves both real and imaginary parts of ITDs),
\item $\Fbar_v(x,\mu)$ -- only valence quarks (only real part of ITDs),
\item $\Fbar_{v2s}(x,\mu)$ -- valence quarks and twice sea quarks (only imaginary part of ITDs).
\end{itemize}

\begin{figure*}[t!]
    \centering
    \includegraphics[width=0.9\textwidth]{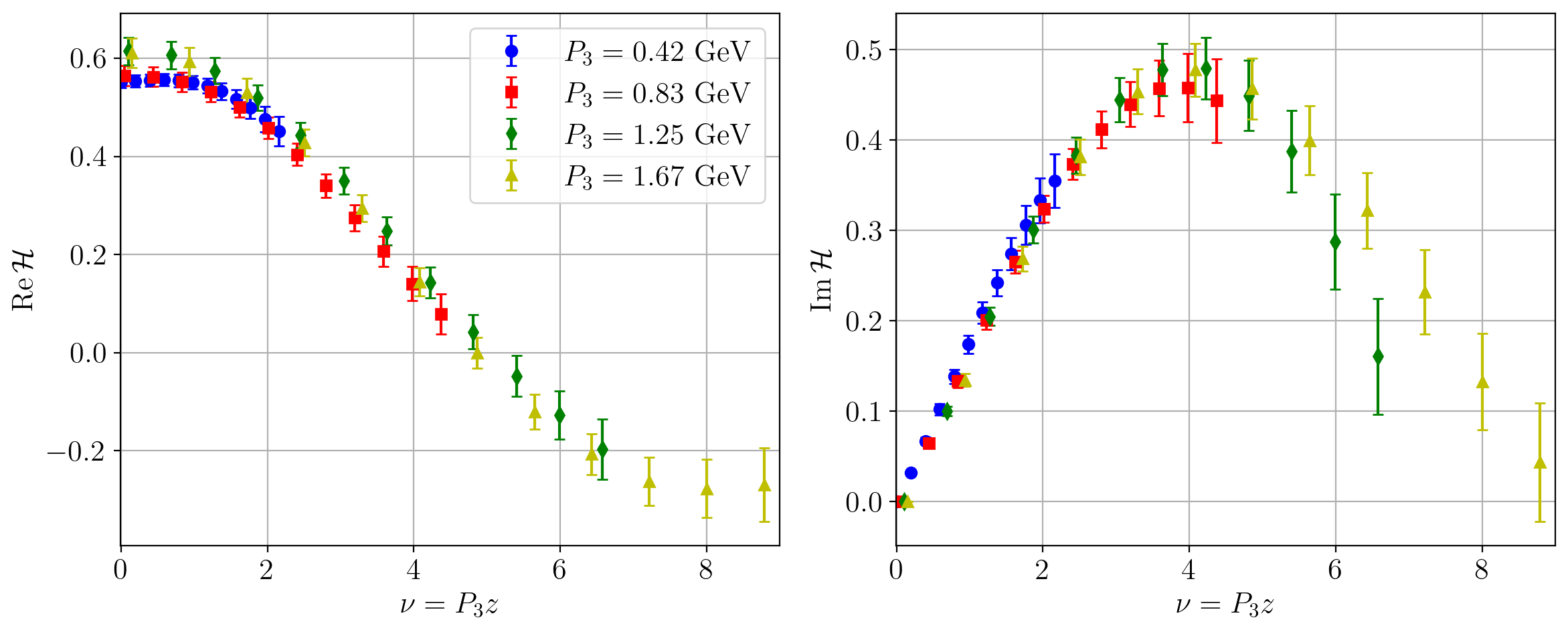}
    \includegraphics[width=0.9\textwidth]{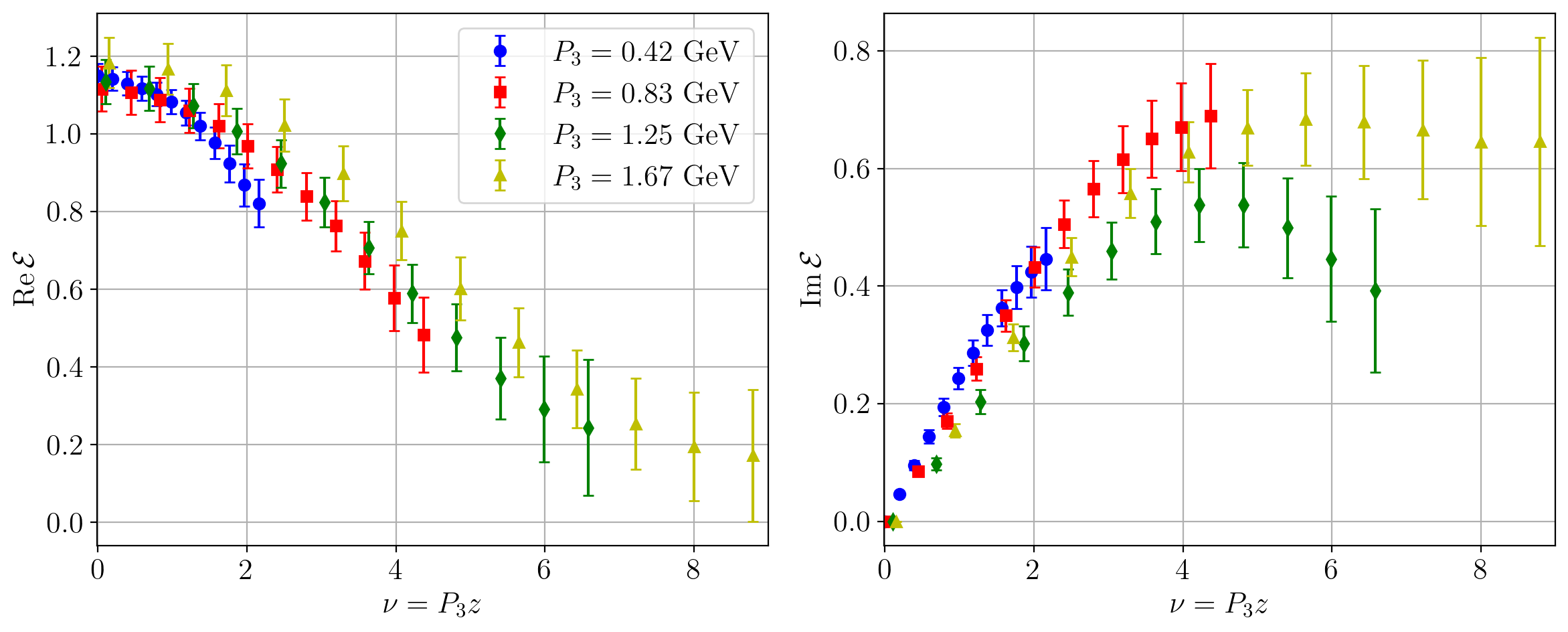}
    \caption{Reduced $H$ (top) and $E$ (bottom) ITDs at $-t=0.65 $ GeV$^2$ with four nucleon momentum values, slightly shifted for better visibility. The left/right panels show the real/imaginary part.}
    \label{fig:reduced_Q2}
\end{figure*}

The inversion of the above Fourier transforms necessitates access to ITDs in an infinite range of continuous Ioffe times. 
Obviously, the lattice data are discrete and available only for a truncated range, up to $\numax=P_3^{\rm max}\zmax$, with $P_3^{\rm max}$ being the maximum boost.
This poses an important limitation for lattice extractions of partonic distributions; see Ref.~\cite{Karpie:2019eiq} for a detailed discussion.
The most common method to address this inverse problem in the context of pseudo-distributions is to reconstruct the momentum-space quantities with a fitting ansatz.
The simplest and most common ansatz for the reconstruction is the standard one that captures the limiting behaviors both for small and large $x$,
\begin{equation}
\label{eq:ansatz}
\Fbar(x) = N x^a (1-x)^b,
\end{equation}
where the fitting parameters are:
\begin{itemize}
\item real part: $a,\,b$; the normalization is fixed by the $\nu=0$ ITD ($\Fbar(\nu=0)$), i.e.\ $\int_0^1 dx \Fbar(x)=\Fbar(\nu=0)$ -- thus, $N=\Fbar(0)/B(a+1,b+1)$, expressed in terms of the Euler beta and gamma functions, $B(x,y)=\Gamma(x)\Gamma(y)/\Gamma(x+y)$,
\item imaginary part: $a,\,b,\,N$; i.e.\ the normalization is also fitted due to no constraint on its value.
\end{itemize}
The parameter $b$ is taken to be non-negative to accommodate the physical restriction that the GPDs vanish at $x=1$.
Likewise, the normalization $N$ is positive.
The fitting procedure involves minimizing the $\chi^2$ function,
\begin{equation}
\label{eq:chi2}
\chi^2_{{\rm Re}/{\rm Im}}=\sum_{\nu=0}^{\numax}\frac{{\rm Re}/{\rm Im}\,\Fbar(\nu,\mu)-{\rm Re}/{\rm Im}\,\Fbar_{\rm fit}(\nu,\mu)}{\sigma_{{\rm Re}/{\rm Im}\,\Fbar(\nu,\mu)}^2}.
\end{equation}
The weights are given by the statistical errors $\sigma_{{\rm Re}/{\rm Im}\,\Fbar(\nu,\mu)}$ of the matched ITDs ${\rm Re}/{\rm Im}\,\Fbar(\nu,\mu)$.
The fitting parameters enter the $\chi^2$ function through the cosine/sine Fourier transform of the fitting ansatz, denoted by ${\rm Re}/{\rm Im}\,\Fbar_{\rm fit}(\nu,\mu)$ and referred to as fitted ITDs.
The fitted ITDs are, obviously, continuous functions of the Ioffe time, and thus, this cosine/sine Fourier transform is well-defined.
The key parameter of the fits is the maximum Ioffe time of the lattice data, $\numax$, chosen according to the practical criterion discussed above.
Nevertheless, we will test the sensitivity of the results to this value by also considering other choices.

To test the robustness of our fit, we also explore functions with additional parameters of the generic form:
\begin{equation}
\label{eq:ansatz2}
\Fbar(x) = N x^a (1-x)^b \,(1+c x^{d_1} (1-x)^{d_2}).
\end{equation}
Specifically, we try the following choices:
\begin{itemize}
\item $c$ fitted, $d_1=0.5$ or 1, $d_2=0$,
\item $c$ fitted, $d_2=0.5$ or 1, $d_1=0$,
\item $c$, $d_1$ fitted, $d_2=0$,
\item $c$, $d_1$, $d_2$ fitted.
\end{itemize}

We note that all statistical analyses are performed using a bootstrap procedure.

\begin{figure*}[p!]
    \centering
    \includegraphics[width=0.9\textwidth]{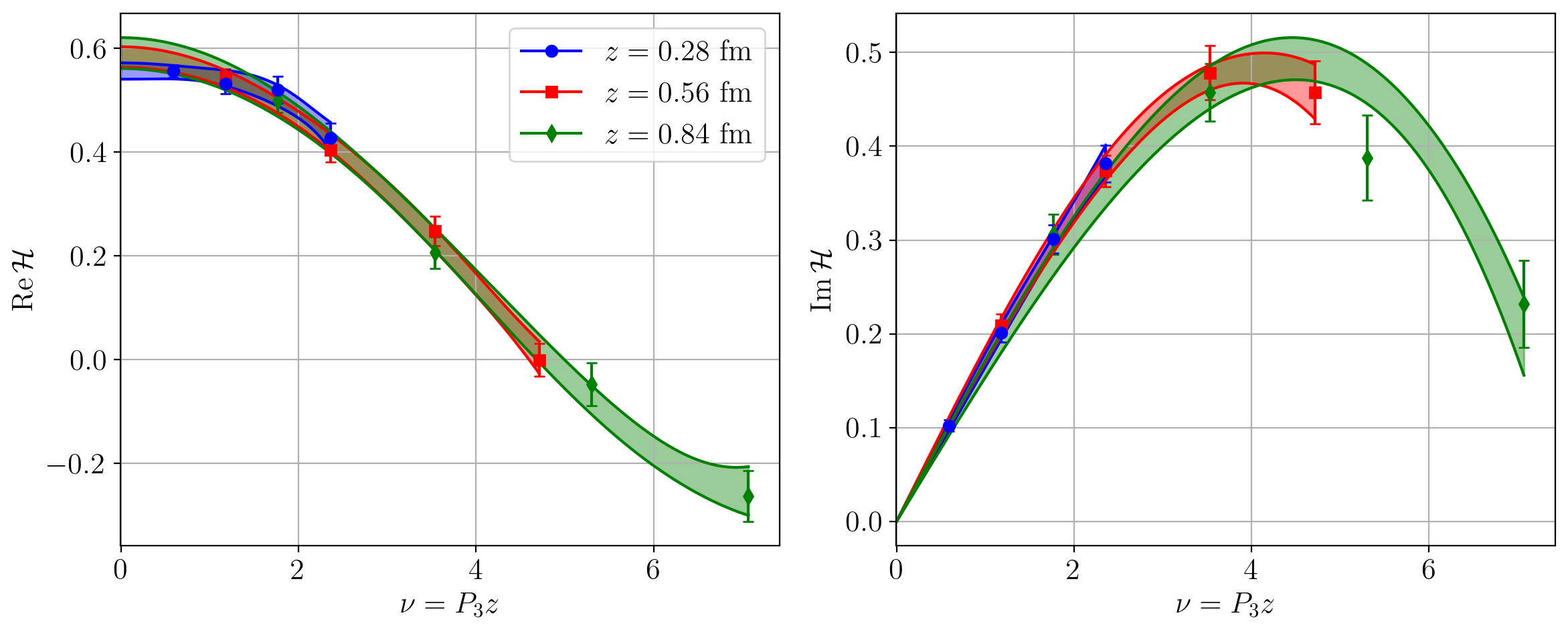}
    \caption{Interpolations of the reduced $H$ ITDs with second order polynomials, at three values of $z$. The momentum transfer is $-t=0.65 $ GeV$^2$. The left/right panels show the real/imaginary part.}
    \label{fig:interpolated_Q2}
\end{figure*}
\begin{figure*}[p!]
    \centering
    \includegraphics[width=0.9\textwidth]{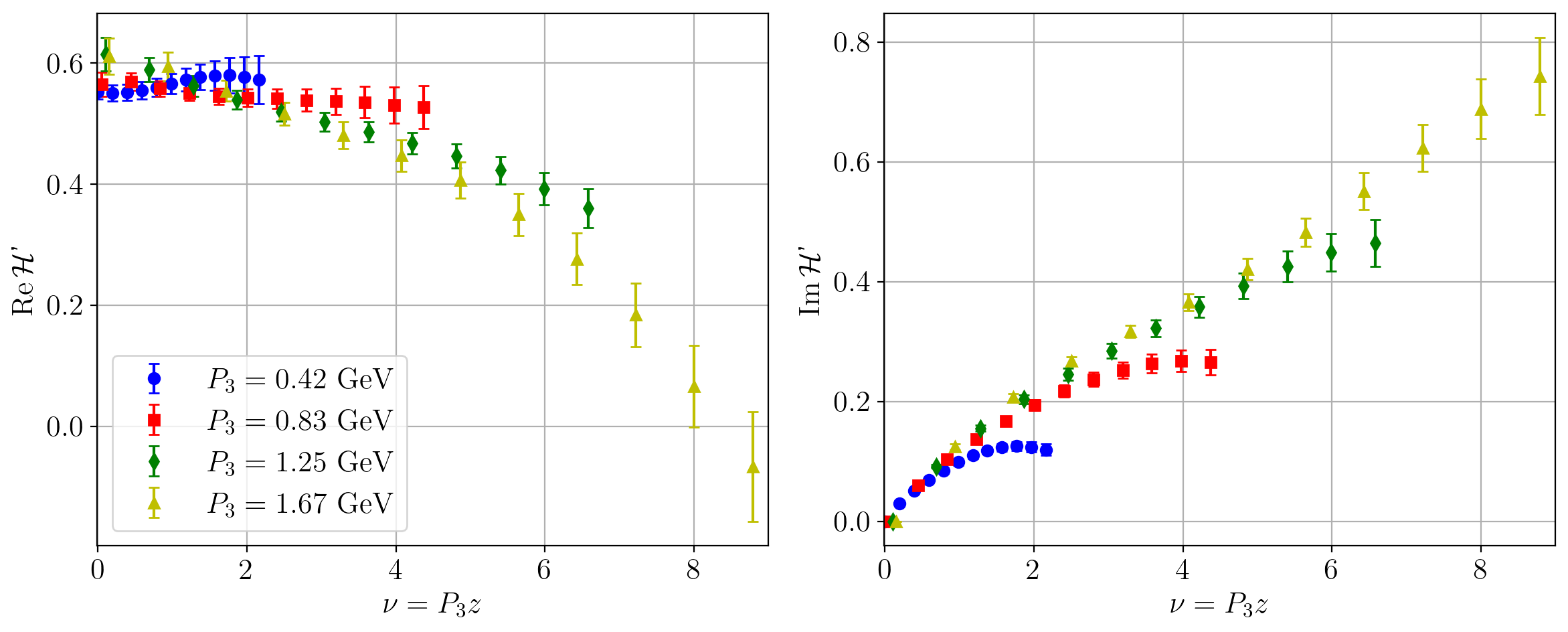}
    \includegraphics[width=0.9\textwidth]{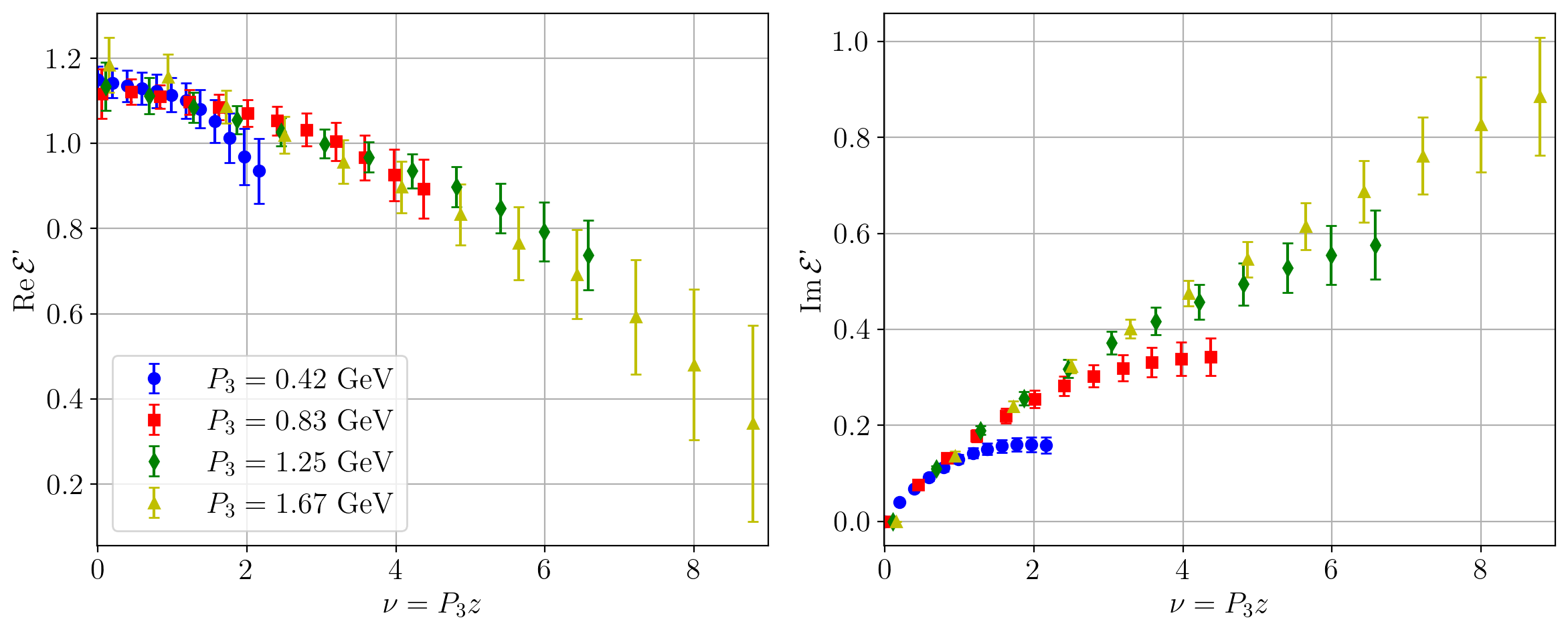}
    \caption{Evolved $H$ (top) and $E$ (bottom) ITDs at $-t=0.65 $ GeV$^2$ with four nucleon momentum values, slightly shifted for better visibility. The left/right panels show the real/imaginary part. The common scale for all evolved ITDs is $\mu=2$ GeV.}
    \label{fig:evolved_Q2}
\end{figure*}

\begin{figure*}[t!]
    \centering
    \includegraphics[width=0.9\textwidth]{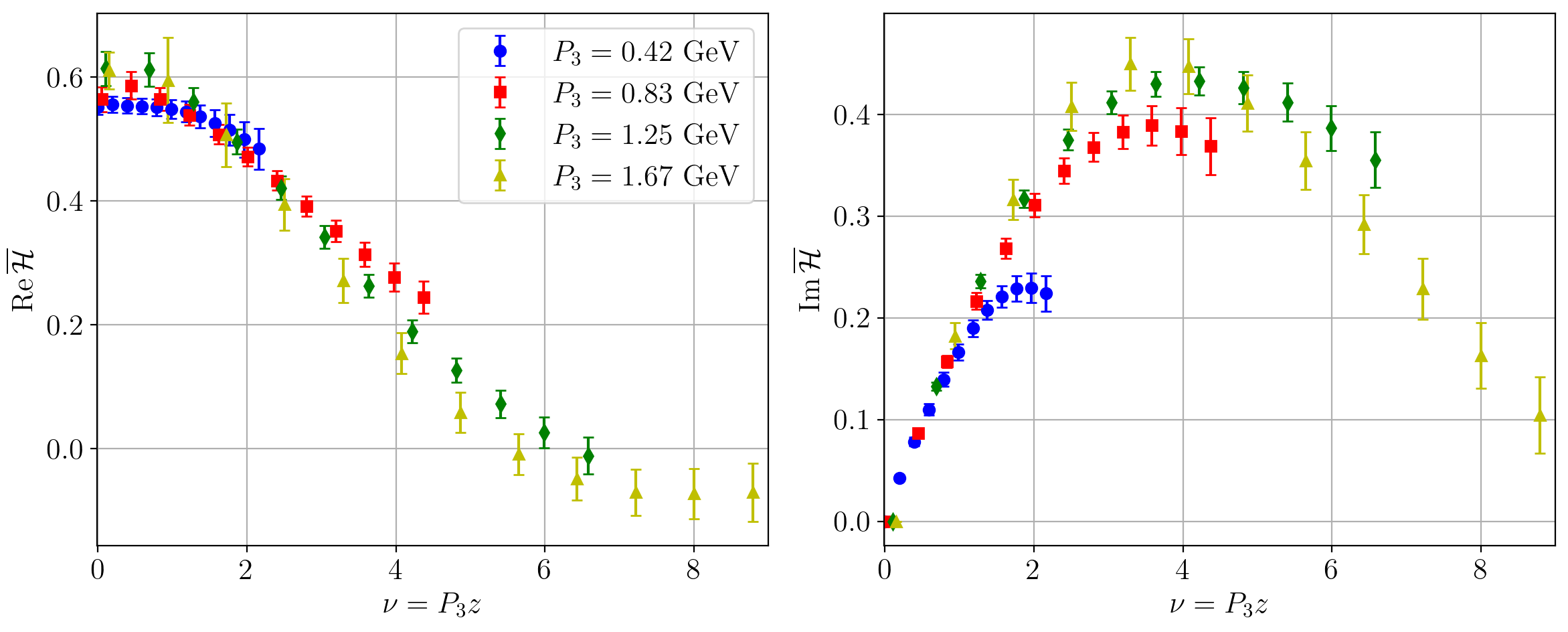}
    \includegraphics[width=0.9\textwidth]{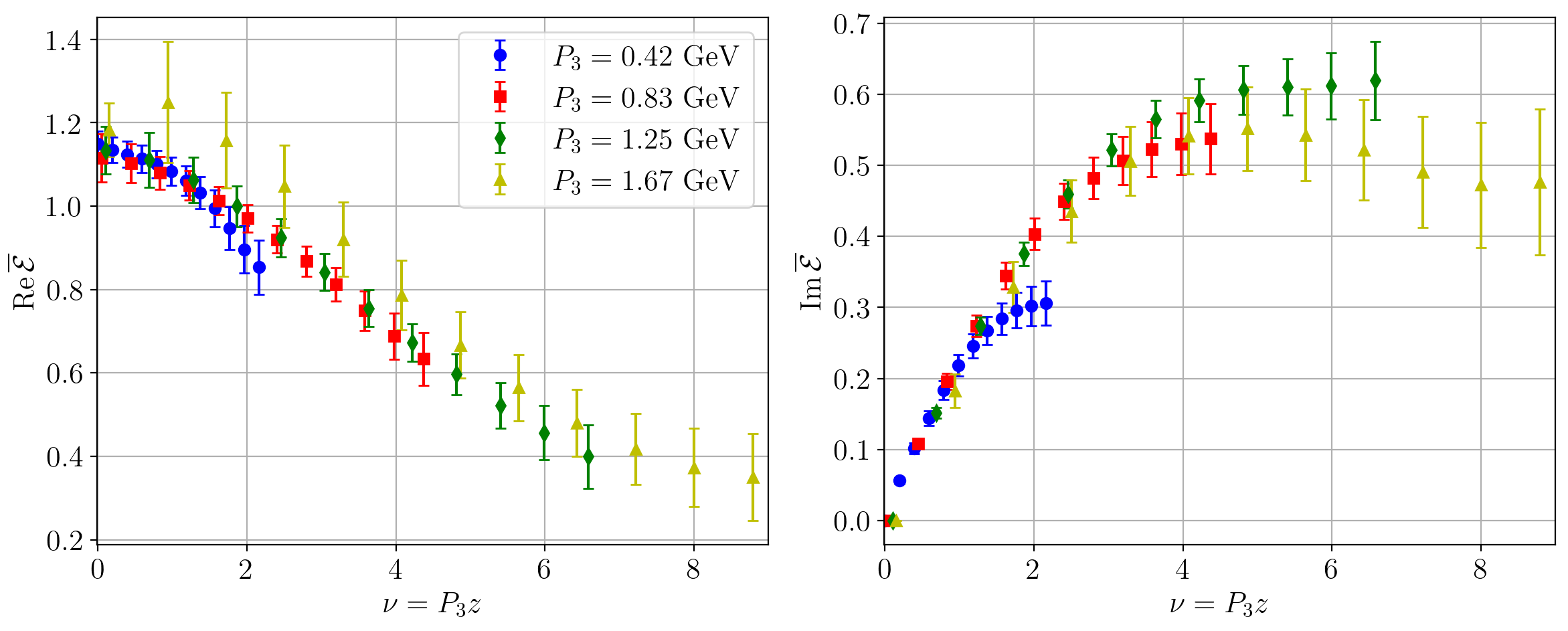}
    \caption{Matched $H$ (top) and $E$ (bottom) ITDs at $-t=0.65 $ GeV$^2$, $\mu=2$ GeV, with four nucleon momentum values, slightly shifted for better visibility. The left/right panels show the real/imaginary part.}
    \label{fig:matched_Q2}
\end{figure*}
\begin{figure*}[t!]
    \centering
    \includegraphics[width=0.9\textwidth]{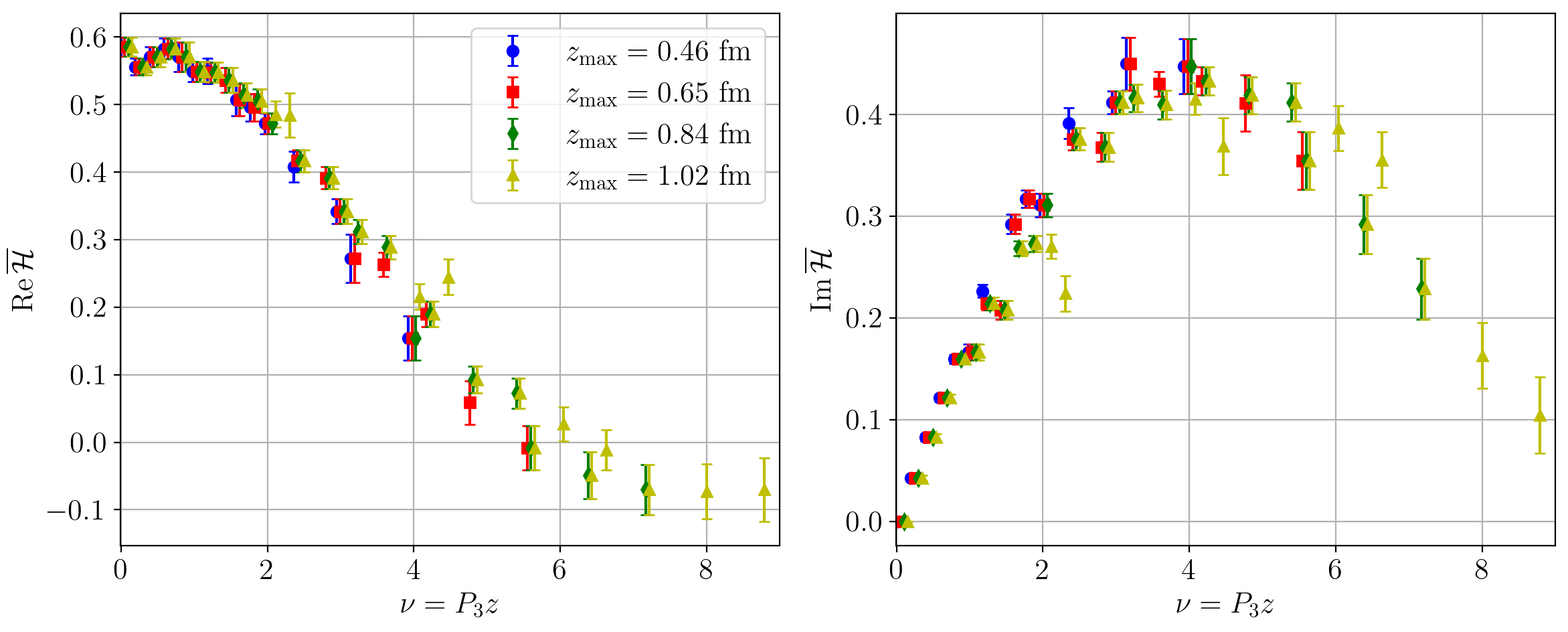}
    \includegraphics[width=0.9\textwidth]{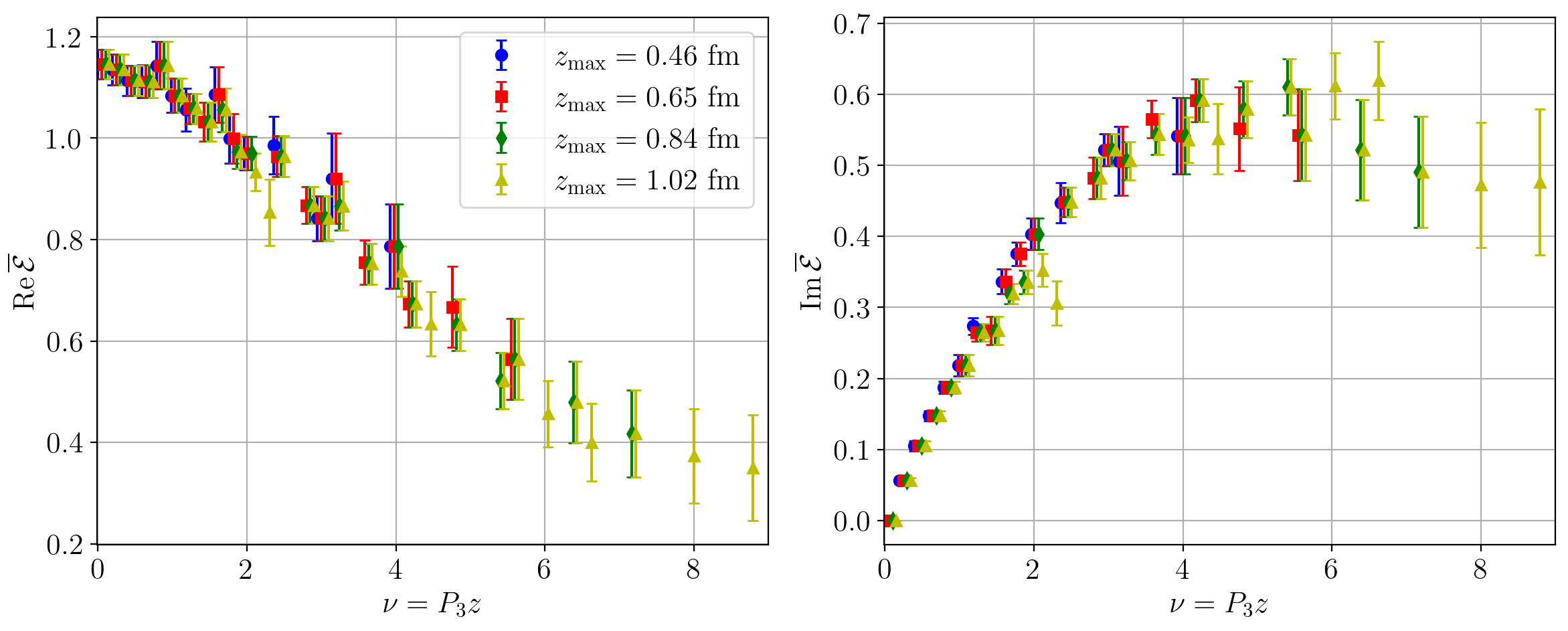}
    \caption{Matched $H$ (top) and $E$ (bottom) ITDs at $-t=0.65 $ GeV$^2$ after averaging data from different combinations $(P_3,z)$ at the same Ioffe time. We show four cases of $\zmax$ and the data are slightly shifted for better visibility. The left/right panels show the real/imaginary part.}
    \label{fig:nu_averaged_matched_Q2}
\end{figure*}
\begin{figure*}[t!]
    \centering
    \includegraphics[width=0.9\textwidth]{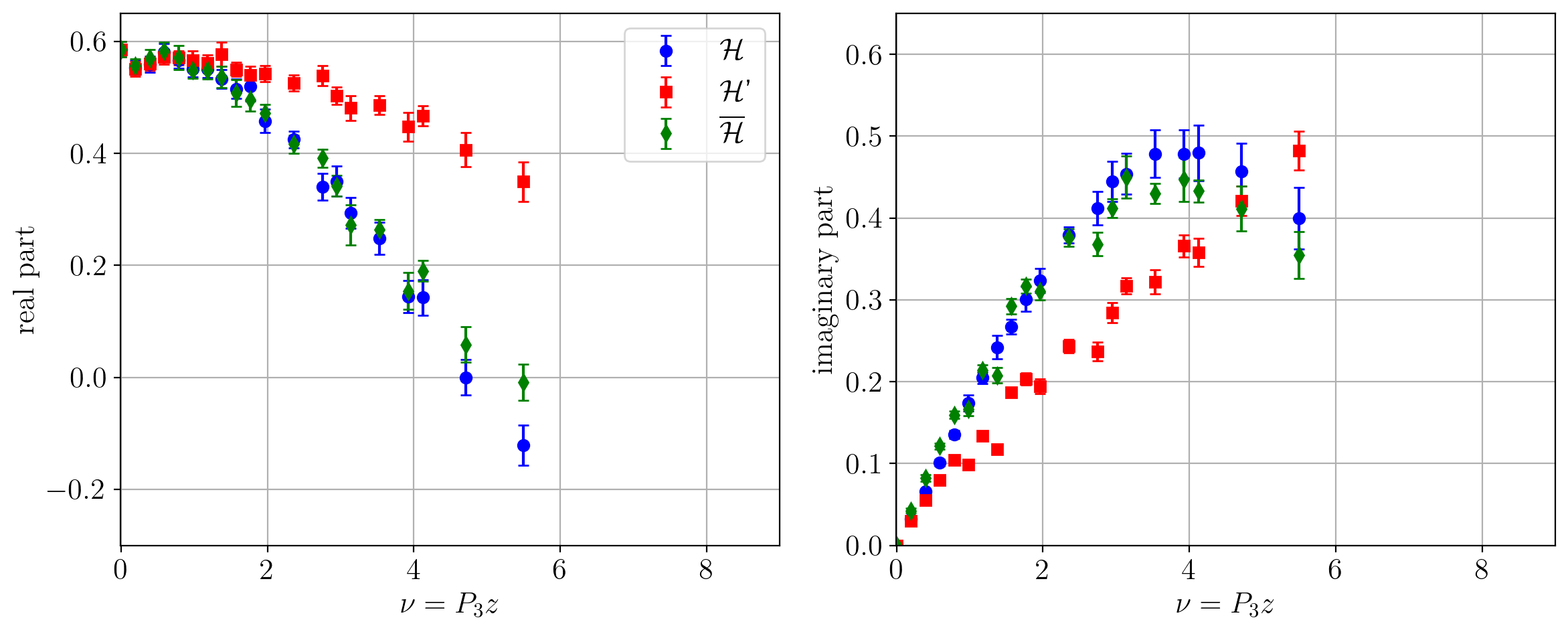}
    \includegraphics[width=0.9\textwidth]{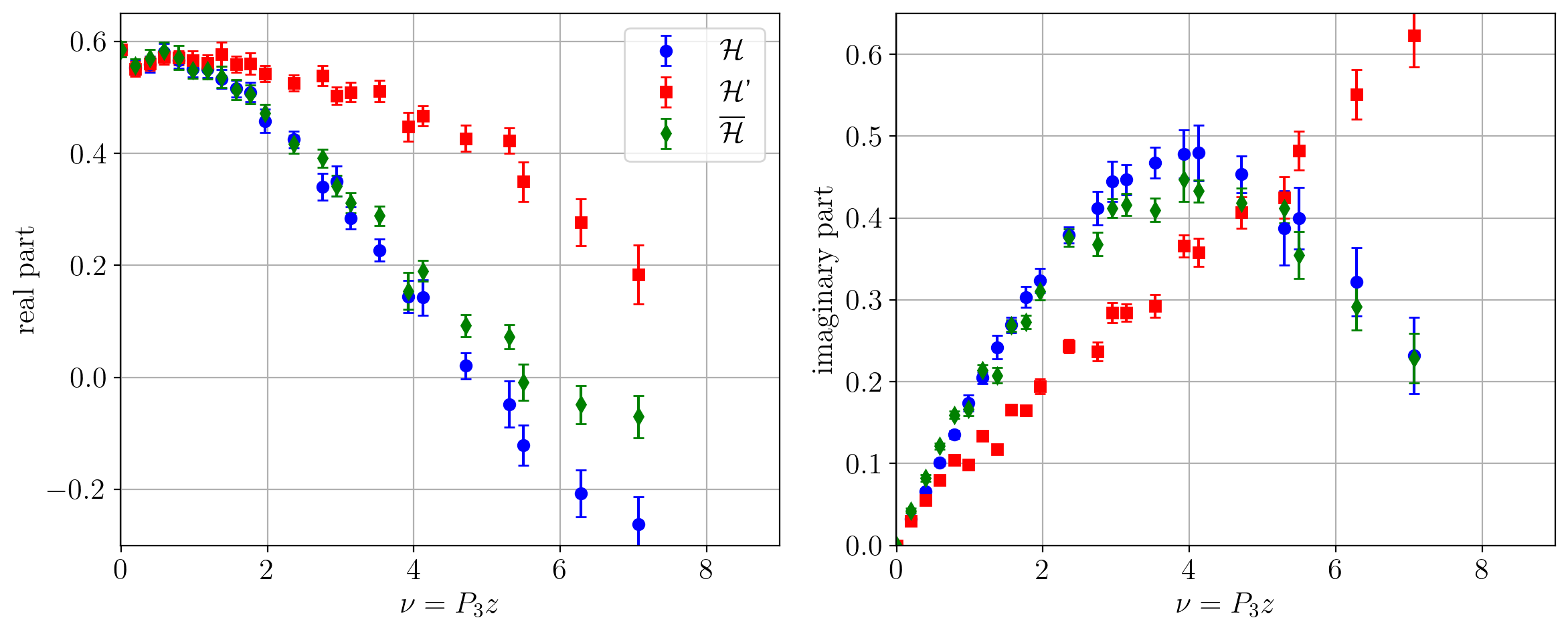}
    \caption{Real (left) and imaginary (right) part of reduced (blue circles), evolved (red squares) and matched (green diamonds) H-ITDs, averaged over the combinations of $(P_3, z)$ with the same Ioffe time. Upper plots are computed at $\zmax=0.65$ fm and bottom plots at $\zmax = 0.84$ fm.}
    \label{fig:nu_averaged_all_Q2_H}
\end{figure*}

\begin{figure*}[t!]
    \centering
    \includegraphics[width=0.9\textwidth]{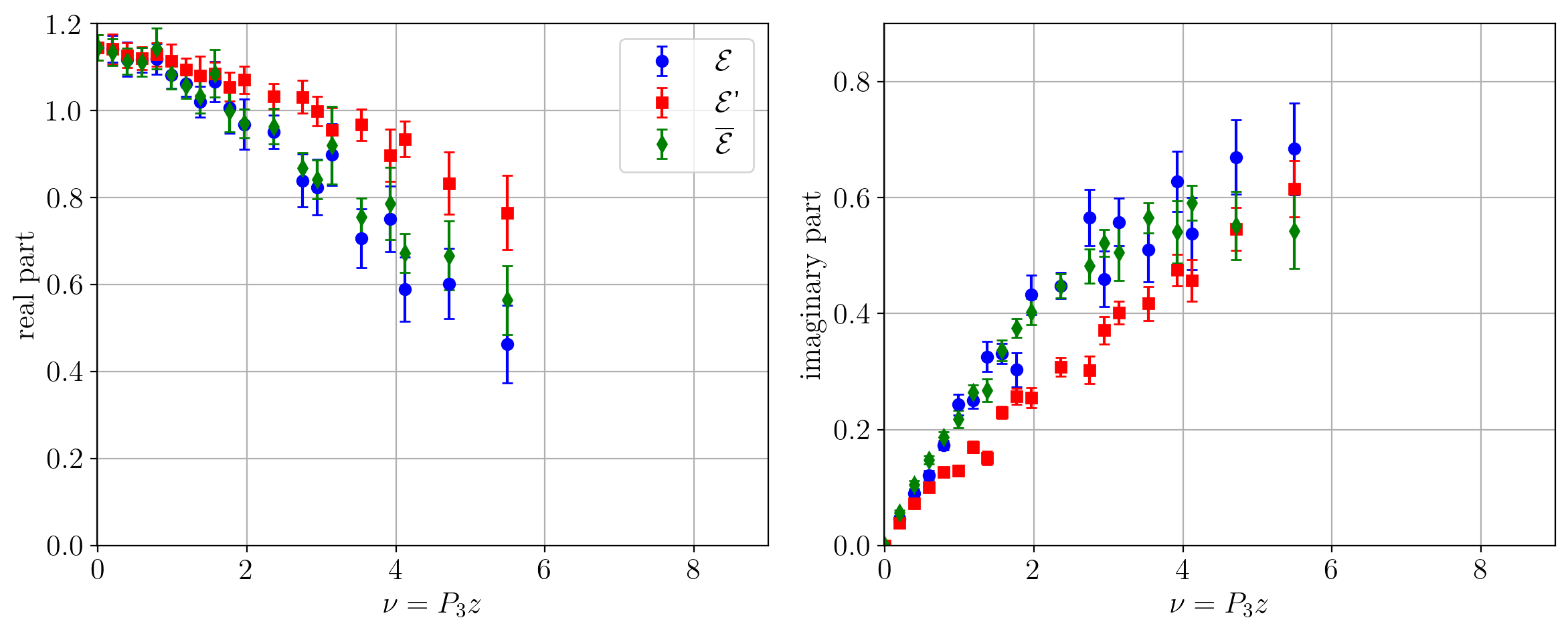}
    \includegraphics[width=0.9\textwidth]{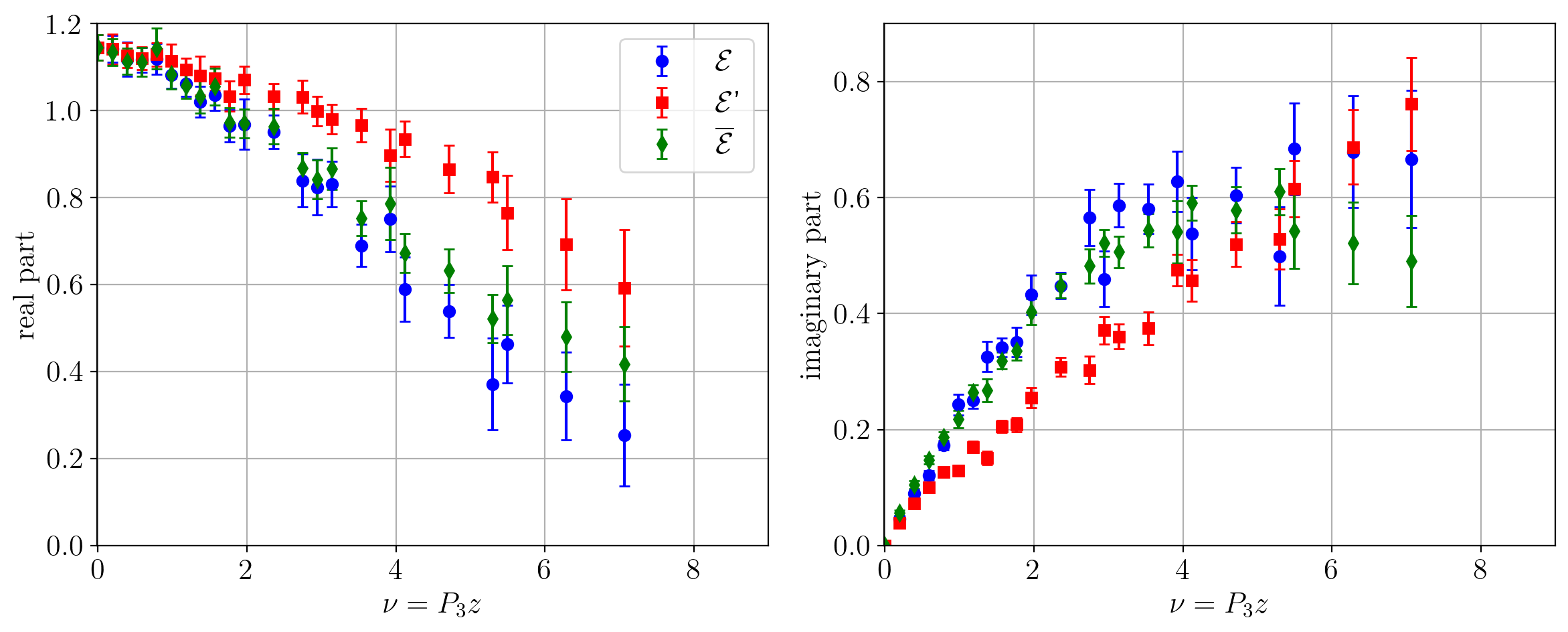}
    \caption{Real (left) and imaginary (right) part of reduced (blue circles), evolved (red squares) and matched (green diamonds) E-ITDs, averaged over the combinations of $(P_3, z)$ with the same Ioffe time. Upper plots are computed at $\zmax=0.65$ fm and bottom plots at $\zmax = 0.84$ fm.}
    \label{fig:nu_averaged_all_Q2_E}
\end{figure*}
\begin{figure*}[t!]
    \centering
    \includegraphics[width=\textwidth]{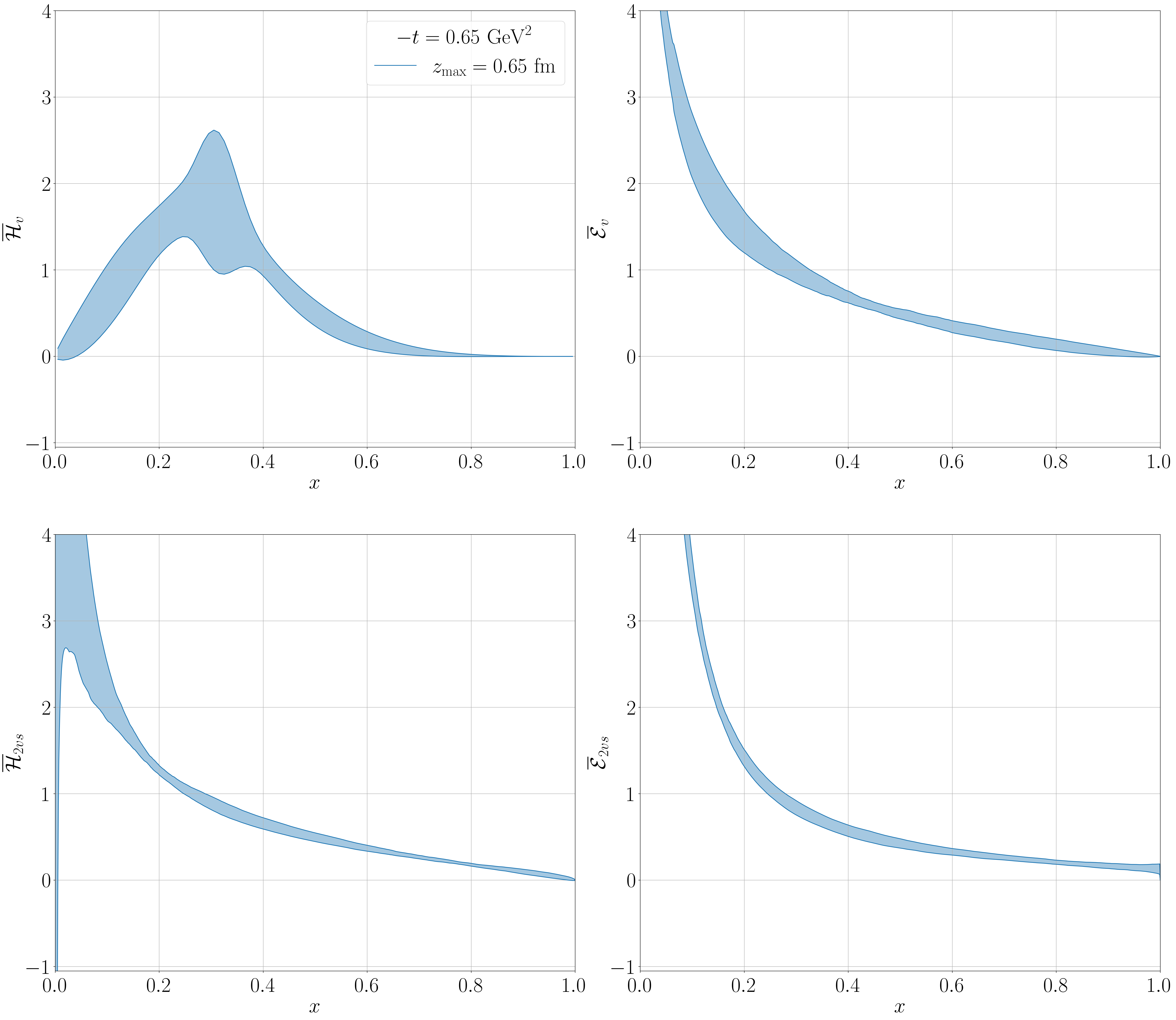}
    \caption{Reconstructed $x$-dependent $H$ (left) and $E$ (right) GPDs at $-t=0.65 $ GeV$^2$ and $\zmax=0.65$ fm. The top/bottom row depicts the fits of the real/imaginary part of matched ITDs (valence/v2s distributions).}
    \label{fig:constraintless_Q2}
\end{figure*}

\begin{figure*}[t!]
    \centering
    \includegraphics[width=\textwidth]{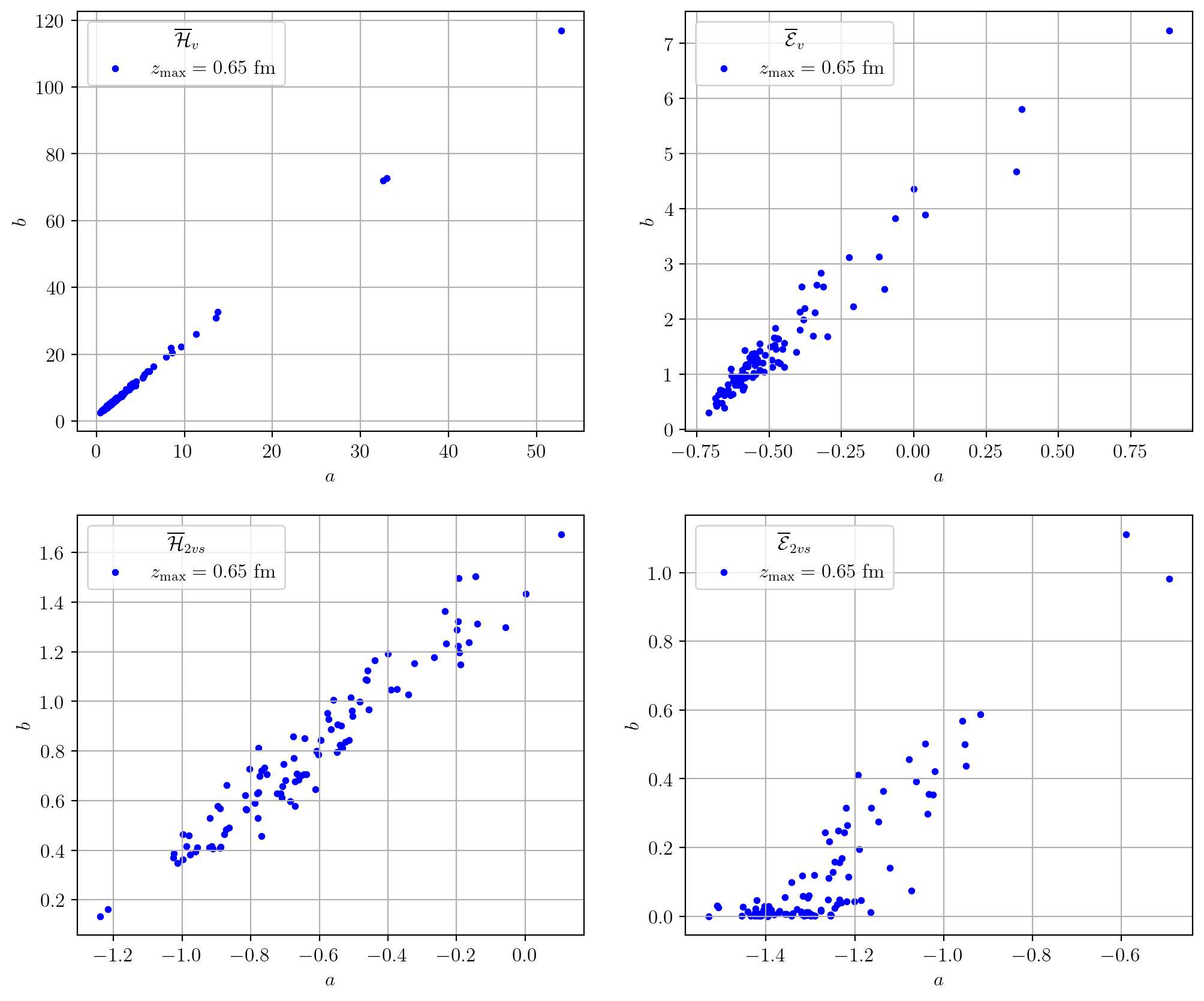}
    \caption{Correlation plots of the $a$ and $b$ fitting parameters in the reconstruction of $H$ (left) and $E$ (right) GPDs at $-t=0.65 $ GeV$^2$ and $\zmax=0.65$ fm. The top/bottom row depicts the fits of the real/imaginary part of matched ITDs (valence/v2s distributions). Each point represents the result of one bootstrap sample.}
    \label{fig:scatter_constraintless_Q2}
\end{figure*}

\begin{figure*}[t!]
    \centering
    \includegraphics[width=\textwidth]{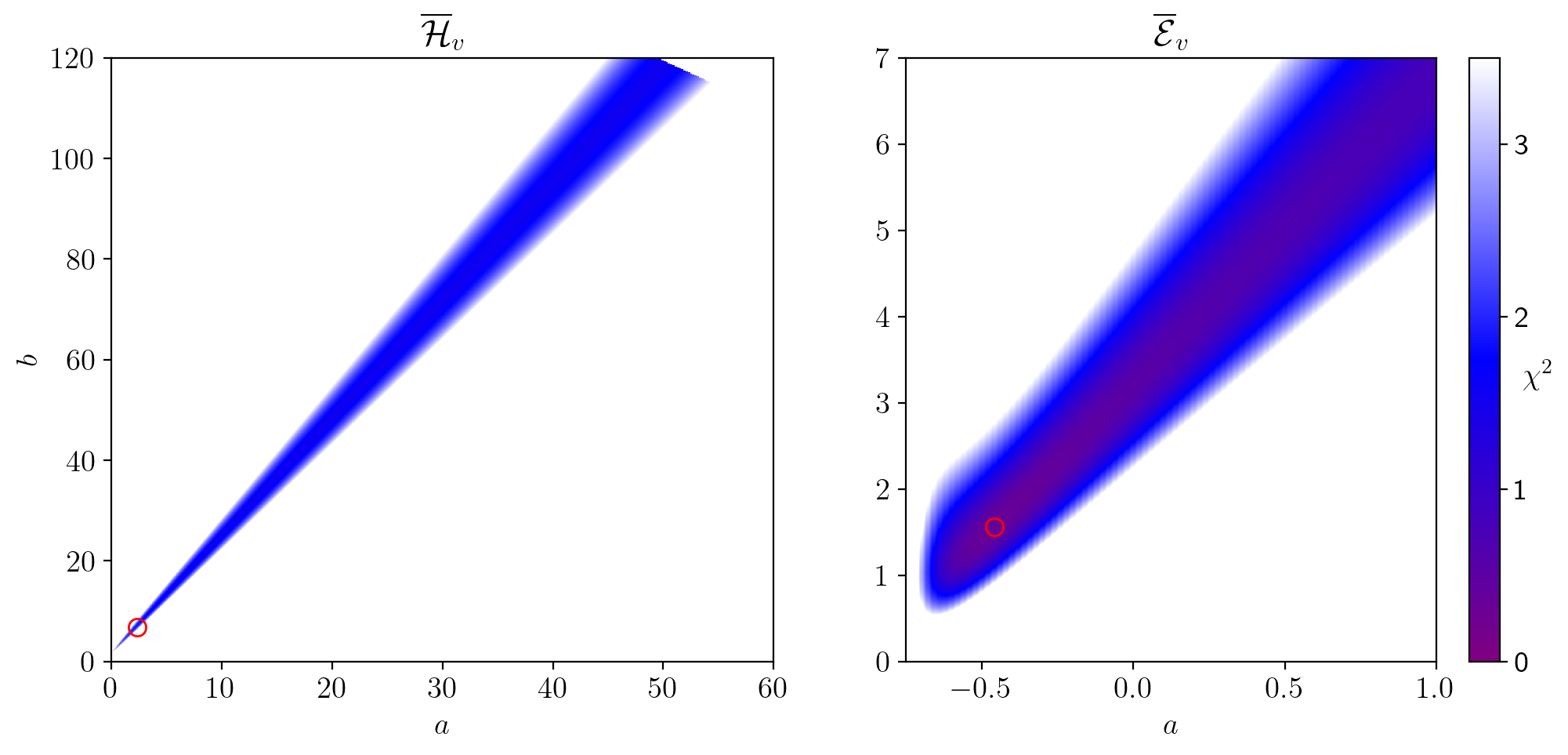}
    \caption{Mapping of the value of the $\chidof$ function (color-coded), defined in Eq.~(\ref{eq:chi2}), for the fitting reconstruction of valence $H$ (left) and $E$ (right) GPDs at $-t=0.65 $ GeV$^2$ and $\zmax=0.65$ fm.}
    \label{fig:samp0_fullab_Q2}
\end{figure*}
\begin{figure*}[t!]
    \centering
    \includegraphics[width=\textwidth]{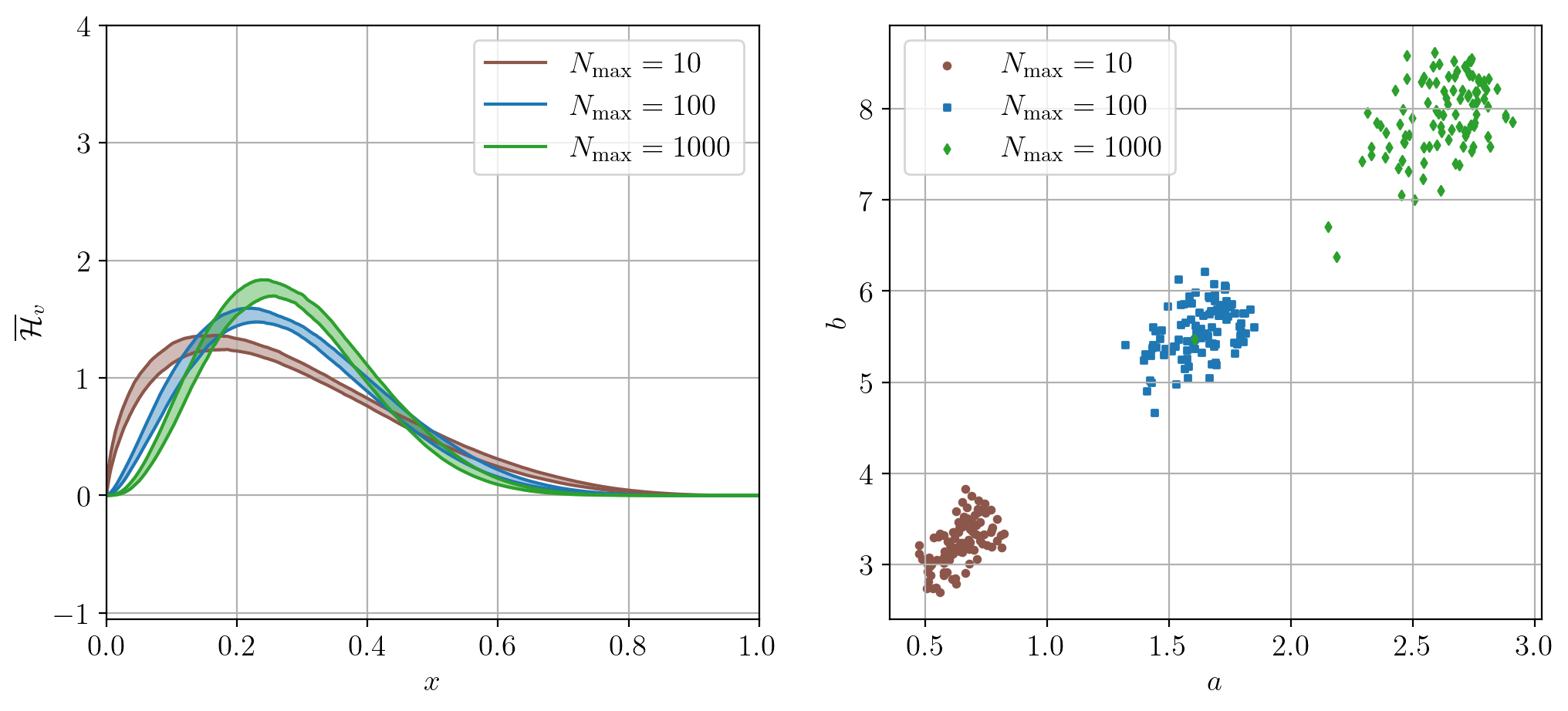}
    \caption{Valence distribution $\overline{\mathcal{H}}_v$ at $-t=0.65 $ GeV$^2$ with $\zmax=0.65$ fm and its corresponding fitting parameters $a$ and $b$ depicted with three different values of $\Nmax$.}
    \label{fig:premax_comparison_Q2}
\end{figure*}

\begin{figure*}[t!]
    \centering
    \includegraphics[width=\textwidth]{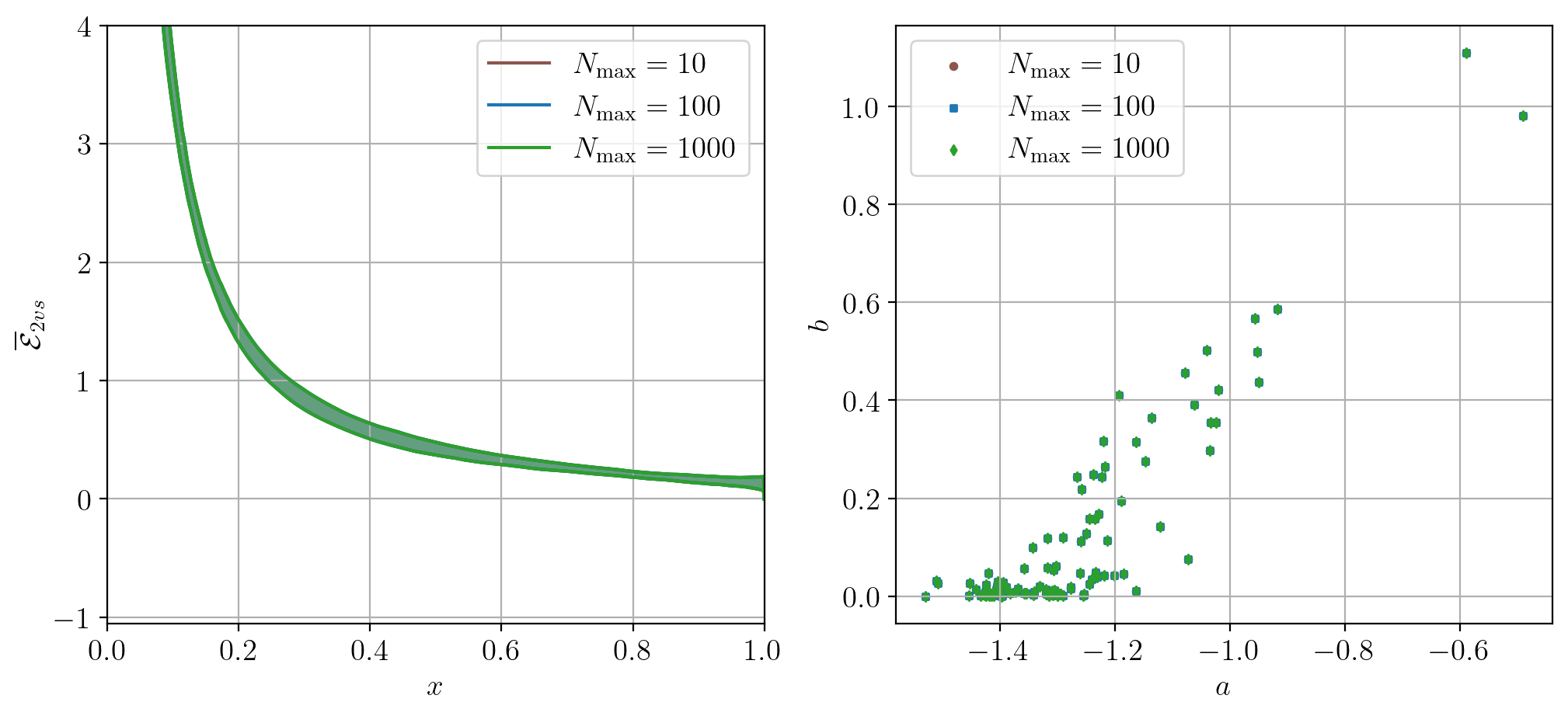}
    \caption{2vs distribution $\overline{\mathcal{E}}_{2vs}$ at $-t=0.65 $ GeV$^2$ with $\zmax=0.65$ fm and its corresponding fitting parameters $a$ and $b$ depicted with three different values of $\Nmax$. All data points overlap with ones without the $\Nmax$ constraint, indicating that this constraint is not in effect.}
    \label{fig:premax_comparison_E_Q2}
\end{figure*}

\begin{figure*}[t!]
    \centering
    \includegraphics[width=\textwidth]{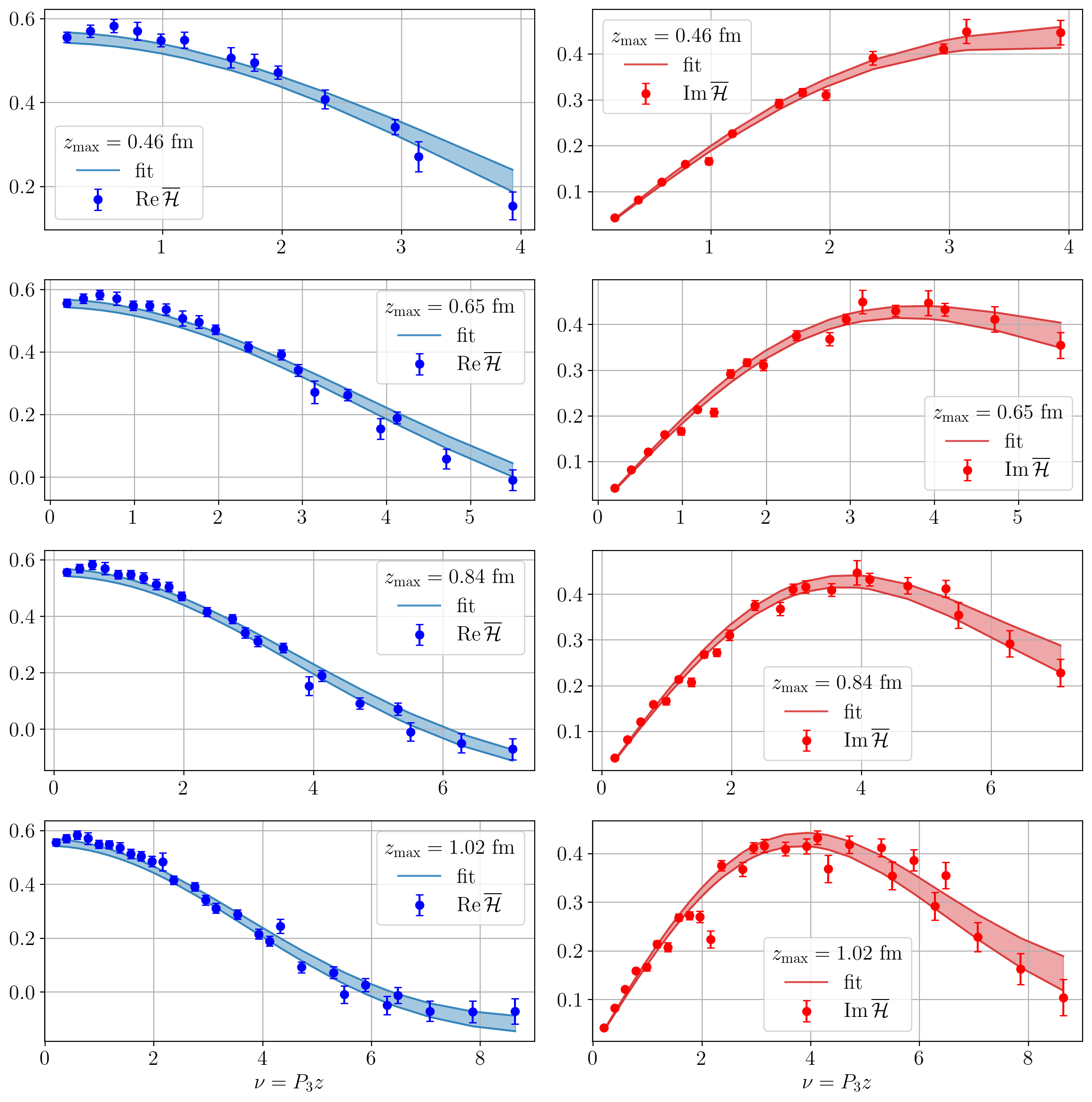}
    \caption{Real (left) and imaginary (right) part of matched ITDs $\Hbar$ and fitted ITDs at $-t=0.65 $ GeV$^2$, $\zmax$ increases from top to bottom.}
    \label{fig:fitted_vs_matched_Q2}
\end{figure*}

\begin{figure*}[t!]
    \centering
    \includegraphics[width=\textwidth]{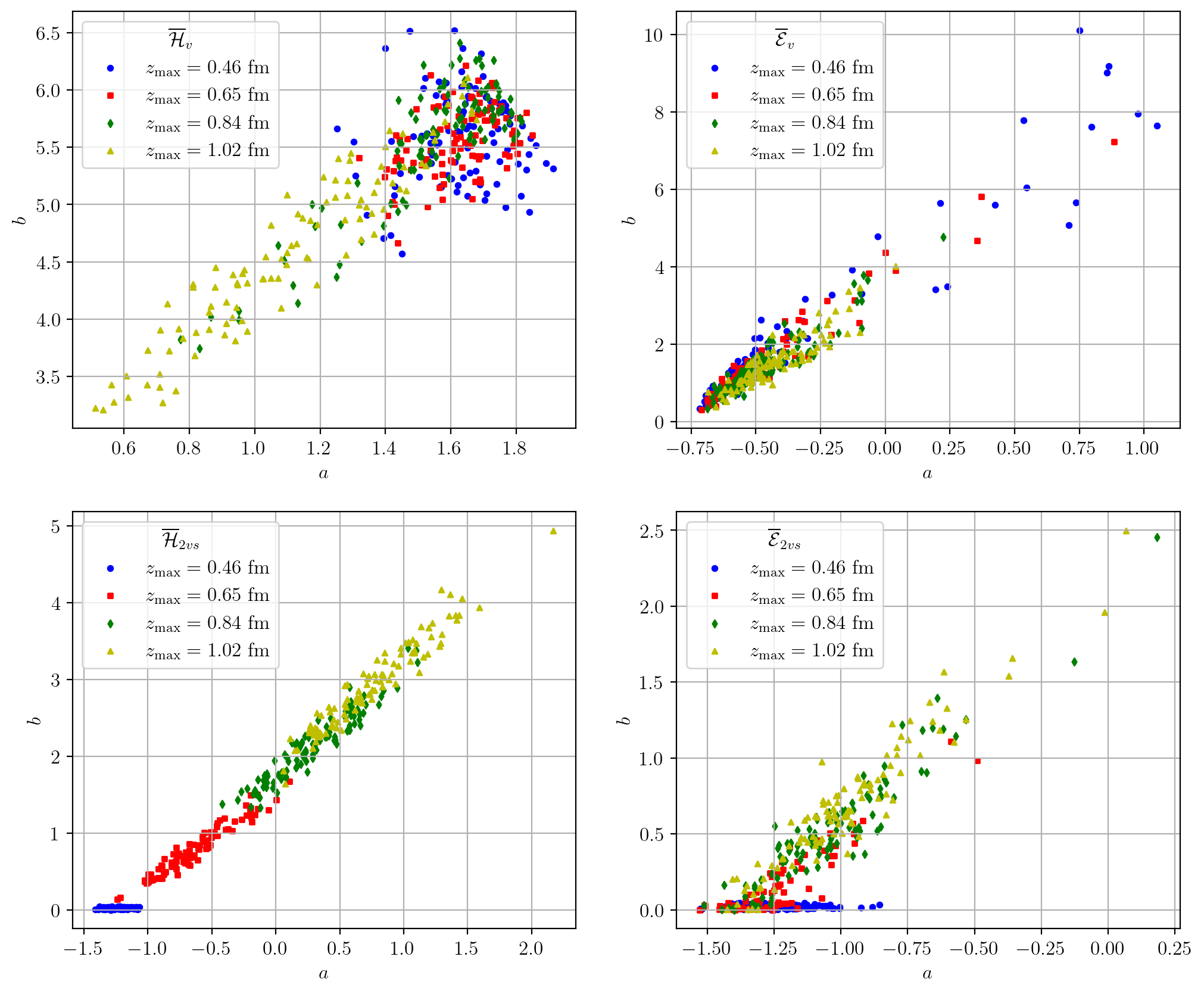}
    \caption{Correlation plots of the $a$ and $b$ fitting parameters in the reconstruction of $H$ (left) and $E$ (right) GPDs at $-t=0.65 $ GeV$^2$ and four values of $\zmax$ represented by different colors. The top/bottom row depicts the fits of the real/imaginary part of matched ITDs (valence/v2s distributions). Each point represents the result of one bootstrap sample. The red points ($\zmax=0.65$ fm) are identical to the points in Fig.~\ref{fig:scatter_constraintless_Q2} for $\Hvs$, $\Eval$ and $\Evs$ ($N\leq\Nmax$ constraint ineffective), while for $\Hval$, the picture with imposed $\Nmax$ is significantly different. }
    \label{fig:fitting_params_scatter_Q2}
\end{figure*}

\begin{figure*}[t!]
    \centering
    \includegraphics[width=\textwidth]{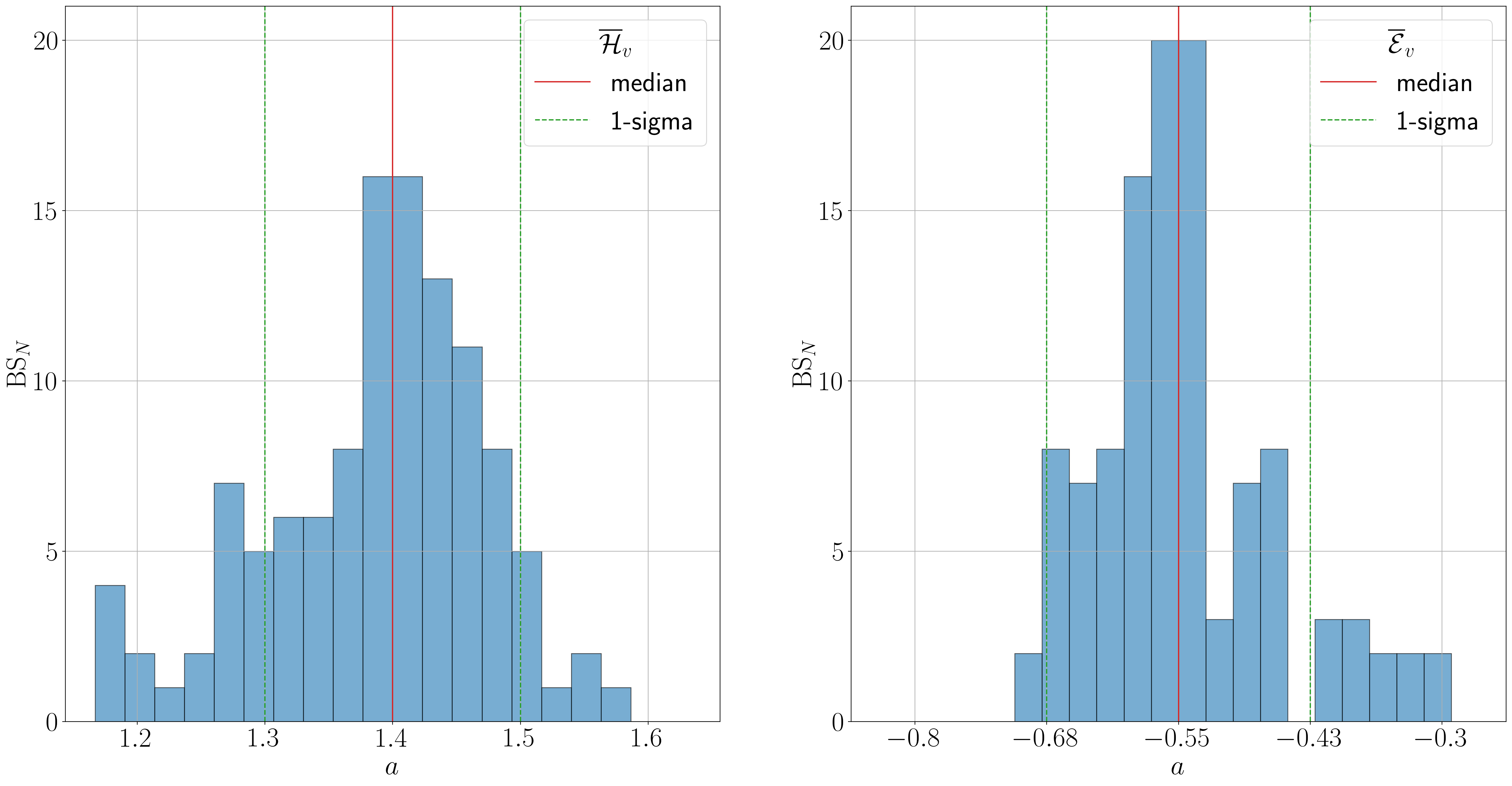}
    \caption{Fitting parameter histograms (numbers of bootstrap samples in a given interval) for the valence distributions $\Hval$ and $\Eval$ at $-t=0.65 $ GeV$^2$ and $\zmax = 0.65$ fm. We show the median of the distribution with a vertical red line. The 1-sigma vertical green lines correspond to the 16th and 84th centiles, such that they coincide with one standard deviation for a Gaussian distribution. }
    \label{fig:fitting_params_histo_Q2}
\end{figure*}

\begin{figure*}[t!]
    \centering
    \includegraphics[width=\textwidth]{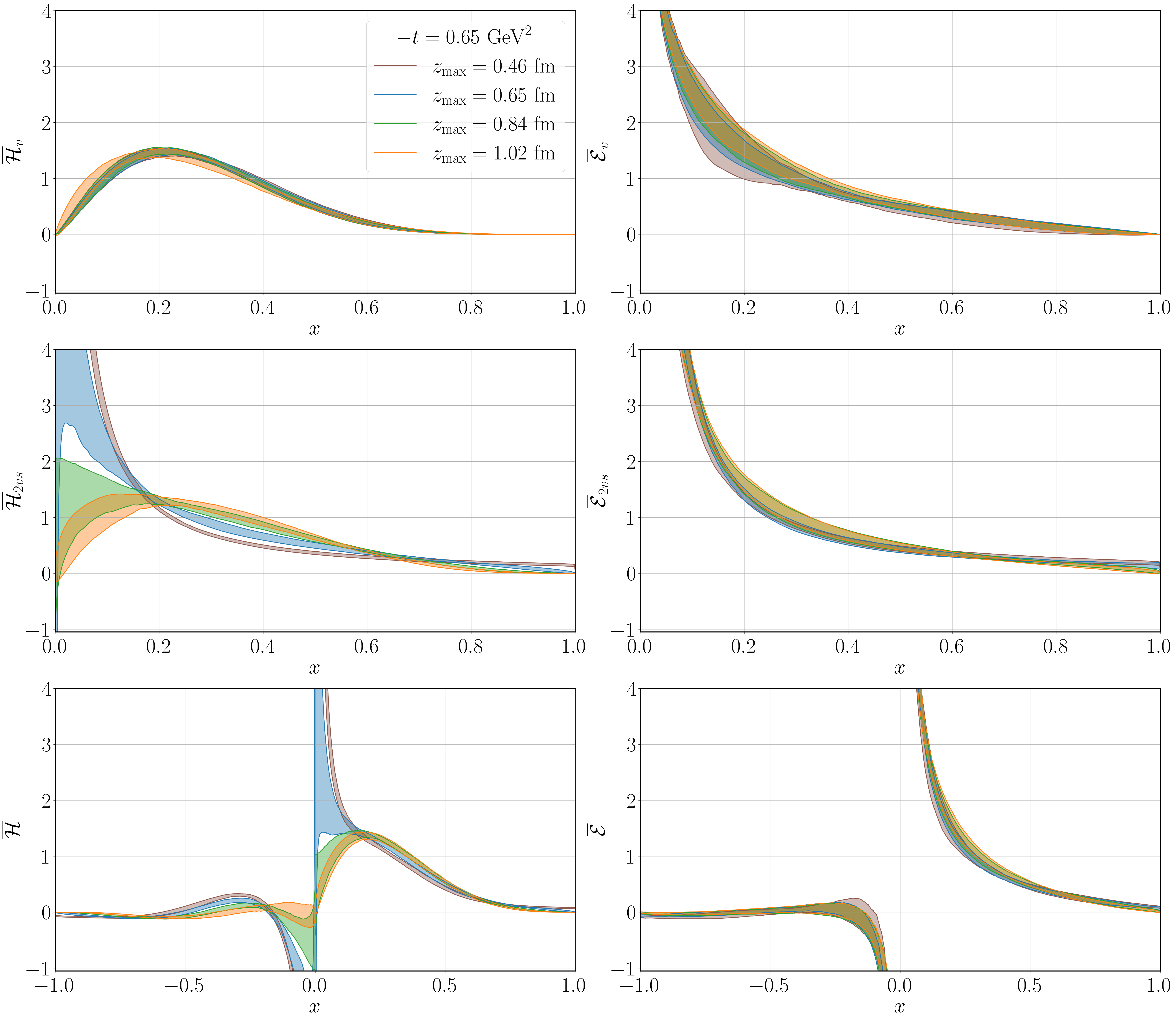}
    \caption{Reconstructed $x$-dependent $H$ (left) and $E$ (right) GPDs at $-t=0.65 $ GeV$^2$ with four different $\zmax$ values. The top row depicts the valence distributions ($\Fval$; from the real part of ITDs), the middle row the valence plus twice sea distributions ($\Fvs$; from the imaginary part of ITDs), and the bottom row the valence plus sea distribution ($\Fbar$; mixing real and imaginary parts of ITDs.}
    \label{fig:reconstructed_Q2}
\end{figure*}

\begin{figure*}[t!]
    \centering
    \includegraphics[width=\textwidth]{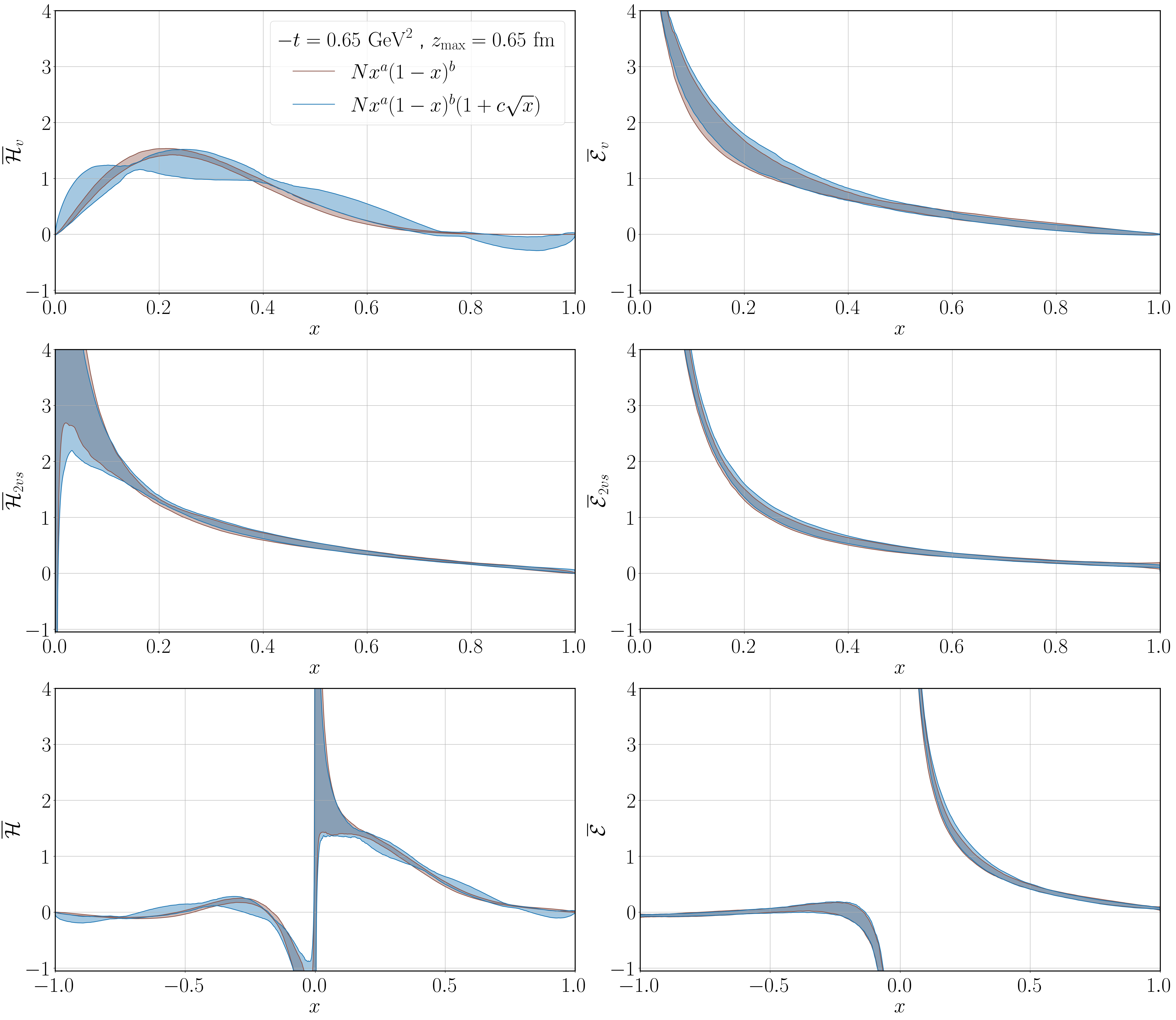}
    \caption{Comparison of fitting reconstruction of $\Hval$ (upper left), $\Eval$ (upper right), $\Hvs$ (middle left), $\Evs$ (middle right, $\Hbar$ (lower left) and $\Ebar$ (lower right) using the simplest fitting ansatz (purple bands) and an ansatz with an additional fitting parameter (blue bands), at $-t=0.65 $ GeV$^2$, $\zmax=0.65$ fm.}
    \label{fig:2par_3par_comparison_Q2}
\end{figure*}
\begin{figure*}[t!]
    \centering
    \includegraphics[width=\textwidth]{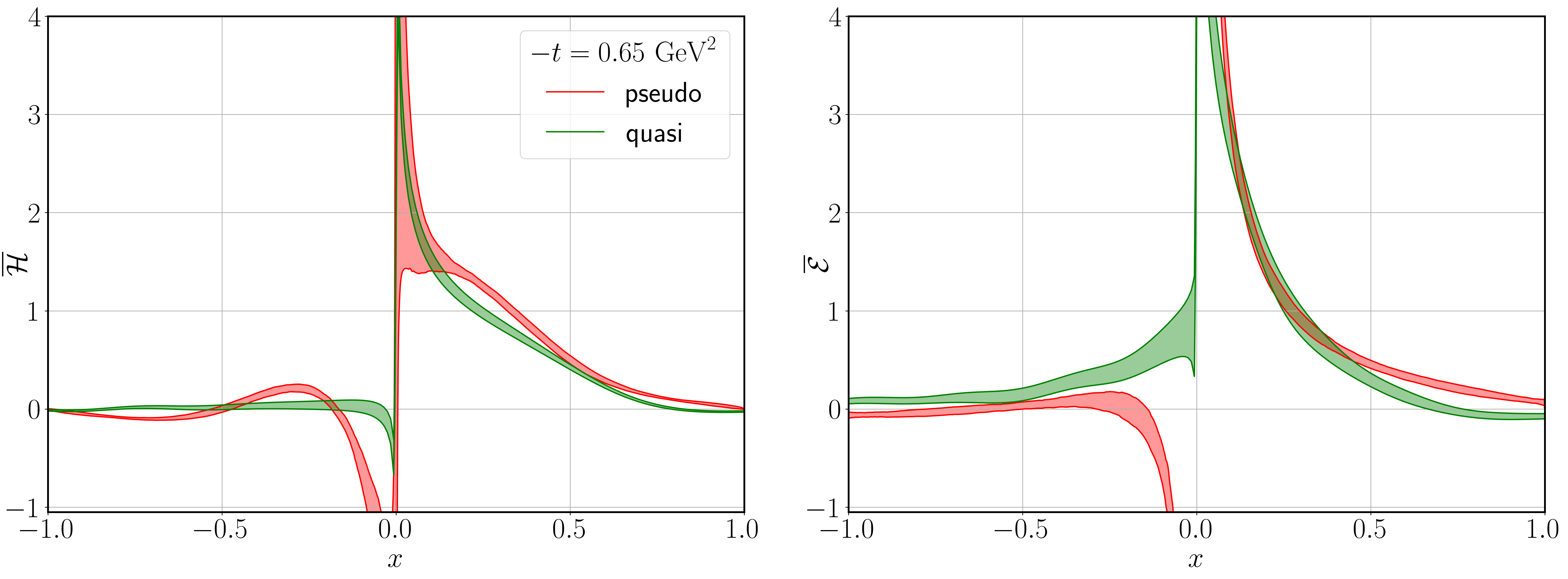}
    \caption{Comparison between reconstructions using the pseudo-distribution approach (red bands) and the quasi-distribution approach (green bands). Left/right panel for $\Hbar$/$\Ebar$, $-t = 0.65$ GeV$^2$, $\zmax=0.65$ fm for pseudo-GPDs.}
    \label{fig:reconstructed_quasicomp_no22_Q2}
\end{figure*}
\begin{figure*}[t!]
    \centering
    \includegraphics[width=\textwidth]{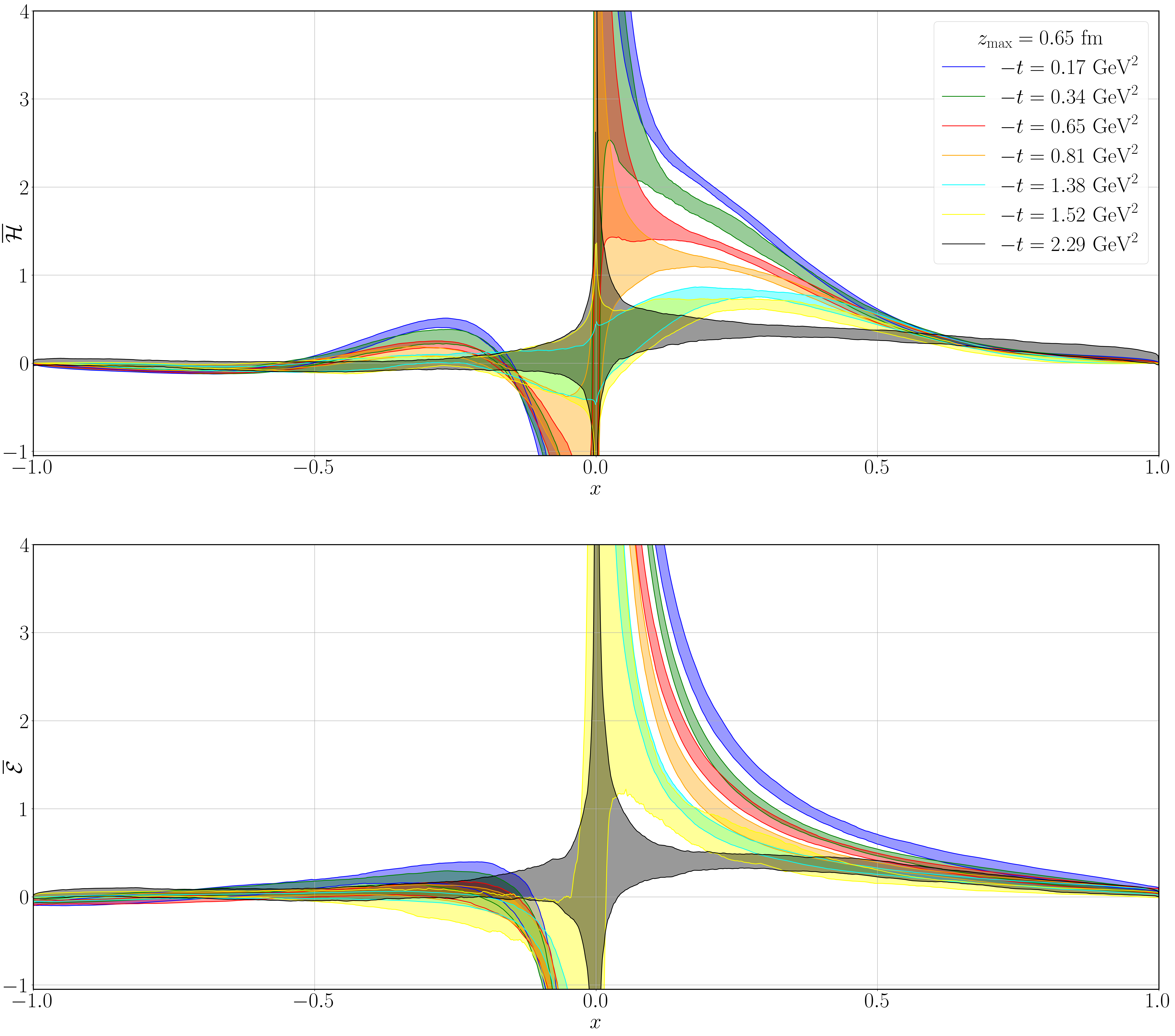}
    \caption{Momentum transfer dependence of the reconstructed $\Hbar$ (top) and $\Ebar$ (bottom) GPDs. All reconstructions use the preferred value of $\zmax=0.65$ fm.}
    \label{fig:reconstructed_tdep_no22_Q2}
\end{figure*}

\section{Results}
\label{sec:results}
In this section, we present a detailed illustration of our analysis.
For our example case, we choose $-t=0.65 $ GeV$^2$, i.e.\ the value of momentum transfer that was thoroughly analyzed in Ref.\ \cite{Bhattacharya:2022aob} within the quasi-GPD framework.
The quasi-distribution analysis employed only one value of the nucleon boost in the asymmetric frame, $P_3=1.25$ GeV.
Here, we reuse the data for this boost, but, as pointed out above, we complement it with the nucleon momenta 0, 0.42, 0.83, and 1.67 GeV.

\subsection{Ioffe time distributions}
We start by showing our reduced ITDs for this case, calculated according to Eq.~(\ref{eq:DR}), see Fig.~\ref{fig:reduced_Q2}.
The plots include data points up to $z\approx1$ fm, of which the largest Wilson line lengths are clearly inappropriate for the matching procedure.
We note that the data at small and intermediate values of $z$ align close to universal curves.
However, we observe some incompatibilities between the data from different $P_3$ at very small $z$ in the real part of the $\mathcal{H}$ function and the imaginary part of the $\mathcal{E}$ function.
In particular, the $z=0$ value in Re$\,\mathcal{H}$ is slightly off between the lowest and the largest boost.
Since this value should be boost-independent, we ascribe the $\approx2$-$\sigma$ difference to a statistical fluctuation.
Concerning statistical errors, we note they are roughly comparable for all boosts in the intermediate-$z$ regime.
Only at small-$z$, the precision in the low-$P_3$ ITDs is somewhat better than in the large-$P_3$ ones.
However, interestingly, at fixed Ioffe times of $v\approx2$, the precision of large-$P_3$ ITDs (originating from small $z$'s) exceeds the one at the lowest boost (from comparatively large $z$).
Finally, we observe that the real part seems to be relatively more precise than the imaginary part, implying a better signal for the valence distribution than the valence plus twice the sea.
Similarly, the $H$ function tends to have smaller errors than the $E$ GPD.

Reduced ITDs for every combination $(P_3,z)$ are subjected to the evolution procedure that brings them from their specific $1/z$ scales to a common scale of $\mu=2$ GeV.
This process involves the creation of the interpolation functions $\mathcal{F}_u(P_3, z)$. 
As an example, we show quadratic fits at three values of $z=0.28,\,0.56$, and 0.84 fm, see Fig.~\ref{fig:interpolated_Q2}.
The curvature encoded in the second-order polynomial is enough to provide a perfect description of the real part and a satisfactory one of the imaginary part.
We tested an addition of a cubic term in the imaginary part, which obviously improves the $\chi^2$ value of the fits, but $\chidof$ of the fits is roughly unchanged with one fewer degree of freedom (DOF).
The impact of including this cubic term on the final results of the evolution is also minor.
Thus, we implement the quadratic interpolations for all cases.

The results of the evolution procedure are shown in Fig.~\ref{fig:evolved_Q2}.
In the real part, the evolution to a common scale has an effect universally increasing the value of the ITD with respect to its value at the scale $1/z$, making the Ioffe time dependence almost flat at small $\nu$, particularly in the $H$ GPD.
The effect in the imaginary part is the opposite up to $\nu\approx5$, and changes sign around this value, effectively removing the maximum at $\nu\approx4$ observed in the imaginary part of reduced ITDs.
Overall, the differences between ITDs at a fixed value of $\nu$, but from different nucleon boosts, tend to increase upon the evolution to a common scale.

The next stage of the perturbative process translates pseudo-distributions in the ratio scheme to light-cone ones in the $\MSb$ scheme, without changing the scale, kept constant at $\mu=2$ GeV.
The matched ITDs are depicted in Fig.~\ref{fig:matched_Q2}. 
As observed in earlier pseudo-PDF analyses, e.g.\ in Refs.\ \cite{Orginos:2017kos,Bhat:2020ktg,Karpie:2021pap,Bhat:2022zrw}, the effect of the matching and the scheme conversion is very close to the evolution effect, but with an opposite sign.
Hence, matched ITDs are rather close to reduced ITDs; see also below after averaging data from different combinations of $(P_3,z)$ and the same $\nu$.

We argued above that this stage of the procedure provides a practical criterion for the choice of $\zmax$ used in the $x$-space dependence reconstruction.
Recalling the argument, we consider $\zmax$ to be the value for which matched ITDs, $\Fbar(P_3,z\leq\zmax)$, do not depend on the initial scale or boost, as long as they correspond to the same Ioffe time.
To establish $\zmax$ robustly, we discuss each of the four cases, real/imaginary parts of $\Hbar/\Ebar$, separately.
We also discuss other values of the momentum transfer, with the corresponding plots shown in the appendix (Figs.\ \ref{fig:matched_Q1}-\ref{fig:matched_Q4}).
\begin{itemize}
\item Re$\,\Hbar$ -- at $-t=0.65$ GeV$^2$, the agreement between fixed-$\nu$ matched ITDs persists up to $z\approx0.8$ fm with the main tension seen between the two intermediate boosts.\footnote{We ignore the tension at very small $z$, evinced already at $z=0$ and ascribed to a statistical fluctuation.}
This conclusion persists for values of $-t\lesssim1.4$ GeV$^2$, while larger momentum transfers evince agreement at even larger values of $z$ due to increased errors.
\item Im$\,\Hbar$ -- at $-t=0.65$ GeV$^2$, tensions start already around $z\approx0.6$ fm, indicating possibly larger HTEs induced by the sea quarks (absent in the real part of ITDs).
The conclusion is again valid for other values of $-t$, with agreement slightly extended for the two largest momentum transfers.
\item Re$\,\Ebar$ -- in this case, the larger statistical errors imply compatible fixed-$\nu$ ITDs even at $z\approx1$ fm, for all momentum transfers.
\item Im$\,\Ebar$ -- similarly to the real part, the errors are somewhat enhanced with respect to the $\Hbar$ function, implying better compatibility between different boosts. Nevertheless, tensions start to be seen around $z\approx0.7$ fm at $-t=0.65$ GeV$^2$, particularly at the lowest boost.
At other values of $-t\lesssim1.4$ GeV$^2$, tensions are seen to develop between $z=0.6$ fm and $z=0.8$ fm, while $-t\gtrsim1.5$ GeV$^2$ implies no tensions even at $z\approx1$ fm.
\end{itemize}
Overall, the behavior observed at $-t=0.65$ GeV$^2$ is representative of all values of the momentum transfer, with only the two largest values of $-t=1.52,\,2.29$ GeV$^2$ extending the viable range of $z$ to at least 1 fm, simply as a consequence of increased statistical noise.
Thus, the value of $\zmax$ robust from the point of view of perturbative matching is seen to be rather universal in our data, with somewhat better agreement between matched ITDs originating from different nucleon boosts seen in the real part, i.e.\ probing only the valence distribution.

The final stage of the coordinate-space analysis is to average matched ITDs that originate from different combinations of the boost and the Wilson line length but correspond to the same Ioffe time.
We test this averaging with four distinct values of $\zmax$, ranging from 0.46 to 1.02 fm, to better see the effects of contaminating the data with exceedingly large $z$ values, see Fig.\ \ref{fig:nu_averaged_matched_Q2}.
The effect of increasing $\zmax$ has a two-fold effect.
The obvious one is to extend the range of covered Ioffe times, but also data at smaller Ioffe times are affected.
For example, the Ioffe time $\nu\approx1.57$ can be obtained from the combinations $(P_3,z/a)=(2\pi/L,8),(4\pi/L,4),(8\pi/L,2)$.
With $\zmax/a=7$, only the last two enter the average, while $\zmax/a=9$ also includes the lowest boost.
Overall, the smoothness of the $\nu$-averaged curves depends on the considered case, according to the above discussion.
The inclusion of $\zmax=1.02$ fm data leads to non-smooth curves in all cases, apart from the real part of $\Ebar$.
All other cases are relatively smooth, with somewhat better behavior of $\zmax=0.84$ fm in the real parts as compared to the imaginary parts.

In the end, we single out two $\zmax$ values that lead to the best compromise between the covered range of Ioffe times and the validity of the perturbative evolution and matching, $\zmax=0.65$ fm and $\zmax=0.84$ fm.
The former can be considered rather conservative for the real parts of matched ITDs, but it is the most proper for the imaginary parts.
The latter, in turn, is somewhat less conservative and can have enhanced HTEs in the distributions, including sea quarks.
Below, we will also compare these two preferred choices for $\zmax$ with one smaller ($z=0.46$ fm) and one larger value ($z=1.02$ fm) to better reflect the influence of this parameter.
We summarize the perturbative process from reduced via evolved to matched ITDs in Fig.~\ref{fig:nu_averaged_all_Q2_H} ($H$ ITD) and Fig.~\ref{fig:nu_averaged_all_Q2_E} ($E$ ITD).
The most conspicuous feature is that the effects of evolution and matching with scheme conversion are almost identical, although with the opposite sign.
Thus, matched ITDs are compatible with reduced ITDs almost in the whole range of Ioffe times, with the most notable deviation from this behavior at $\nu\gtrsim5$ and only in the real part of the $H$ function (a tendency towards this behavior is obscured in the $E$ function due to larger errors).
The matched ITDs, $\Hbar$ and $\Ebar$, are approximately aligned on universal curves, with some irregularities observed predominantly in the imaginary parts.
The latter again indicates a potential problem when including the effects of sea quarks.
We also again note an artefact of the incompatibility of small-$z$ data in the real part of $\mathcal{H}$, between the boosts $\{2,4\}\pi/L$ and $\{6,8\}\pi/L$, manifesting in the irregularity at the lowest Ioffe times.
In practice, this leads to an inflation of the $\chidof$ function in the fitting reconstruction of $\Hbar$, from not being able to capture the difference between data at Ioffe times originating from only the lowest boosts ($\nu=\{2,4\}a/L$) and the ones including the two larger ones (($\nu=\{6,8\}a/L$).
We emphasize that this issue is not indicative of large HTEs at these small Ioffe times, but it should be attributed to a statistical fluctuation and kept in mind in the $x$-dependence reconstruction when evaluating the quality of the fits.

\subsection{$x$-dependent distributions}
The final step of the analysis is to reconstruct the $x$-dependence of the GPDs from matched ITDs averaged over data from different nucleon momenta, utilizing Wilson line lengths up to the selected $\zmax$.
We follow the strategy discussed in Section \ref{sec:pseudo} and start with the simplest fitting ansatz, including the parameters $a,\,b$.
We show examples of such fits for $\zmax=0.65$ fm, depicted in Fig.~\ref{fig:constraintless_Q2}.
A clear problem is encountered in the reconstruction of the valence part of the $H$ GPD (upper left panel).
To understand the problem, we show the values of the $(a,\,b)$ fitting parameters for all bootstrap samples, see Fig.~\ref{fig:scatter_constraintless_Q2}.

This plot reveals a striking behavior for $\Hval$ -- the values of $a$ and $b$ are almost perfectly correlated and span a huge range of values, from slightly positive up to $a\approx50$ and $b\approx120$, with the corresponding $N$ prefactor of the fit getting up to around $10^{50}$.
As such, the GPD is strongly suppressed at small and large $x$, with visibly positive values only in a relatively short range of intermediate $x$. It is still subject to much larger errors than other distributions.
The fitting reconstruction for this case is, thus, numerically ill-defined.
This is further illustrated in the left panel of Fig.~\ref{fig:samp0_fullab_Q2}, where we map out the underlying $\chidof$ function for one of the bootstrap samples.
The plot depicts a very shallow minimum of this function, with the global minimum shown with a red circle.
However, the value of $\chidof$ is almost constant along the direction of the correlation between the $a$ and $b$ parameters.
In practice, for different bootstrap samples, it implies that small differences between them are translated into huge differences in the position of the global minimum of $\chidof$.
Below, we propose a solution to this problem based on an additional constraint on the value of the prefactor $N$, a quantity that depends on both fitting parameters of the valence distribution.
However, it needs to be remembered that results utilizing such a constraint have to be considered with care, and the problem's occurrence calls for better lattice data to be obtained in the future.
We note that the problem is present only at small values of $\numax$ and only in the valence distribution.
Concerning the former,  better lattice data means that $\numax$ needs to be extended without increasing $\zmax$, i.e.\ it shows the need for data at larger nucleon boosts.
However, interestingly, the problem does not occur for any other distribution than $\Hval$, as shown in the other panels of Fig.~\ref{fig:scatter_constraintless_Q2} ($\Eval$ -- upper right, $\Hvs$ -- lower left, $\Evs$ -- lower right).
For all these cases, the correlation between the $a$ and $b$ parameters is still significant but visibly smaller than in the case of $\Hval$, and it does not imply a huge range of values for these parameters for different bootstrap samples.
The $\chidof$ function, shown for $\Eval$ in the right panel of Fig.~\ref{fig:samp0_fullab_Q2}, reveals again a rather shallow global minimum, but robust enough to be limited to the range $a\in(-0.7,-0.4)$ for 75\% of bootstrap samples and to $a\in(-0.7,0)$ for 95\% of samples. 

In view of the observed correlation between the $a$ and $b$ parameters, we propose the following strategy to decorrelate the parameters.
The essence of the problem is that small differences between bootstrap samples translate to huge differences in the pairs $(a,b)$ corresponding to the global minimum, with small differences in the value of this function along the direction of the correlation.
Thus, instead of a single global minimum, there is, in practice, a direction of minima of the $\chidof$ function with very close values of this function but considerably different physical implications.
The formal global minimum for some bootstrap samples implies a large-$x$ behavior of $(1-x)^{100}$, small-$x$ suppression of $x^{50}$ and a value of the prefactor of $10^{50}$.
This is clearly nonphysical, and we propose to constrain these coefficients for fitting results that are accepted.
Since the fitting parameters are strongly correlated, we choose to impose this constraint on the prefactor $N$, looking for a minimum of the $\chidof$ function in the parameter subspace such that $N\leq\Nmax$.

In Figs.~\ref{fig:premax_comparison_Q2}, \ref{fig:premax_comparison_E_Q2}, we show the effect of introducing this $\Nmax$ constraint for the ill-behaved case of $\Hval$ and the well-behaved case of $\Evs$, respectively, using three values of $\Nmax=10,\,100,\,1000$.
Starting with the well-behaved case, it can be clearly seen that the $\Nmax$ constraint is not in effect, and all results coincide with the unconstrained one, shown in the lower right panels of Figs.~\ref{fig:constraintless_Q2}, \ref{fig:scatter_constraintless_Q2}.
Meanwhile, the ill-behaved case is significantly affected by the constraint.
The almost perfect correlation covering $a\lesssim50$, $b\lesssim120$ and $N\lesssim10^{50}$ is replaced by a clump of $(a,b)$ pairs with $N\leq\Nmax$ by construction.
It is interesting to see the implication of the different $\Nmax$ constraints on the maximum allowed values of the $a$ and $b$ parameters:
\begin{itemize}
\item $\Nmax=10$ implies $a\lesssim0.8$, $b\lesssim 3.8$,
\item $\Nmax=100$ implies $a\lesssim1.8$, $b\lesssim 6.2$,
\item $\Nmax=1000$ implies $a\lesssim2.8$, $b\lesssim 8.7$.
\end{itemize}
The constraint on the small- and large-$x$ behavior of the GPD may be too restrictive with $\Nmax=10$, but seems physically reasonable at $\Nmax=100$, in view of typical values of this coefficient, as well as the powers of $1-x$ at large $x$, found e.g.\ in global fits of PDFs~\cite{Accardi:2016qay}. Theoretically~\cite{Brodsky:1973kr,Brodsky:1974vy,Sivers:1982wk}, it is also expected that the parameter $b$ is within the range covered by the choice $\Nmax=100$.
We assume that results for zero-skewness GPDs should not be drastically different.
For the remainder of the paper, we choose $\Nmax=100$ and present results according to this choice.
We note that the result from $\Nmax=100$ is compatible with the unconstrained one within the much larger uncertainties of the latter.
Thus, the constraint effectively acts by replacing the fitting results for samples with $N>\Nmax$ by the result corresponding to the minimum of the $\chidof$ function restricted to $N\leq\Nmax$, which leads to a very slight increase of the value of this function.
Nevertheless, the constraint's existence is a warning about the limited kinematic range of the data used in the reconstruction.
We emphasize that, in the future, a clean solution to this issue should consist of obtaining improved lattice data for these cases, meaning predominantly larger nucleon boosts, allowing for the inclusion of a broader range of Ioffe time at fixed $\zmax$. 
Luckily, the current problem is restricted to only a few cases.
We summarize it as follows:
\begin{itemize}
\item $\Hval$ -- the constraint $N\leq\Nmax$ is effective for $\zmax=0.46,\,0.65$ fm and ineffective for $\zmax=0.84,\,1.02$ fm,
\item $\Hvs$ -- the constraint is ineffective at all $\zmax$,
\item $\Eval$ -- the constraint is ineffective at all $\zmax$,
\item $\Evs$ -- the constraint is ineffective at all $\zmax$.
\end{itemize}

Now, we discuss the results of our fits that employ $\Nmax=100$, emphasizing again that this constraint is relevant only for $\Hval$ at $\zmax=0.46,\,0.65$ fm.
The fits at the level of ITDs are displayed in Fig.~\ref{fig:fitted_vs_matched_Q2}, for all four considered values of $\zmax$, with fits of the real/imaginary part shown in the left/right panel.
For each case, we present the matched ITDs as discrete data points and the fitted ITDs as a continuous band, being the inverse Fourier transform of the fitting ansatz.
Overall, the fits provide a good description of the data, particularly for the two preferred intermediate values of $\zmax$.
The two extreme $\zmax$ are inappropriate for different reasons --
the smallest one restricts the Ioffe time range too much, while the largest one includes data severely contaminated by HTEs, which may cause large deviations of some points from the fitting band.

Before we present the $x$-dependent distributions, we discuss again the correlation plots of the $a$ and $b$ parameters, see Fig.~\ref{fig:fitting_params_scatter_Q2}.
The upper left panel summarizes the situation for $\Hval$.
The two lower values of $\zmax$ are the ones for which the constraint $N\leq\Nmax$ is in effect and behave qualitatively in a similar manner, with clumping of the $(a,\,b)$ pairs.
The two larger $\zmax$ values lead to behavior more akin to the other distributions, shown in the upper right and lower panels.
In all cases, the correlation between $a$ and $b$ is clearly visible, but the values of both fitting parameters are restricted to comparatively narrow intervals.
A different kind of behavior is manifested for the distributions related to the imaginary part of ITDs (lower panels), for $\zmax=0.46$ fm (both $\Hvs$ and $\Evs$) and $\zmax=0.65$ fm (only $\Evs$).
The limited Ioffe time range, which at the lowest $\zmax$ ($\numax\approx4$) does not even capture the presence of the maximum of the imaginary part, leads to the fits favoring negative values of the $b$ parameter for most of the bootstrap samples and triggers the constraint $b\geq0$.
Thus, most of the samples are attributed $b=0$ by the fitting routine.
Finally, we also show histograms of the $a$ parameter for $\Hval$ and $\Eval$ at $\zmax=0.65$ fm.
In both cases, i.e.\ with and without the $\Nmax$ constraint being in effect, the distributions of the fitting parameter are approximately Gaussian.
We show the median of the distributions with a vertical red line and also vertical green lines corresponding to the 16th and 84th centiles, such that around 68\% of realizations of $a$ are between these lines.
In the ideal Gaussian case, the error defined by such centiles would coincide with the standard deviation of the Gaussian distribution.
For $x$-dependent PDF extraction, we choose to define the central values and the errors based on the median and these centiles to avoid distortion of the average and the error defined as standard deviation by some fitting outliers induced by numerical instabilities in minimizing the $\chidof$ function.

Our $x$-dependent distributions are summarized in Fig.\ \ref{fig:reconstructed_Q2}.
In the upper left panel, we display the fits of $\Hbar_v$ at four values of $\zmax$.
We recall that our preferred choice is $\zmax=0.65$ fm, but the reconstructed GPDs are consistent between all $\zmax$ and almost ideally overlap for the three lower choices.
In particular, the agreement between $\zmax=0.65$ fm (that hits the $\Nmax$ constraint) and $\zmax=0.84$ fm (where $N$ is always naturally below $\Nmax$) allows us to conclude a very robust reconstruction.
The situation is considerably different for $\Hbar_{v2s}$ (middle left panel).
Here, we observe significant dependence on $\zmax$.
Specifically, the result at $\zmax=0.46$ fm is affected by almost all samples having $b=0$, which leads to artificially suppressed errors and the unphysical behavior of $\Hbar_{v2s}(x=1)>0$.
At larger $\zmax$, the fitting parameter $b$ is robustly non-zero (hence, $\Hbar_{v2s}(x=1)=0$), but there are still significant differences between $\zmax=0.65$ fm and $\zmax=0.84$ fm for several ranges of $x$.
As we have argued above, the larger of these two $\zmax$ values is likely contaminated by HTEs in the data from $z/a=8,\,9$.
Hence, we consider the distribution obtained at $\zmax=0.65$ fm to be more reliable.
Finally, extending the Ioffe time range even further with data at $z/a=10,\,11$ ($\zmax=1.02$ fm) mostly has the effect of suppressing the errors, without tension in central values with respect to $\zmax=0.84$ fm.
Comparing $\Hbar_v$ and $\Hbar_{v2s}$, the different behavior can be attributed to the effects of sea quarks, absent in $\Hbar_v$, which, apparently, induce additional HTEs.
However, it is interesting to consider also the distribution that combines the real and imaginary parts, $\Hbar$, shown in the lower left panel.
At $x>0$, it sums the contributions from valence and sea quarks, while at $x<0$, the valence part is identically zero, i.e.\ $\Hbar(x<0)=-\Hbar_s(x>0)$.
The suppressed sea-quark contribution for $x>0$ eases the tension between different $\zmax$ choices, leading to compatible results for $\zmax\leq0.84$ fm in the range of $x\gtrsim0.1$.
The antiquark (fully sea) part is, obviously, still characterized by tensions between different $\zmax$ and, thus, should be considered unreliable.
Overall, we conclude robust reconstruction of the full $\Hbar$ distribution in the range $x\gtrsim0.1$, and we attribute the problems at small-$x$ and $x<0$ to potentially large HTEs related to sea quarks.

Now, we analyze the results of the $x$-dependence of $\Ebar$.
For both $\Ebar_v$ and $\Ebar_{v2s}$ (upper and middle right panels of Fig.\ \ref{fig:reconstructed_Q2}), the outcomes are qualitatively similar to the case of $\Hbar_v$, with enlarged statistical errors.
The robustness of the reconstruction with respect to $\zmax$ is particularly visible in $\Ebar$ (lower right panel), which indicates compatible distributions at all $\zmax$ in the whole range of $x$.
This brings the natural question of why the picture for $\Ebar$ seems to be significantly more robust as compared to $\Hbar$.
In the latter case, we attributed the difficulties in some ranges of $x$ to enhanced HTEs induced by sea quarks.
This interpretation is lent credence in view of the convergence properties of the LI definition of $H$ and $E$ GPDs mentioned in Section \ref{sec:setup}.
Namely, it is consistent with the result reported in Ref.~\cite{Cichy:2023dgk}, where the $x$-dependent $H$ and $E$ GPDs were reconstructed following the quasi-distribution approach.
We observed that the standard and LI definitions give almost identical results for the $H$ case, whereas the $E$ function behaves in a substantially different manner.
Specifically, using the standard definition leads to incompatible results at $P_3=0.83,\,1.25,\,1.67$ GeV, while the LI variant evinces total agreement for all these boosts for all $x$.
Moreover, this is not the case for the LI $H$ function, showing incompatibility between the two largest boosts in the range between approximately $x=0.2$ and $x=0.5$.
We can, thus, conclude that the interplay of HTEs in the LI $E$ function is more favorable than in the $H$ GPD -- even if some cancellation of HTEs is largely accidental in the former, it translates to a more robust reconstruction for $E$.

Apart from the simplest fits employing the functional form $Nx^a(1-x)^b$, we also attempted to include more fitting parameters, according to Eq.~(\ref{eq:ansatz2}). We found that no additional parameters are relevant in any of the scenarios discussed below Eq.~(\ref{eq:ansatz2}), i.e.\ they turn out to be statistically insignificant or on the verge of statistical significance in some cases employing one additional fitting parameter.
We exemplify the results of including a $1+c\sqrt{x}$ correction to the leading functional form in Fig.~\ref{fig:2par_3par_comparison_Q2}, for $\zmax=0.65$ fm, with other cases (differing by $\zmax$ and/or $-t$) evincing similar behavior.
It is clear that the modification with respect to fits including only $a,\,b$ does not alter the reconstruction, the only visible effect being some enhancement of the error and introduction of a kind of oscillatory behavior in $\Hval$.
Hence, for our final results, we choose the simplest fits with only two parameters describing the shape of the GPDs.
We note that upon increased precision of the lattice data, additional fitting parameters will likely be required for a robust reconstruction.

The example of $-t=0.65$ GeV$^2$, discussed in detail above, is found to be representative of all other considered values of the momentum transfer.
The final $\Hbar$ and $\Ebar$ distributions for all $-t$ can be found in the appendix (Figs.~\ref{fig:appen_reconstructions_1}, \ref{fig:appen_reconstructions_2}).
We establish the following general conclusions.
\begin{itemize}
\item The $\Hbar$ GPD is robustly reconstructed in the positive-$x$ part, at least down to $x\approx0.1$, with some values of $-t$ evincing even better compatibility between $\zmax$ (particularly at large $-t$ that has enhanced errors and suppressed values). The leftover incompatibilities usually involve the two extreme values of $\zmax$, with generically good agreement between our preferred choices, $\zmax=0.65$ fm, and $\zmax=0.84$ fm.
\item The antiquark part of $\Hbar$ ($x<0$) is subject to enhanced HTEs, and the final results are compatible with zero.
\item The positive-$x$ $\Ebar$ is reconstructed with practically no dependence on $\zmax$ at any momentum transfer, likely due to the better convergence properties of the LI definition of the $E$ GPD, i.e.\ smaller contamination by HTEs.
\item The antiquark part of $\Ebar$ ($x<0$) likewise evinces perfect agreement between different $\zmax$. The values are, nevertheless, compatible with zero for $x\lesssim-0.1$, with indications of divergence at small negative $x$ for momentum transfers $-t\lesssim1$ GeV$^2$.
\end{itemize}
Before we show full $t$-dependence of GPDs from our data, it is interesting to compare the outcomes of our analyses at $-t=0.65$ GeV$^2$ from pseudo- and quasi-distribution approaches.

\subsection{Pseudo-GPDs vs.\ quasi-GPDs}
The pseudo- and quasi-distribution approaches utilize the same bare lattice data.
Thus, given the considerable computational cost to obtain such data, it is natural to follow both prescriptions.
In this paper, we analyze these data independently from our previous quasi-GPD work.
However, as discussed below, the approaches can be combined to augment each other in regions where their applicability may be questionable.

The treatment of the bare data is considerably different in both methods and implies distinct systematic effects.
Below, we summarize and discuss the main differences.
\begin{itemize}
\item Bare MEs are renormalized in a variant of the RI/MOM scheme (quasi) and in a ratio scheme (pseudo).
\item Renormalized MEs (ITDs) are first subjected to $x$-dependence reconstruction using the Backus-Gilbert method (quasi) and they are matched to light-cone ITDs still in coordinate space (pseudo).
\item In the next step, quasi-GPDs, already in momentum space, are matched to their light-cone counterparts. Light-cone ITDs in coordinate space are subjected to $x$-dependence reconstruction using a fitting ansatz. 
\end{itemize}
Note that some of the above differences are not intrinsic to the approaches but follow from standard practice.
For example, RI/MOM or ratio renormalization for pseudo/quasi could be used in both cases if appropriate matching equations are available.
Also, GPDs in the pseudo-distribution approach could be reconstructed with the Backus-Gilbert method (see Ref.~\cite{Bhat:2020ktg}), but the limited range of Ioffe times makes it less preferable.
Conversely, fitting ansatz reconstruction of quasi-GPDs is hindered by their non-canonical support in $x$, leaving the form of the ansatz unclear.
With the above differences, systematic effects in pseudo- and quasi-reconstructed distributions are significantly different.
Thus, a comparison of the final outcomes allows us to estimate the size of these systematics.

In Fig.~\ref{fig:reconstructed_quasicomp_no22_Q2}, we show a comparison of $H$ and $E$ GPDs extracted using both pseudo- and quasi-distribution procedures.
Once again, we use $-t = 0.65$ GeV$^2$ as an example, with the quasi-GPD data taken from Ref.~\cite{Bhattacharya:2022aob}.
Both reconstructed GPDs use the same bare matrix elements, with the quasi one limited to only one nucleon boost, $P_3=1.25$ GeV.

Starting with the antiquark part, one is forced to conclude that, at present, no definite conclusions can be drawn for the sea-quark distributions.
This agrees with our statements above, based exclusively on the pseudo-GPD method.
In the positive-$x$ part, the differences between quasi and pseudo are much smaller.
In $\Hbar$, the most prominent differences appear around $x=0.25$.
Interestingly, in $\Ebar$, we observe perfect agreement up to $x\approx0.5$, but clear differences for larger $x$.
Generally, we find the level of agreement in the positive $x$-region encouraging.

Nevertheless, it is of utmost importance to robustly quantify all sources of systematic effects and to introduce further refinements in the analysis procedures and in the lattice data.
The latter are the bare input in both approaches, and they can be improved in several ways.
The obvious aspect is to access data at different lattice parameters to estimate the traditional sources of lattice systematics.
For example, data at different lattice spacings will allow one to control discretization effects and ultimately extrapolate them out by taking the continuum limit.
Similarly, the physical quark masses limit can be reached by an extrapolation or, preferably, with simulations directly at the physical point.
However, the key aspect is to be able to access larger nucleon boosts since power corrections in the matching of both pseudo- and quasi-distributions can be naturally suppressed by larger $P_3$.
Unfortunately, while acquiring data at additional lattice spacings or pion masses poses no principal problem apart from computational time demand, reaching larger $P_3$ is an exponentially hard problem, i.e.\ the signal-to-noise ratio decays exponentially with each additional unit of lattice momentum.
Thus, for the time being, only moderate increases in $P_3$ are conceivable.

Another group of refinements involves analyzing the bare lattice data and concerns their renormalization and matching to the light cone.
There has been a lot of theoretical effort to improve these aspects, particularly in quasi-distributions.
It has been argued that more robust renormalization needs to be applied to avoid undesirable effects.
For example, the RI/MOM renormalization applied in the case of nonlocal operators is bound to introduce non-perturbative contaminations at large $z$ when converting the renormalization functions from the RI/MOM scheme to the $\MSb$ one.
A remedy for this can be the usage of hybrid schemes \cite{Ji:2020brr}, e.g., \ with the ratio scheme used at small $z$ and the power divergence at larger $z$ subtracted in a different manner, e.g., \ from the static potential \cite{Gao:2021dbh}.
A self-renormalization approach has also been proposed \cite{LatticePartonCollaborationLPC:2021xdx}, which extracts the required renormalization factors from suitable bare matrix elements that can be matched to a continuum scheme at short distance.
Further refinements involve subtraction of the linear renormalon \cite{Holligan:2023rex,Zhang:2023bxs} and resummations of logarithms both at small $x$ and large $x$ \cite{Gao:2021hxl, Su:2022fiu, Ji:2023pba}.
Analogous improvements can be done in short-distance factorization relevant for pseudo-distributions \cite{Gao:2021hxl,Braun:2024snf}.

In this paper, we have treated the pseudo- and quasi-GPD methods as alternatives to each other.
However, the optimal way of using both frameworks might be to combine them in a single analysis.
This was postulated in Ref.~\cite{Ji:2022ezo}.
In this reference, it was emphasized that factorization, either in momentum or in coordinate space, leads to different applicability for both approaches.
Quasi-distributions provide a way to directly calculate the $x$-dependence in an intermediate-$x$ region (from 0.15-0.20 to 0.80-0.85), with power corrections in the matching being enhanced both around $x=0$ and $x=1$.
Pseudo-distributions, instead, can probe light-cone correlations up to some Ioffe time $\numax$ above which power corrections again dominate.
Then, going to $x$-space necessitates some modeling of the large-$\nu$ behavior, typically done by employing some fitting ansatz.
Thus, one can imagine combining the two approaches by supplementing the information missing in one of them with the other.
Ref.~\cite{Ji:2022ezo} advocates using the information from coordinate-space factorization to provide constraints at small- and large-$x$.
However, another possibility is to extend the range of Ioffe times accessible in coordinate space by employing information from momentum-space factorization.
Developing such ideas, together with the above-mentioned improvements in lattice data and analysis methods, is likely to produce the most robust results for partonic distributions in the future.
When the pseudo- and quasi-distribution approaches are used independently, the discussed refinements should also bring them toward convergence by eliminating important sources of systematics.
In particular achieving larger boosts on the lattice will contribute towards extending the range of validity of both approaches, i.e.\ broadening the range of $x$ reliably extracted from quasi-distributions and the range of Ioffe times accessible via pseudo-distributions.

\subsection{$t$-dependence of GPDs}
We finalize by presenting the $t$-dependence from our analyses.
The upper/lower panel of Fig.~\ref{fig:reconstructed_tdep_no22_Q2} displays our results for $\Hbar$/$\Ebar$, with the invariant momentum transfer ranging from $-0.17$ GeV$^2$ to $-2.29$ GeV$^2$.\footnote{We chose to skip data at $-t=1.24$ GeV$^2$, which is numerically ill-behaved at the stage of fitting reconstruction, particularly for $\Ebar$ (see this case in the appendix). While the results at this $-t$ show no tension with other momentum transfers, the inclusion of this uncertainty-enhanced case in the $t$-dependence plot obscures the general picture.}
As expected, we observe monotonous suppression of both GPDs with increasing $-t$ at $x>0$, with the $E$ GPD being suppressed somewhat faster than its $H$ counterpart.
We also note that the picture for the $E$ GPD is better-behaved as compared to $H$.
Above, we speculated that this is related to the better convergence properties of the LI definition of $E$, with the accidentally more favorable interplay of HTEs, i.e., \ their probable cancellation between different amplitudes.
Overall, the picture summarized in Fig.~\ref{fig:reconstructed_tdep_no22_Q2} is akin to the analogous picture obtained with the quasi-GPD method, see Fig.~6 of Ref.~\cite{Cichy:2023dgk} for direct comparison.
While at the quantitative level, there are some differences between results from quasi- and pseudo-distributions (again, somewhat larger for the $H$ GPD), the striking similarities between the outcomes of both approaches promise good prospects for the future.
At the same time, we repeat that better lattice data are needed.
Thus, it needs to be kept in mind that the current statements are predominantly qualitative and quantitative and have to be postponed until robust estimation and subtraction of several systematic effects are achieved.

\section{Conclusions and prospects}
\label{sec:summary}
In this paper, we extracted, for the first time, the zero-skewness flavor non-singlet GPDs $H$ and $E$ of the nucleon from Radyushkin's pseudo-distribution approach.
We utilized lattice data at a single lattice spacing with a pion mass of about $260 \, \textrm{MeV}$.
We employed data previously used for an analogous analysis in Ji's quasi-GPD framework \cite{Bhattacharya:2022aob} at $P_3=1.25$ GeV, but we extended it significantly, adding three lower nucleon boosts and one larger, $P_3=1.67$ GeV.
We profited significantly from the acquisition of the data in asymmetric frames of reference \cite{Bhattacharya:2022aob}, which allowed us to obtain results for several momentum transfers in a single calculation.
Moreover, we used the Lorentz-invariant definition of pseudo-GPD matrix elements in coordinate space proposed in Ref.~\cite{Bhattacharya:2022aob}, which turned out to lead to very robust results, particularly for the GPD $E$.

The pseudo-distribution approach is a multistep procedure which starts from the same bare matrix elements as quasi-distributions.
These matrix elements are renormalized in a ratio scheme (leading to objects termed reduced or pseudo-ITDs), perturbatively evolved to a common scale (evolved ITDs), and matched to light-cone distributions in coordinate space (matched ITDs).
Finally, the reconstruction of the $x$-dependence follows, using fitting ansatzes similar to ones employed in standard global fits of experimental data.
Some remarks are in order to emphasize important aspects of the analysis.
The ratio scheme removes logarithmic and power divergences in the bare lattice data and plausibly reduces some of the higher-twist effects, allowing us to extend the lengths of Wilson lines beyond the naive perturbative values, $z\lesssim0.2-0.3$ fm.
In practice, HTEs start to become visible around $z=0.7-0.8$ fm.
We emphasized the importance of a reliable choice of $\zmax$, adopting a practical criterion involving agreement of matched ITDs extracted from different combinations of the nucleon boost ($P_3$) and the Wilson line length ($z$) but corresponding to the same Ioffe time $\nu=P_3z$.
We noted that the imaginary parts of ITDs are somewhat more sensitive to HTEs than the real parts.
In the end, we singled out two plausible choices for $\zmax$, $0.65$ fm, and $0.84$ fm, which appear to be the best compromise between the reliability of the perturbative procedure and the attained range of Ioffe times.
Upon both choices, the final $x$-dependent distributions $\Hbar$ and $\Ebar$, the sums of the valence and the sea parts of $H$ and $E$ GPDs, are compatible within uncertainties in the positive-$x$ region.
The negative-$x$ region is found to be much less reliable, and the results are compatible with zero for most of this region.
These conclusions hold for a wide range of momentum transfers employed in our analyses, from $-t=0.17$ GeV$2$ to $-t=2.29$ GeV$2$.
The reconstruction of the $x$-dependence is found to be robust for all cases.
The most troublesome cases appear when the Ioffe time range is limited, particularly for $\zmax=0.46$ fm and, to some extent, also $\zmax=0.65$ fm.
We identified problems of two kinds, summarized below.
The valence distribution $\Hval$ at small $\zmax$ suffers from enhanced correlations between the fitting parameters, which led us to adopt a constraint on the value of the prefactor of the fitting ansatz, $Nx^a(1-x)^b$, where $N\leq\Nmax$.
However, results, when this constraint is effective at $\zmax=0.65$ fm, are fully compatible with the ones at $\zmax=0.84$ fm without the constraint playing any role, establishing minor practical importance of the problem.
The other problem observed in the reconstruction appears for the distributions involving the imaginary parts of ITDs, again at small $\zmax$.
The insufficient range of Ioffe times, reaching below or only close to the maximum of the imaginary part, renders the fitting parameter $b$ zero for many bootstrap samples, thus failing to describe the decay of the GPDs to zero at $x=1$.
Luckily, the issue is comparatively mild at $\zmax=0.65$ fm and does not affect positive-$x$ GPDs $\Hbar$ and $\Ebar$ within our precision.
Overall, the two above problems are not severe at the present stage but call for improved lattice data in the future.
Most importantly, there is a need to extend the Ioffe time range at moderate values of $\zmax$, which requires precise data at larger nucleon boosts.
Obviously, this is difficult in view of the exponentially decaying signal-to-noise ratio when increasing $P_3$.

Overall, our work clearly demonstrates the feasibility of the pseudo-distribution approach to extracting GPDs from the lattice. 
In the future, several systematic effects need to be addressed in order to obtain fully reliable results.
The present analysis involved data at a single lattice spacing and a single volume -- thus, one needs to investigate and quantify discretization and finite volume effects.
Quark mass effects are also potentially contaminating the results.
As hinted at above, also HTEs need to be reduced, which can be achieved with data at larger nucleon boosts.
Finally, other ways of reconstructing the $x$-dependence should be explored, such as those based on machine learning.

\begin{acknowledgements}
We thank Yong Zhao for interesting discussions about the improvements of quasi- and pseudo-distribution approaches and their complementarity.
The work of S.~B. has been supported by the Laboratory Directed Research and Development program of Los Alamos National Laboratory under project number 20240738PRD1.
S.~B. has also received support from the U.~S. Department of Energy through the Los Alamos National Laboratory. Los Alamos National Laboratory is operated by Triad National Security, LLC, for the National Nuclear Security Administration of U.~S. Department of Energy (Contract No. 89233218CNA000001).
K.~C.\ and N.~N. are supported by the National Science Centre (Poland) grant OPUS no.\ 2021/43/B/ST2/00497.
M.~C. acknowledges financial support from the U.S. Department of Energy, Office of Nuclear Physics, Early Career Award under Grant No.\ DE-SC0020405.
M.C. acknowledges support from the project EXCELLENCE/0421/0043 “3D-Nucleon,” co- financed by the European Regional Development Fund and the Republic of Cyprus through the Research and Innovation Foundation.
The work of A.~M. has been supported by the National Science Foundation under grant number PHY-2110472.
M.~C. and A.~M. acknowledge partial support by the U.S. Department of Energy, Office of Science, Office of Nuclear Physics under the Quark-Gluon Tomography (QGT) Topical Collaboration with Award DE-SC0023646. 
F.~S.\ was funded by the NSFC and the Deutsche Forschungsgemeinschaft (DFG, German Research Foundation) through the funds provided to the Sino-German Collaborative Research Center TRR110 “Symmetries and the Emergence of Structure in QCD” (NSFC Grant No. 12070131001, DFG Project-ID 196253076 - TRR 110). 
This research was supported in part by PLGrid Infrastructure (Prometheus supercomputer at AGH Cyfronet in Cracow).
Computations were also partially performed at the Poznan Supercomputing and Networking Center (Eagle/Altair supercomputer), the Interdisciplinary Centre for Mathematical and Computational Modelling of the Warsaw University (Okeanos supercomputer), the Academic Computer Centre in Gda\'nsk (Tryton supercomputer) and facilities of the USQCD Collaboration, funded by the Office of Science of the U.S. Department of Energy. 
This research used resources of the National Energy Research Scientific Computing Center, a DOE Office of Science User Facility supported by the Office of Science of the U.S. Department of Energy under Contract No. DE-AC02-05CH11231 using NERSC award NP-ERCAP0022961.
The gauge configurations have been generated by the Extended Twisted Mass Collaboration on the KNL (A2) Partition of Marconi at CINECA, through the Prace project Pra13\_3304 ``SIMPHYS".
Inversions were performed using the DD-$\alpha$AMG solver~\cite{Frommer:2013fsa} with twisted mass support~\cite{Alexandrou:2016izb}. 
\end{acknowledgements}

\appendix
\section{Matched ITDs for all momentum transfers}
\begin{figure*}[t!]
    \centering
    \includegraphics[width=0.9\textwidth]{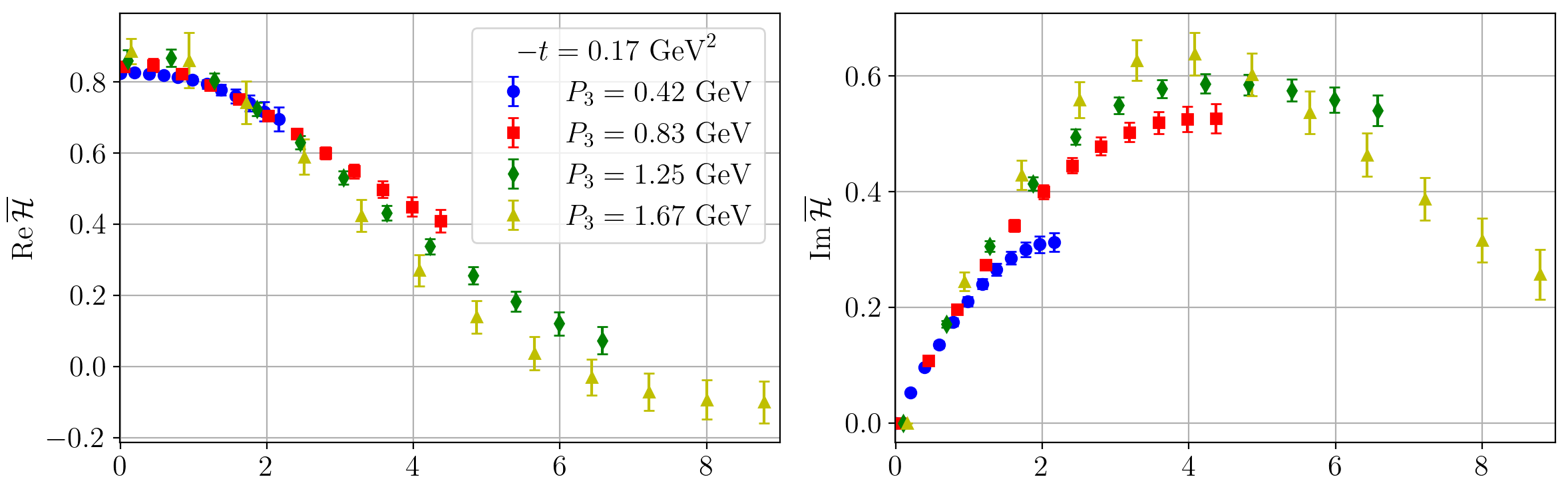}
    \includegraphics[width=0.9\textwidth]{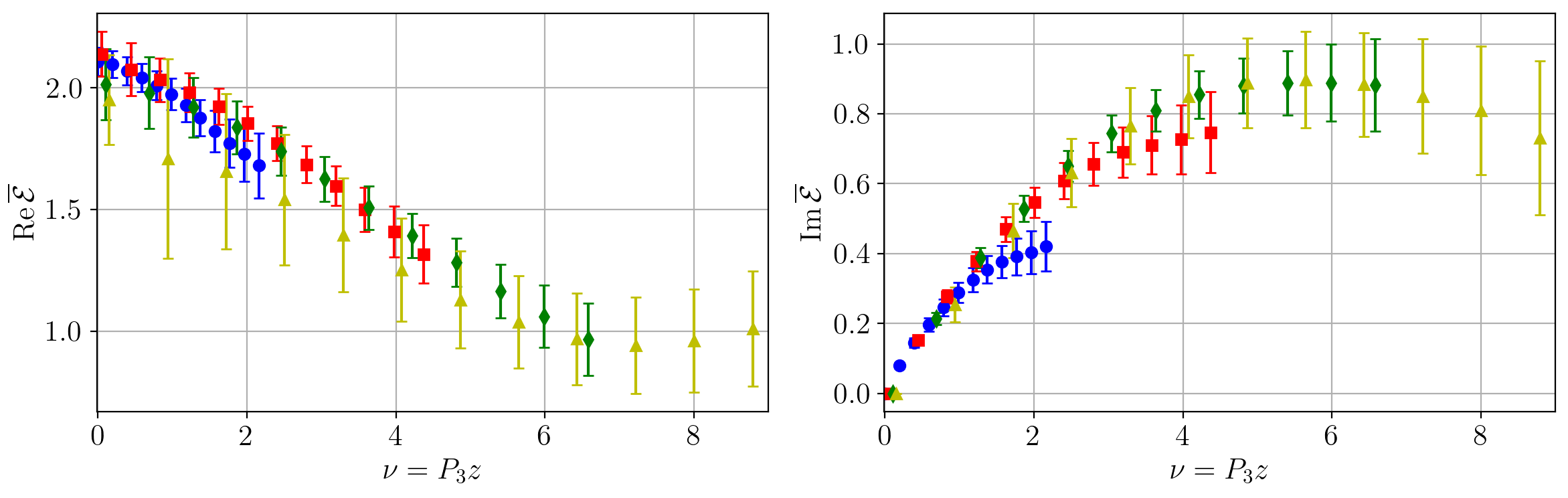}
    \caption{Matched $H$ (top) and $E$ (bottom) ITDs at $-t=0.17 $ GeV$^2$ with four nucleon momentum values, slightly shifted for better visibility. The left/right panels show the real/imaginary part.}
    \label{fig:matched_Q1}
\end{figure*}

\begin{figure*}[t!]
    \centering
    \includegraphics[width=0.9\textwidth]{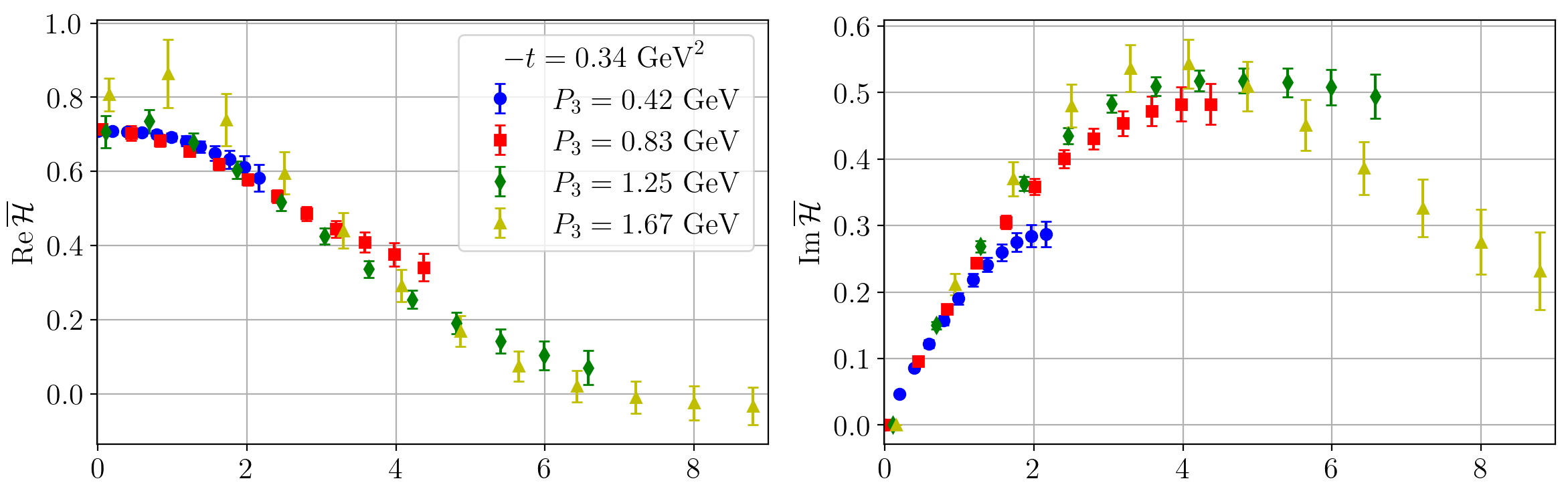}
    \includegraphics[width=0.9\textwidth]{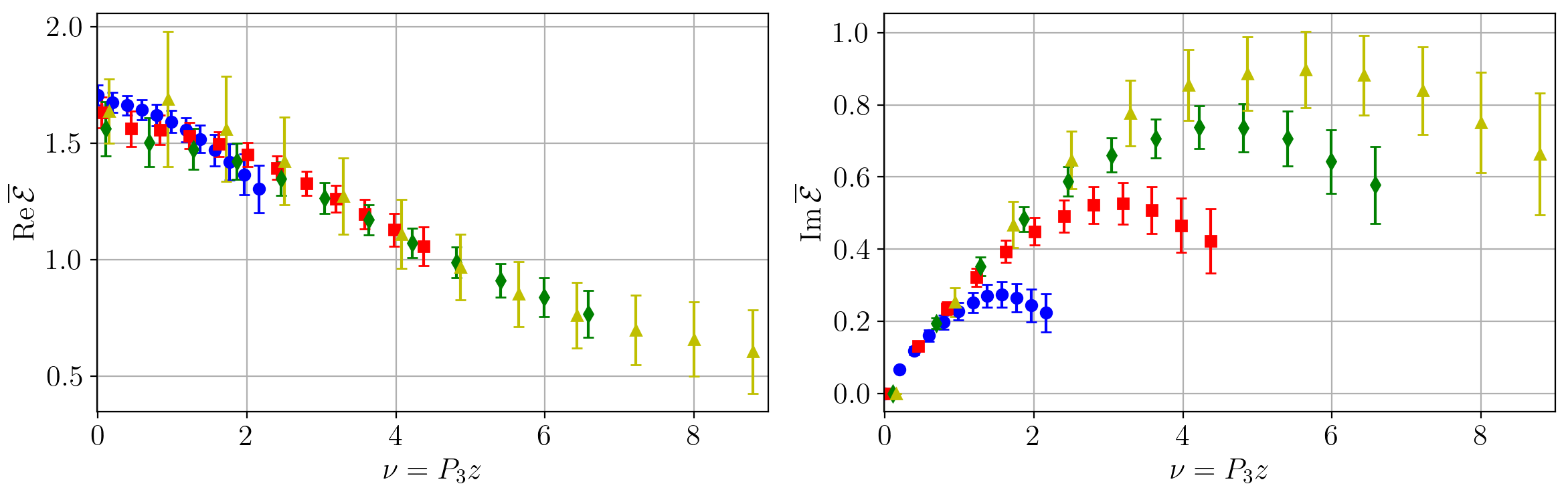}
    \caption{Matched $H$ (top) and $E$ (bottom) ITDs at $-t=0.34 $ GeV$^2$ with four nucleon momentum values, slightly shifted for better visibility. The left/right panels show the real/imaginary part.}
    \label{fig:matched_Q11}
\end{figure*}

\begin{figure*}[t!]
    \centering
    \includegraphics[width=0.9\textwidth]{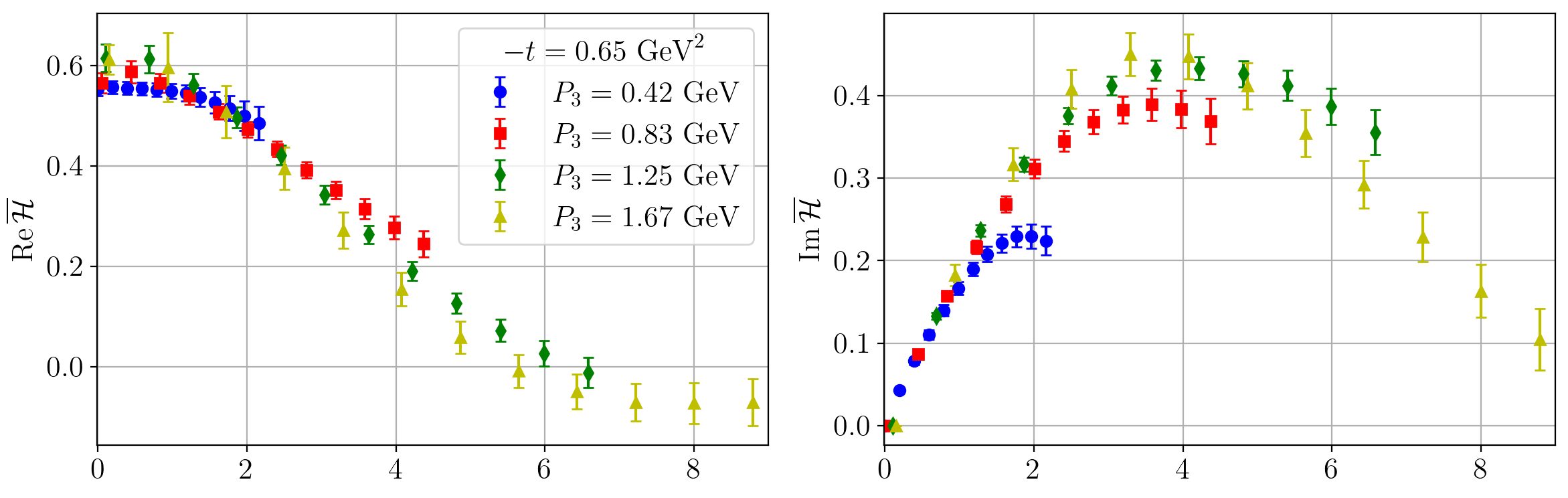}
    \includegraphics[width=0.9\textwidth]{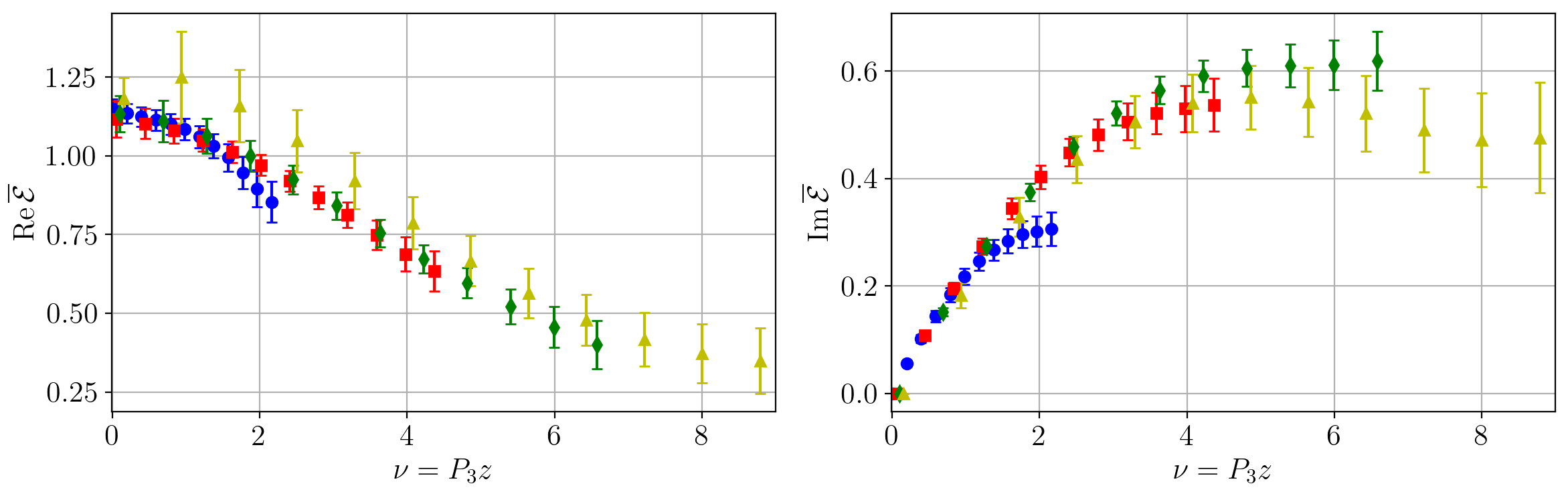}
    \caption{Matched $H$ (top) and $E$ (bottom) ITDs at $-t=0.65 $ GeV$^2$ with four nucleon momentum values, slightly shifted for better visibility. The left/right panels show the real/imaginary part.}
    \label{fig:matched_Q2a}
\end{figure*}

\begin{figure*}[t!]
    \centering
    \includegraphics[width=0.9\textwidth]{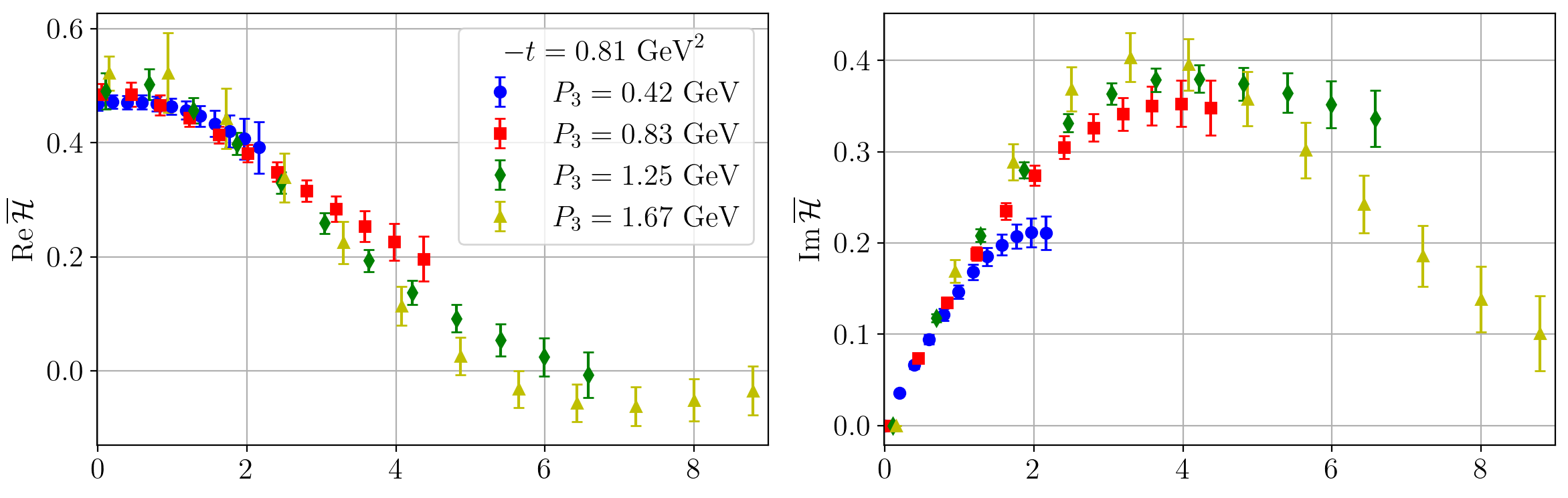}
    \includegraphics[width=0.9\textwidth]{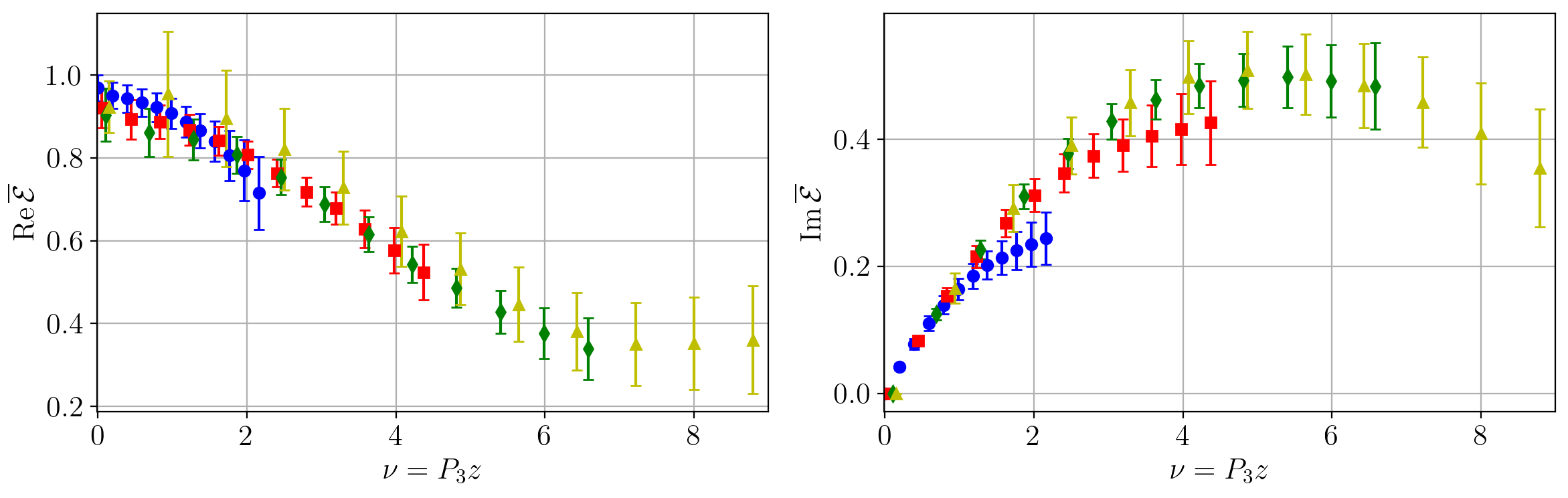}
    \caption{Matched $H$ (top) and $E$ (bottom) ITDs at $-t=0.81 $ GeV$^2$ with four nucleon momentum values, slightly shifted for better visibility. The left/right panels show the real/imaginary part.}
    \label{fig:matched_Q12}
\end{figure*}

\begin{figure*}[t!]
    \centering
    \includegraphics[width=0.9\textwidth]{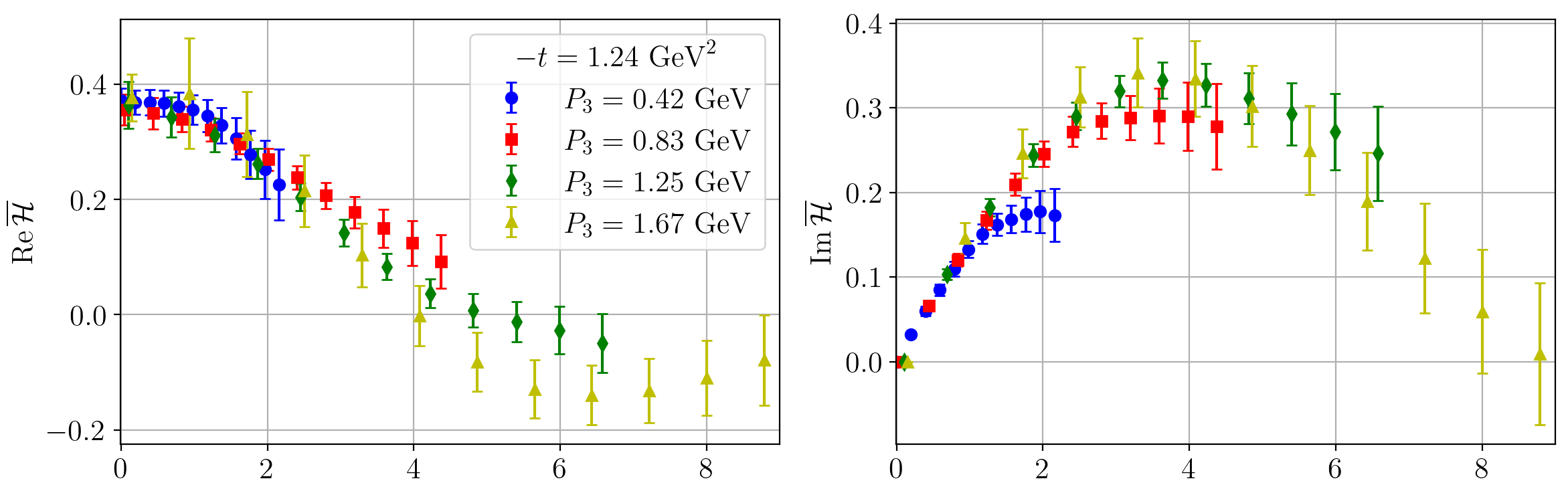}
    \includegraphics[width=0.9\textwidth]{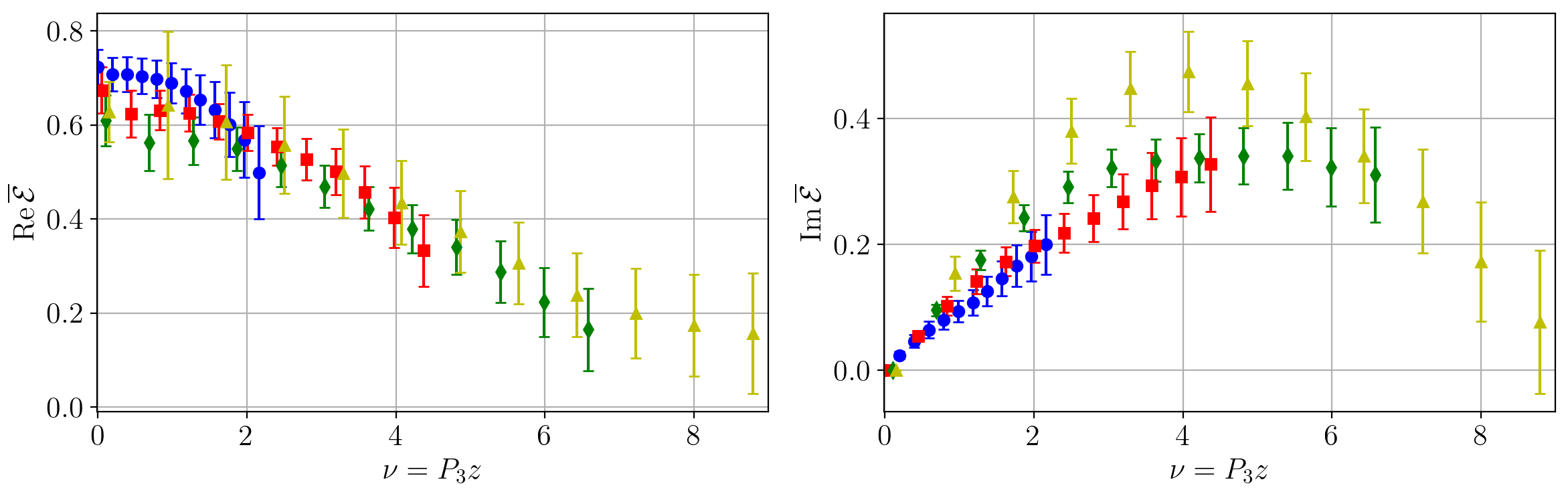}
    \caption{Matched $H$ (top) and $E$ (bottom) ITDs at $-t=1.24 $ GeV$^2$ with four nucleon momentum values, slightly shifted for better visibility. The left/right panels show the real/imaginary part.}
    \label{fig:matched_Q22}
\end{figure*}

\begin{figure*}[t!]
    \centering
    \includegraphics[width=0.9\textwidth]{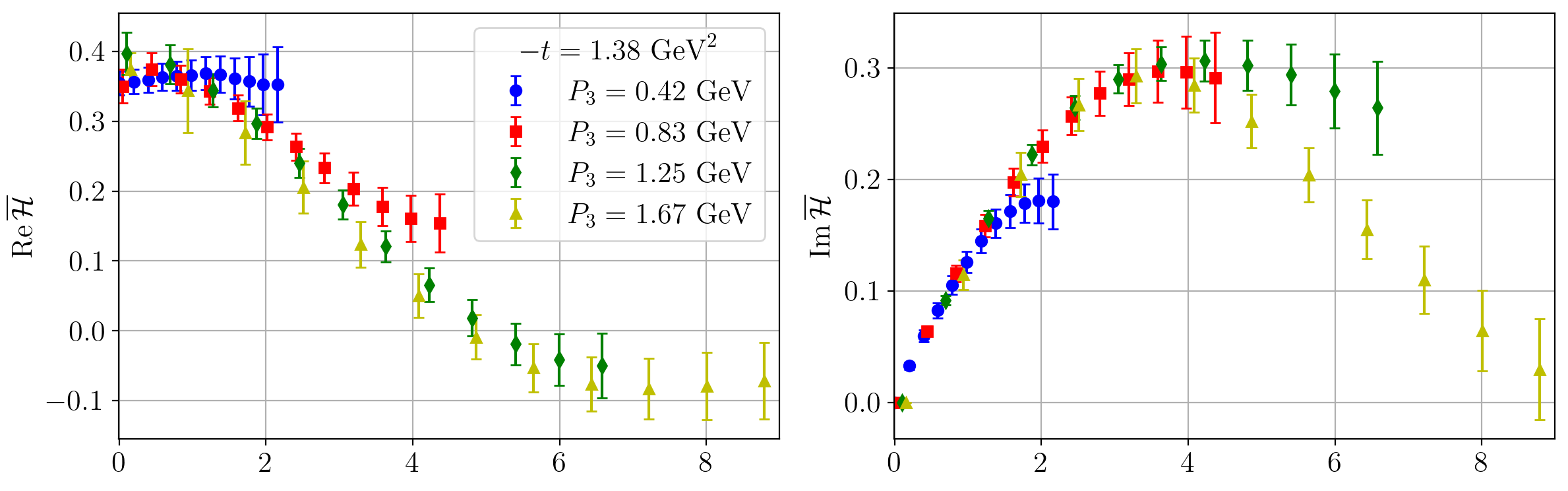}
    \includegraphics[width=0.9\textwidth]{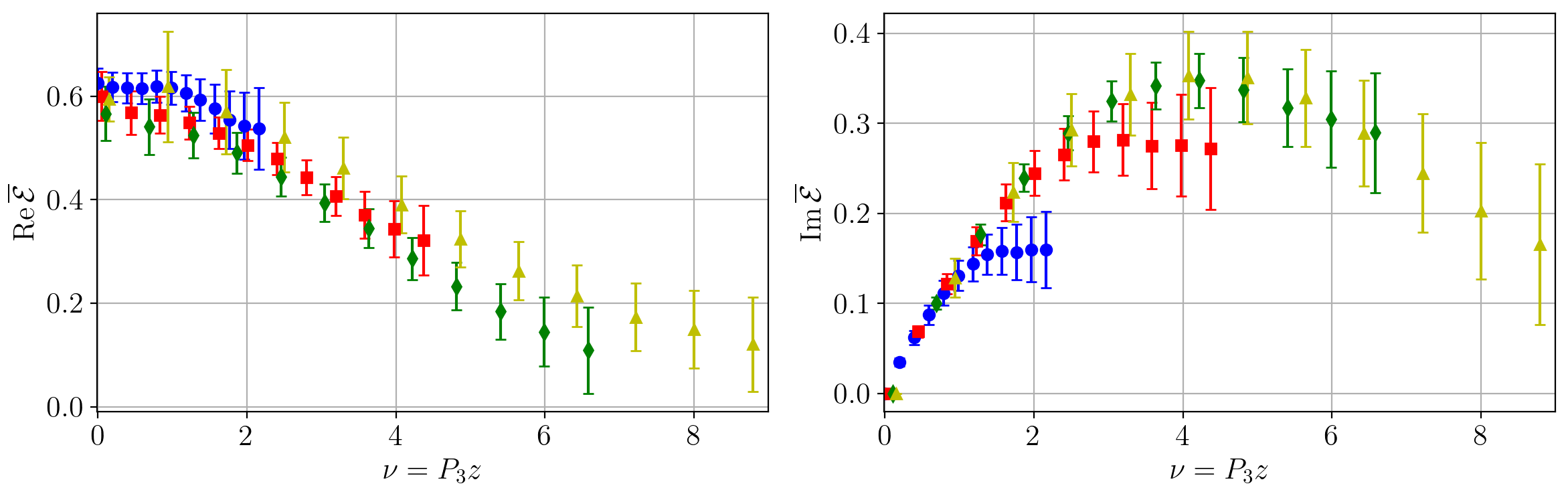}
    \caption{Matched $H$ (top) and $E$ (bottom) ITDs at $-t=1.38 $ GeV$^2$ with four nucleon momentum values, slightly shifted for better visibility. The left/right panels show the real/imaginary part.}
    \label{fig:matched_Q3}
\end{figure*}

\begin{figure*}[t!]
    \centering
    \includegraphics[width=0.9\textwidth]{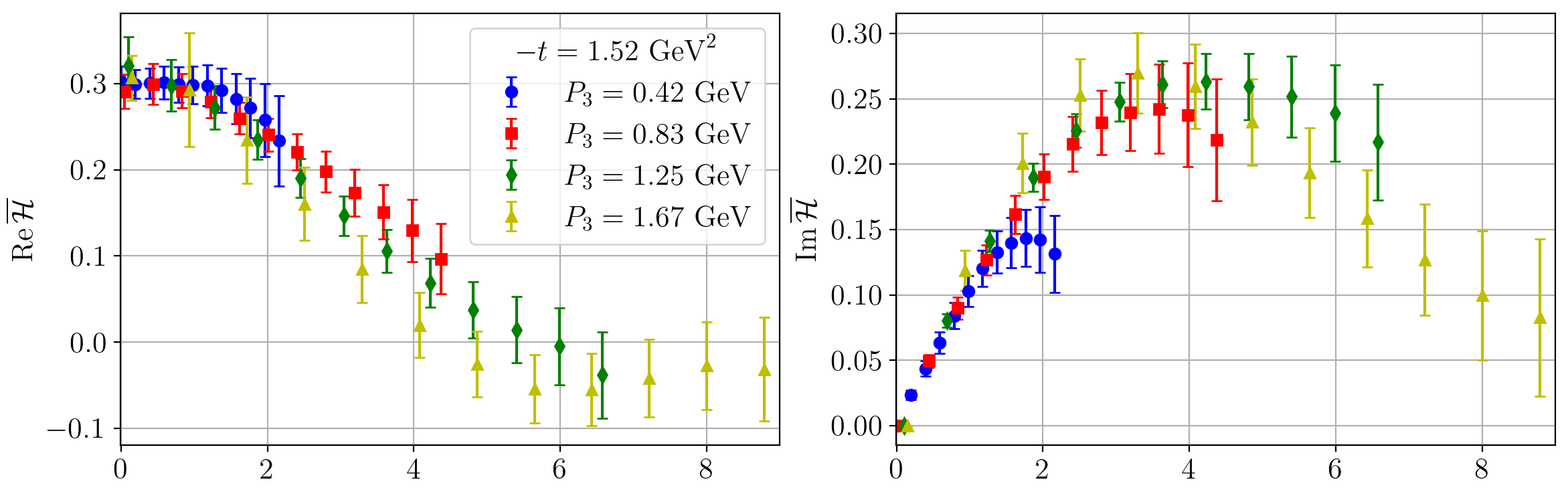}
    \includegraphics[width=0.9\textwidth]{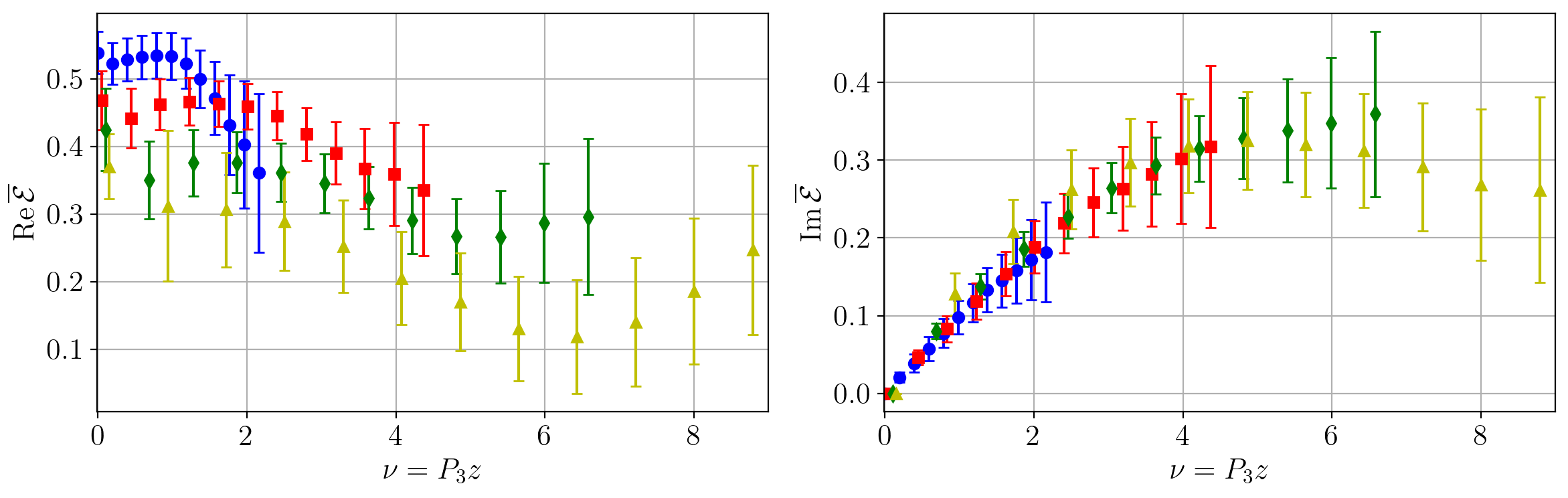}
    \caption{Matched $H$ (top) and $E$ (bottom) ITDs at $-t=1.52 $ GeV$^2$ with four nucleon momentum values, slightly shifted for better visibility. The left/right panels show the real/imaginary part.}
    \label{fig:matched_Q13}
\end{figure*}

\begin{figure*}[t!]
    \centering
    \includegraphics[width=0.9\textwidth]{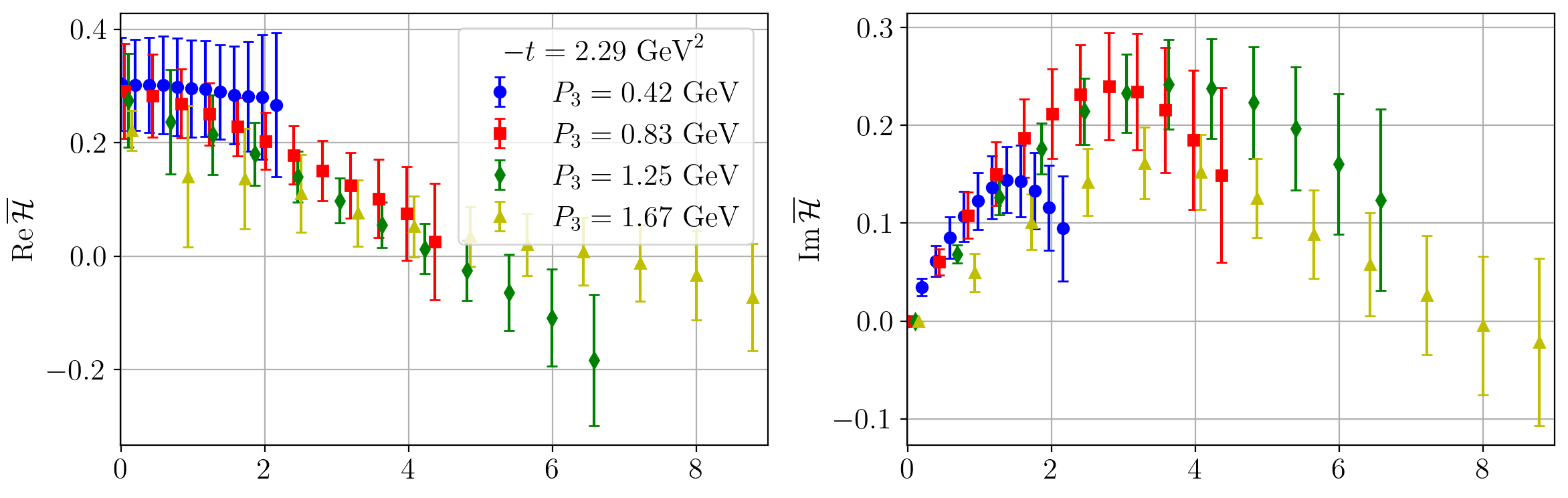}
    \includegraphics[width=0.9\textwidth]{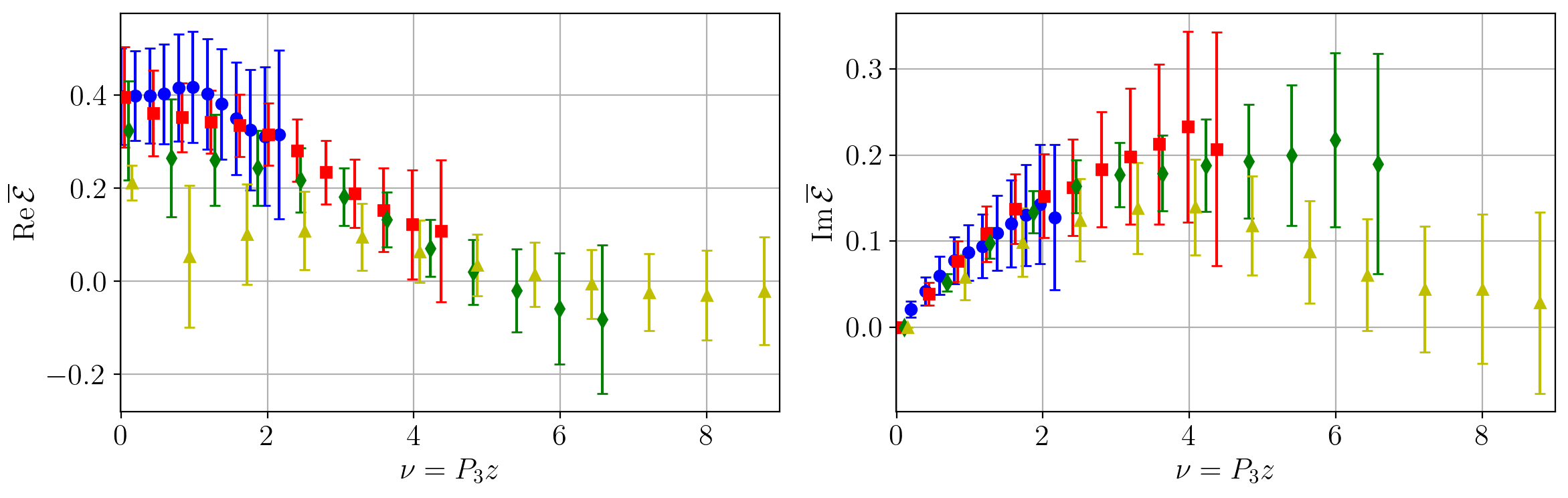}
    \caption{Matched $H$ (top) and $E$ (bottom) ITDs at $-t=2.29 $ GeV$^2$ with four nucleon momentum values, slightly shifted for better visibility. The left/right panels show the real/imaginary part.}
    \label{fig:matched_Q4}
\end{figure*}

In this appendix, we show matched ITDs for all employed momentum transfers, see Figs.~\ref{fig:matched_Q1}-\ref{fig:matched_Q4}.
In each case, we present the data separately for each nucleon boost $P_3$.
The inspection of all plots allows us to draw conclusions about the proper values of $\zmax$ that can be used in the reconstruction of the $x$-dependence of the GPDs, see the main body of the text.

\section{All full reconstructions}
\begin{figure*}[t!]
    \centering
    \includegraphics[width=1.0\textwidth]{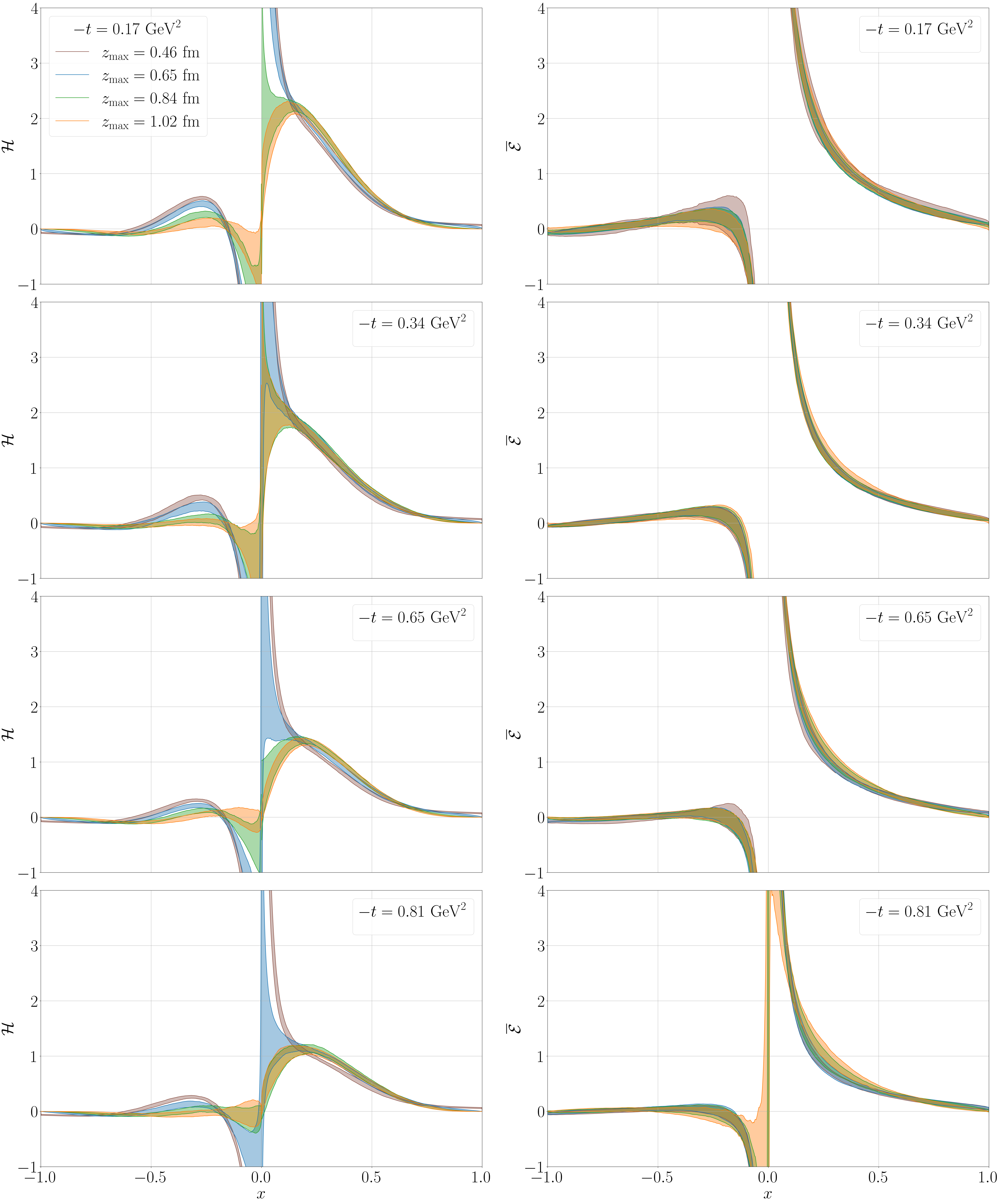}
    \caption{Full H (left) and E (right) reconstructed GPDs from$-t=0.17 $ GeV$^2$ to $-t=0.81 $ GeV$^2$}
    \label{fig:appen_reconstructions_1}
\end{figure*}

\begin{figure*}[t!]
    \centering
    \includegraphics[width=1.0\textwidth]{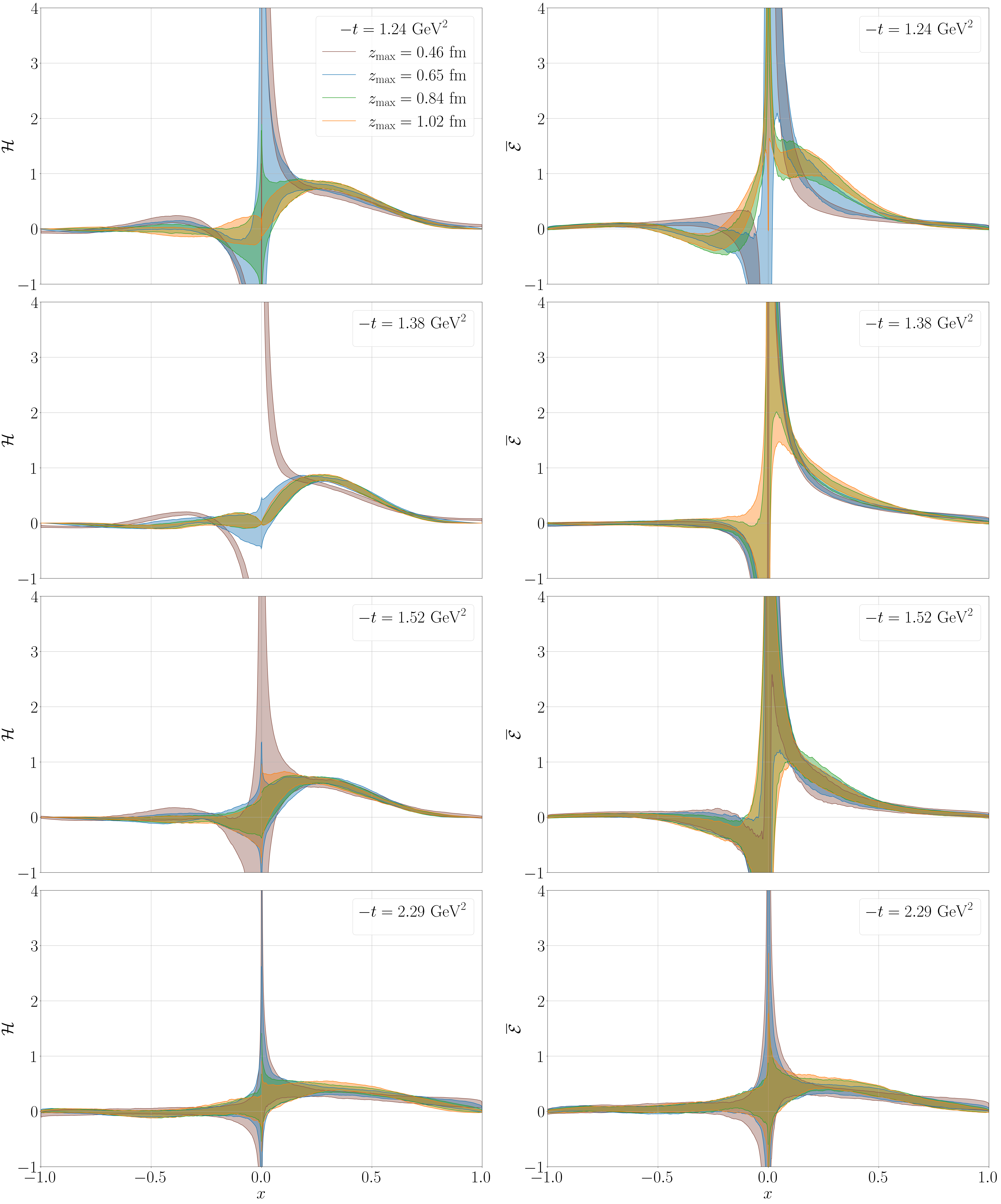}
    \caption{Full H (left) and E (right) reconstructed GPDs from$-t=1.24 $ GeV$^2$ to $-t=2.29 $ GeV$^2$}
    \label{fig:appen_reconstructions_2}
\end{figure*}

In Figs.~\ref{fig:appen_reconstructions_1}, \ref{fig:appen_reconstructions_2}, we present the $\Hbar$ and $\Ebar$ in momentum space, for all our momentum transfers and four values of $\zmax$.
As discussed in the main body of the text, all cases evince agreement between our preferred choices, $\zmax=0.65$ fm, and $\zmax=0.84$ fm, for $\Hbar(x\gtrsim0.1)$ and $\Ebar$ practically at all $x$.

\bibliography{references.bib}

\end{document}